\documentclass[aps, prd,twocolumn,superscriptaddress,nofootinbib,preprintnumbers]{revtex4-1}
\usepackage[utf8]{inputenc}
\usepackage[english]{babel}
\usepackage{natbib}
\usepackage{graphicx}
\usepackage{amsmath,amssymb}
\usepackage{hyperref}
\usepackage{braket}
\usepackage{float}
\usepackage[dvipsnames]{xcolor}
\usepackage{soul}
\usepackage{rotating}
\usepackage{multirow}
\usepackage{mathtools}
\usepackage{makecell}

\usepackage[caption = false]{subfig}
\usepackage{txfonts}
\usepackage[margin=0.75in]{geometry}
\usepackage{tabularx}

\hypersetup{
	colorlinks=true,
	linkcolor=red,
	citecolor=blue,
}

\usepackage{xcolor}

\DeclareMathAlphabet\mathbfcal{OMS}{cmsy}{b}{n}

\newcommand{\beq}{\begin{equation}}
\newcommand{\eeq}{\end{equation}}
\newcommand{\beqa}{\begin{eqnarray}}
\newcommand{\eeqa}{\end{eqnarray}}

\definecolor{darkgreen}{rgb}{0.0, 0.5, 0.0}
\definecolor{darkcyanxf}{RGB}{0.0, 139.0, 139.0}

\newcommand{\msun}[1]{$M_{\odot}$}

\newcommand{\avg}[1]{\langle #1 \rangle}

\newcommand{\mcal}{\textsc{metacalibration}}

\newcommand{\vest}{\mbox{\boldmath $e$}}

\newcommand{\vecg}{\mbox{\boldmath $\gamma$}}

\newcommand{\vecgest}{\mbox{\boldmath $\gamma^{\mathrm{est}}$}}

\defcitealias{Battaglia:2012}{B12}
\defcitealias{G20}{G20}
\newcommand{\G}{\citetalias{G20}}
\defcitealias{paperB}{paper II}

\usepackage{lineno}
\begin{document}
\title[Cosmology with mass map moments]{Dark Energy Survey Year 3 results:  cosmology with moments of weak lensing mass maps}

% \date{Last updated \today}

\label{firstpage}

\begin{abstract}
We present a cosmological analysis using the second and third moments of the weak lensing mass (convergence) maps from the first three years of data (Y3) data of the Dark Energy Survey (DES). The survey spans an effective area of 4139 square degrees and uses the images of over 100 million galaxies to reconstruct the convergence field. The second moment of the convergence as a function of smoothing scale contains information similar to standard shear 2-point statistics.  The third moment, or the skewness, contains additional non-Gaussian information. The data is analysed in the context of the $\Lambda$CDM model, varying 5 cosmological parameters and 19 nuisance parameters modelling astrophysical and measurement systematics. Our modelling of the observables is completely analytical, and has been tested with simulations in our previous methodology study. 
We obtain a 1.7\% measurement of the amplitude of fluctuations parameter $S_8\equiv \sigma_8 (\Omega_m/0.3)^{0.5} = 0.784\pm 0.013$. The measurements are shown to be internally consistent across redshift bins, angular scales, and between second and third moments. In particular, the measured third moment is consistent with the expectation of gravitational clustering under the $\Lambda$CDM model. The addition of the third moment improves the constraints on $S_8$ and $\Omega_{\rm m}$ by $\sim$15\% and $\sim$25\% compared to an analysis that only uses second moments.  We compare our results with {\it Planck} constraints from the Cosmic Microwave Background (CMB), finding a $2.2$ \textendash $2.8\sigma$ tension in the full parameter space, depending on the combination of moments considered. The third moment independently is in $2.8\sigma$ tension with {\it Planck}, and thus provides a cross-check on analyses of 2-point correlations.

%Our results are comparable with those of Kilo-Degree Survey (KiDS), the Hyper-Suprime Camera (HSC), and are found to  with Planck at the level of {X $\sigma$}.  
\end{abstract}

\preprint{DES-21-333}
\preprint{FERMILAB-PUB-21-333-AE}

\author{M.~Gatti}\email{marcogatti29@gmail.com}
\affiliation{Department of Physics and Astronomy, University of Pennsylvania, Philadelphia, PA 19104, USA}
\author{B. Jain}
\affiliation{Department of Physics and Astronomy, University of Pennsylvania, Philadelphia, PA 19104, USA}
\author{C. Chang}
\affiliation{ Department of Astronomy and Astrophysics, University of Chicago, Chicago, IL 60637, USA}
\affiliation{ Kavli Institute for Cosmological Physics, University of Chicago, Chicago, IL 60637, USA}
\author{M. Raveri}
\affiliation{Department of Physics and Astronomy, University of Pennsylvania, Philadelphia, PA 19104, USA}
\author{D. Z\"urcher}
\affiliation{Department of Physics, ETH Zurich, Wolfgang-Pauli-Strasse 16, CH-8093 Zurich, Switzerland}
\author{L. Secco}
\affiliation{Department of Physics and Astronomy, University of Pennsylvania, Philadelphia, PA 19104, USA}
\affiliation{Kavli Institute for Cosmological Physics, University of Chicago, Chicago, IL 60637, USA}
\author{L. Whiteway}
\affiliation{Department of Physics \& Astronomy, University College London, Gower Street, London, WC1E 6BT, UK}
\author{N. Jeffrey}
\affiliation{Department of Physics \& Astronomy, University College London, Gower Street, London, WC1E 6BT, UK}
\affiliation{Laboratoire de Physique de l'Ecole Normale Sup\'erieure, ENS, Universit\'e PSL, CNRS, Sorbonne Universit\'e, Universit\'e de Paris, Paris, France}
\author{C. Doux}
\affiliation{Department of Physics and Astronomy, University of Pennsylvania, Philadelphia, PA 19104, USA}
\author{T. Kacprzak}
\affiliation{Department of Physics, ETH Zurich, Wolfgang-Pauli-Strasse 16, CH-8093 Zurich, Switzerland}
% ...
\collaboration{DES Collaboration}

\maketitle

%\blfootnote{Affiliations are listed at the end of the paper.}
\setcounter{footnote}{1}

%\maketitle  %SWITCH TO PRD

%%%%%%%%%%%%%%%%%%%%%%%%%%%%%%%%%%%%%%%%%%%%%%%%%%

%%%%%%%%%%%%%%%%% BODY OF PAPER %%%%%%%%%%%%%%%%%%

\section{Introduction}
Gravitational lensing is one of the cleanest probes for studying the mass distribution in the Universe. General relativity predicts that the trajectories of photons emitted by distant galaxies are bent as they pass through regions of space-time perturbed by the mass distribution between the galaxy and the observer \citep{Einstein1936}. When studying the light emitted by distant galaxies, the level of distortion induced by the mass distribution of the Universe, or large scale structure (LSS), is usually small, at the percent level {\textendash} the regime of weak gravitational lensing. By collecting observations and measuring the shapes of many galaxies, statistical tools can be used to infer the mass distribution of the Universe
\citep{VanWaerbeke2013, Vikram2015, Chang2015, Liu2015, Chang2018, Oguri2018,y3-massmapping}. Ongoing and future surveys (DES, \citealt{DES2016}; Kilo-Degree Survey KIDS, \citealt{Kuijken2015}; Hyper Suprime-Cam HSC, \citealt{Aihara2018}; {Vera C. Rubin Observatory's Legacy Survey}, \citealt{Abell2009}; Euclid, \citealt{Laureijs2011}) are currently measuring (or planning to measure) the shapes of tens to hundreds of millions of galaxies, spanning thousands of square degrees of the sky. In particular, DES recently measured 100 million galaxies spanning $\sim$5000 square degrees of the southern hemisphere \citep{y3-shapecatalog}, and created the largest map of the mass distribution of the universe from a galaxy survey \citep{y3-massmapping}.

For a given cosmological model, the statistical properties of the mass distribution can be predicted over time. Second-order statistics, such as correlation functions \citep{Troxel2017, Hildebrandt2017, Hikage2019,y3-cosmicshear1,y3-cosmicshear2}, the power spectrum \citep{Hamana2020}, or the wavelet-like COSEBIs (complete orthogonal sets of E/B-integrals) \citep{Asgari2021}, are standard tools used to exploit the Gaussian information of the mass maps. However, a weak lensing mass map contains information beyond that captured by second order statistics, as its probability distribution function (PDF) has non-Gaussian features induced by gravitational evolution. In particular, the PDF of the mass distribution in the late Universe is {roughly} approximated by a log-normal \citep{Hubble1934,Coles1991,Wild2005}, a fact that has also been investigated for the weak lensing convergence field with DES data \citep{Clerkin2017}. 

Higher order statistics are appealing, as their use can improve constraints on cosmological parameters \citep{Vafaei2010,Petri2015,G20,Zuercher2021} over standard 2-point statistics, or can help discriminate between extended models such as modified gravity theories \citep{Cardone2013,Peel2018}. Numerous tools have been developed to extract the non-Gaussian information from mass maps. Higher order statistics commonly used with weak lensing include shear peak statistics \citep{Dietrich2010, Kratochvil2010, Liu2015, Kacprzak2016, Martinet2018, Peel2018, Shan2018, ajani_peaks,Zuercher2021}, higher moments of the weak lensing convergence field  \citep{VanWaerbeke2013,Petri2015,Vicinanza2016,Chang2018,Vicinanza2018,Peel2018, G20}, three-point correlation functions or bispectra \citep{Takada2003, Takada2004, Semboloni2011, Fu2014}, Minkowski functionals \citep{Kratochvil2012,Petri2015,Vicinanza2019,Parroni2020}, and machine-learning methods \citep{Ribli2018, Fluri2018,Fluri2019,jeffrey_lfi}. Many of these have recently been applied to data \citep{Liu2015, Kacprzak2016,Martinet2018,Fluri2019,jeffrey_lfi}, often performing well in terms of cosmological constraints. The theoretical modelling of some of these statistics is often complex, and large suites of N-body simulations, spanning the parameter space considered in the analysis, are used to model the observables.

This work focuses on the use of second and third moments of weak lensing mass maps to constrain cosmology. Moments have been studied in the past, and have been measured both in data and simulations \citep{Jain1997,Gaztanaga1998,Fosalba2008,Vafaei2010, VanWaerbeke2013,Petri2015,Pujol2016,Chang2018}, although they have not been used to place constraints on cosmological parameters. Tests using simulations have shown improvements to cosmological constraints arising from using moments of order higher than second \citep{Vafaei2010,Petri2015,G20}. The methodology used in this paper has been developed and tested using simulations in a companion paper, \cite{G20} (hereafter \G{}). Although the methodology can be applied to any dataset, the analysis in \G{} was geared towards the first three years of data of DES. The modelling of second and third moments developed in \G{} is based on theoretical predictions, therefore it does not rely on large suites of N-body simulations (though the predictions are tested against simulations); moreover, observational systematics errors such as photometric redshift uncertainties or intrinsic alignment are modelled and marginalised during the analysis. This work applies that methodology to the first three years of data (Y3) from DES, presenting the cosmological constraints, discussing a number of observational systematic null tests, and comparing the results with constraints from other DES Y3 probes and/or external datasets (e.g. \emph{Planck}).

The paper is organised as follows: \S \ref{sect:data} describes the data and simulations used in this work; \S \ref{sect:model} provides a short description of the theoretical modelling of the observables used in the analysis (the second and third moments of the convergence field); \S \ref{sect:like_tt} describes the likelihood and the covariance used in the cosmological parameter inference, and discusses the priors adopted in the analysis; \S \ref{sect:unblinding} summarises the pre-unblinding tests; \S \ref{sect:results} presents the cosmological results, along with a number of internal consistency tests and comparisons with results from other DES analyses or from analyses using external data sets; \S \ref{sect:summary} summarises our findings.

\section{Data and simulations}\label{sect:data}
\begin{figure}
\includegraphics[width=0.45\textwidth]{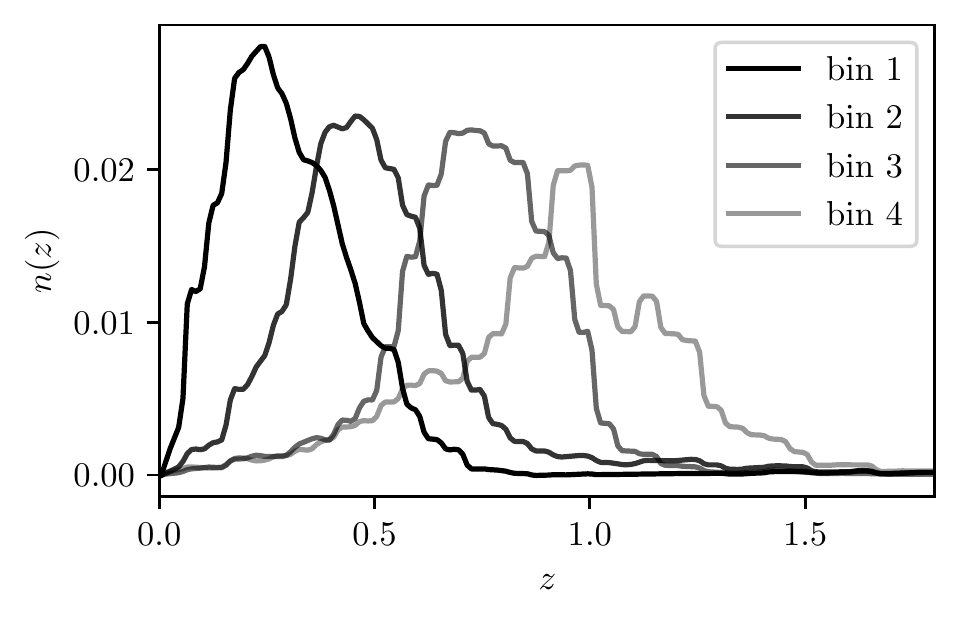}
\caption{Redshift distributions as estimated in data for the four DES Y3 tomographic bins \citep{y3-sompz}.}
\label{fig:nz_comp}
\end{figure}
\subsection{Data}
The main goal of our analysis is to measure second and third moments of the convergence field and use them to estimate cosmological parameters. To this aim, we use the weak lensing catalogue from the first three years (Y3) of the DES \citep{y3-shapecatalog}. 

DES \citep{DES2016} is a six-year survey that spans $\sim 5000~\mathrm{\deg}^2$ of the southern hemisphere. Images have been taken in $grizY$ filters by the $570$~megapixel Dark Energy Camera \citep[DECam,][]{Flaugher2015}, mounted on the Cerro Tololo Inter-American Observatory (CTIO) four-meter Blanco telescope in Chile.  The raw images were processed by the DES Data Management (DESDM) team  \citep{Sevilla2011,Morganson2018,DES_DR1}. Full details about the image processing are provided in \cite{,Morganson2018,DES_DR1}.

The DES Y3 weak lensing sample is described in detail in \cite{y3-shapecatalog} and builds upon the Y3 Gold catalogue \citep{y3-gold}. It is created using the \textsc{METACALIBRATION} algorithm \citep{HuffMcal2017, SheldonMcal2017}, which infers the galaxy ellipticities starting from noisy images of the detected objects in the \textit{r, i, z} bands. The \textsc{METACALIBRATION} algorithm was used previously in the DES Y1 analysis \citep{ZuntzY1}. \textsc{METACALIBRATION} uses an approximate estimator of the shear field and self-calibrates it using the response of the estimator to shear as well as to selection effects. A number of selection cuts are designed to remove objects in the catalogue potentially affected by systematic effects \citep{y3-shapecatalog}. An inverse variance weight is also assigned to galaxies in order to enhance the overall signal-to-noise. The final DES Y3 shear catalogue has 100,204,026 objects, with a weighted $n_{\rm eff}=5.59$~galaxies~arcmin$^{-2}$, over an effective area of 4139 square degrees.

Although the \textsc{METACALIBRATION} self-calibration procedure removes most of the multiplicative bias, for the DES Y3 weak lensing sample there is a known residual additional multiplicative bias at the level of $2$ or $3$ per cent \citep{y3-imagesims}. This residual bias stems mostly from a shear-redshift-dependent detection bias due to blending of galaxy images, for which the \textsc{METACALIBRATION} implementation adopted in DES Y3 is unable to account \citep{SheldonMetadetect2019}. We do not calibrate for this factor at the catalogue level, but we do marginalise over it in the analysis. In \cite{y3-shapecatalog} the weak lensing sample has also been tested for additive biases (e.g. due to point-spread-function residuals). In particular, the catalogue is characterised by a non-zero mean shear whose origin is unknown and which is subtracted at the catalogue level before performing any analysis. 

The weak lensing sample is divided into four tomographic bins of roughly equal number density using the SOMPZ method \citep{y3-sompz}; SOMPZ, in combination with constraints from clustering redshifts \citep{y3-sourcewz}, also provides redshift distribution estimates {(see Fig. \ref{fig:nz_comp})}. The $n(z)$'s are further tweaked to take into account the redshift-dependent effects of blending \citep{y3-imagesims}. During the cosmological analysis, additional constraints on the redshift distributions are provided by shear ratios \citep{y3-shearratio}. Shear ratios are ratios of small-scale galaxy-galaxy lensing measurements obtained using different source samples (in this case, different weak lensing tomographic bins) and a common lens sample. Not only do they improve constraints on redshift distributions, but they also help constraining both intrinsic alignment parameters and cosmological parameters.

A two-stage blinding scheme was implemented for all DES Y3 cosmological analyses in order to avoid intentional or unintentional confirmation bias. First, the weak lensing sample was blinded by means of a multiplicative factor, in a fashion similar to what was adopted in the Y1 analysis \citep{ZuntzY1}. In particular, the ellipticities {e} of the catalogue were transformed via $|${{\boldmath $\eta$}}$|$ $\equiv 2 {\rm arctanh} |\textbf{e}| \rightarrow f$ $|${{\boldmath $\eta$}}$ |$, with a hidden value 0.9$<f<$1.1. After all the catalogue and map-based systematic tests were passed \citep{y3-shapecatalog,y3-massmapping}, the hidden value was revealed and the catalogue unblinded. This work ignores this first level of blinding, as when we started analysing the DES Y3 data the catalogue had already been validated and unblinded. The second level of blinding, which follows the work of \cite{y3-blinding}, was applied to the summary statistics under examination; in this case, it was applied to the measured second and third moments of the convergence field. 
{In particular, to each element $\hat{d}_i$ of the observable vector (i.e., both second and third moments), the following transformation was applied}:
\begin{equation}
\hat{d}_i^{\rm blinded} = \hat{d}_i + d_i\left( \Theta_{\rm ref} +\Delta \Theta \right) -d_i\left( \Theta_{\rm ref}  \right).
\end{equation}
{In the above equation, $d_i$ is a theory data vector computed at a given cosmology $\Theta$; $\Theta_{\rm ref}$ is a fiducial cosmology (we used the DES Y1 3x2pt cosmology from \citep{desy1_results}), and $\Delta \Theta$ is a blind shift in the cosmological parameters (drawn from a distribution three times larger than the DES Y1 3x2pt posterior).}

A number of systematic tests were performed on blinded data vectors (see \S~\ref{sect:unblinding}) before proceeding to inspect the unblinded cosmological results.

\subsection{Simulations}

Covariance matrices for our measurement are generated: a) for our fiducial covariance, using lognormal realisations from \textsc{FLASK} \citep{Xavier2016}, b) for testing, using the N-body simulation hereafter called `T17' \cite{Takahashi2017}, and c) also for testing, using the N-body simulation \textsc{PKDGRAV} \citep{potter2017pkdgrav3}. Moreover, both T17 and PKDGRAV simulations are used to validate our modelling (Appendix~\ref{sect:validation}). Such a validation was already performed in \G{}, but using only T17 simulations; we repeat that here, for both sets of N-body simulations, with updated analysis choices.

\subsubsection{\textsc{FLASK} realisations}\label{sect:FLASK}
We use the \textsc{FLASK} (Full-sky Lognormal Astro-fields Simulation Kit) software \citep{Xavier2016} to rapidly generate full-sky, log-normal realisations of the convergence field. \textsc{FLASK} assumes the convergence field to be described by a zero-mean shifted log-normal distribution, where the parameters of the log-normal probability distribution function (PDF) are chosen to match the variance and skewness of the input. We use here the 1000 independent \textsc{FLASK} realisations produced for the validation of the DES Y3 3$\times$2pt covariance \citep{y3-covariances}. The lognormal approximation for the covariance has been shown to be sufficient to not bias the recovery of the cosmological parameters in \G{}. The cosmological parameters of the input power spectra used for the \textsc{FLASK} realisations are $\Omega_{\rm m} = 0.3$, $\sigma_8 = 0.82355$, $\Omega_{\rm b} = 0.048$, $n_{\rm s} = 0.97 $, $h_{100} = 0.69$ and $\Omega_{\nu}h = 0.00083$. We also assumed DES Y3 redshift distributions. The \textsc{FLASK} convergence realisations were provided in maps using the Hierarchical Equal Area isoLatitude Pixelation scheme (\textsc{HEALPIX}, \citep{GORSKI2005}) with resolution \textsc{NSIDE} = 4096. In order to create a simulated weak lensing galaxy catalogue, we then used the position, shape noise (obtained by randomly rotating each galaxy), and weight of the galaxies of the fiducial DES Y3 weak lensing catalogue; depending on the position of each individual galaxy, we sampled the simulated shear maps and added shape noise accordingly. This procedure allows us to generate 1000 independent simulated shear catalogues.

\subsubsection{T17 simulations}

The first set of N-body simulations used in this work are the T17 \citep{Takahashi2017} simulations. The set consists of 108 full-sky lensing convergence and shear maps, spanning a wide redshift range (between z = 0.05 and 5.3) at intervals of 150 $h^{ - 1}$ Mpc comoving distance. The N-body simulations assume a WMAP 9 cosmology ($\Omega_{\rm m} = 0.279$, $\sigma_8 = 0.82$, $\Omega_{\rm b} = 0.046$, $n_{\rm s} = 0.97 $, $h = 0.7$), and were run using L-GADGET2 \citep{springel2005}. Initial conditions were generated using 2LPTIC \citep{Crocce2006}.

The simulations begin with 14 boxes in steps of 450 $h^{ - 1}$ Mpc, with total side lengths of L = 450, 900, 1350, ..., 6300 $h^{ - 1}$ Mpc. There are six independent copies at each box size and 2048$^3$ particles per box. Lens plane snapshots are taken at intervals of 150 $h^{ - 1}$ Mpc comoving distance. The expected accuracy of the average matter power spectra from the simulations (compared to predictions from the revised \textsc{HALOFIT}, \citep{Takahashi2012}) is  within $5$ per cent for $k < 1$ $h$ Mpc$^{ - 1}$ at $z < 1$, for $k < 0.8$ $h$ Mpc$^{ - 1}$ at $z < 3$, and for $k < 0.5$ $h$ Mpc$^{ - 1}$ at $z < 7$ \citep{Takahashi2017}. Weak lensing quantities for each simulation were estimated using the multiple plane ray-tracing algorithm \textsc{GRayTrix} \citep{Hamana2015}, and shear and convergence \textsc{HEALPIX} maps with resolution \textsc{NSIDE} = 4096 are provided.

For each of the 108 simulations, we cut out four independent (i.e., non-overlapping) regions corresponding to the DES Y3 footprint. We then stacked the convergence and shear snapshots at different redshift to produce convergence and shear maps for the four weak lensing tomographic bins. This gave us 432 independent realisations of the shear field for each tomographic bin. In order to create a simulated weak lensing galaxy catalogue, we used the position, shape noise (obtained by randomly rotating each galaxy), and weight of the galaxies of the fiducial DES Y3 weak lensing catalogue; depending on the position of each individual galaxy, we sampled the simulated shear maps and added shape noise accordingly. We ended up with 432 independent simulated shear catalogues from T17 simulations. %Last, in order to increase the number of realisations, we augmented the catalogs generating 432 additional different realisations of shape noise (similar to the approach used in \citealt{Zuercher2021b}), for a final number of 864 simulated catalogs.

\subsubsection{PKDGRAV simulations}

The second set of N-body simulations is the DarkGridV1 suite, produced using the \textsc{PKDGRAV3} code \citep{potter2017pkdgrav3} and described in detail in \cite{Zuercher2021,Zuercher2021b}. In particular, we use 50 independent realisations at the fixed cosmology $\Omega_{\rm m} = 0.26$, $\sigma_8 = 0.84$, $\Omega_{\rm b} = 0.0493$, $n_{\rm s} = 0.9649 $, $h = 0.673$. {All simulations include three massive neutrino species with a mass of $m_{\nu}=0.02$ eV per species \cite{Zuercher2021b}}. The simulations were obtained using 14 replicated boxes in each direction ($14^3$ replicas in total) so as to span the redshift interval between z = 0 and z = 3. Each individual box contains $768^3$ particles and has a side-length of 900 $h^{ - 1}$ Mpc. Such a configuration is known to yield a field variance that is too small at very large scales \citep{Fluri2019}; however, such scales are not considered in this work. For each simulation, lens planes are provided at $\sim 87$ redshifts between $z=3.0$ and $z=0.0$, equally spaced in proper time. Lensing quantities (shear and convergence) were obtained under the Born approximation. For each simulation, we cut out four independent DES Y3 footprints and thereby created 200 independent catalogues in a fashion similar to the T17 simulations.

\section{theoretical modelling}
\label{sect:model}

We provide here a short summary of the theoretical modelling of our observables. Further details are provided in \G{}. 

Our cosmological analysis relies on the theoretical modelling of the second and third moments of convergence maps, which is based on cosmological perturbation theory \citep{VanWaerbeke2001,Scoccimarro2001,Bernardeau2002}. Consider three  convergence maps, obtained from different tomographic bins (labelled $i$, $j$, $k$) of the weak lensing catalogue (the equations below apply for more tomographic bins as well, taken two or three at a time). The maps are smoothed by a top-hat filter of smoothing length $\theta_0$. The second and third moments are then given by:
\begin{linenomath} % See https://tex.stackexchange.com/questions/461186
\begin{multline}
\label{eq:ksmM2}
\langle \kappa^2_{\theta_0}\rangle^{i,j, {\rm EE/BB}}  = \int d\chi \frac{q^i(\chi) q^j(\chi)}{\chi^2} \times \\ \sum_{\ell} \frac{2\ell+1}{4 \pi} f_{\ell}^{-1} W_{\ell}(\theta_0)^2 \sum_{\ell'}  M_{\ell \ell'}^{\rm EE/BB ,EE}P_{\mathrm{NL}} (\ell'/\chi,\chi) F_{\ell'}^2 f_{\ell'},
\end{multline}
\end{linenomath}
\begin{linenomath} % See https://tex.stackexchange.com/questions/461186
\begin{multline}
\label{eq:ksmM3}
\langle \kappa^3_{\theta_0}\rangle^{i,j,k,EE/BB}  = \int d\chi \frac{q^i(\chi) q^j(\chi)q^k(\chi)}{\chi^4} \times \\
S_3\left[ \sum_{\ell} \frac{2\ell+1}{4 \pi} f_{\ell}^{-1} W_{\ell}(\theta_0)^2 \sum_{\ell'}  M_{\ell \ell'}^{EE/BB ,EE}P_{\mathrm{NL}} (\ell'/\chi,\chi) F_{\ell'}^2 f_{\ell'}\right]^2.
\end{multline}
\end{linenomath}
Here the lensing kernel term $q^i$ is given by:
\begin{equation}
\label{eq:wlk}
q^i(\chi) = \frac{3H_0^2\Omega_{\rm m}}{2c^2}\frac{\chi}{a(\chi)} \int_{\chi}^{\chi_h} d\chi' n^i(z(\chi')) dz/d\chi'\frac{\chi'-\chi}{\chi},
\end{equation}
where $\chi$ is comoving distance, $\chi_h$ is the horizon comoving distance, $H_0$ the Hubble constant at the present time, $c$ the speed of light,  $n^i(z)$ the normalised redshift distribution of a given tomographic bin, and $a(\chi)$ the scale factor. Furthermore, in Eqs. \ref{eq:ksmM2} and \ref{eq:ksmM3}, $W_{\ell}(\theta_0)$ represents the top-hat filter of smoothing length $\theta_0$ in harmonic space, defined as:

\begin{equation}
\label{eq:filter}
W_{\ell}(\theta_0) = \frac{P_{\ell-1}({\rm cos}(\theta_0))-P_{\ell+1}({\rm cos}(\theta_0))}{(2\ell+1)(1-{\rm cos}(\theta_0))},
\end{equation}
where $P_{\ell}$ are Legendre polynomials of order $\ell$. Other terms in Eqs. \ref{eq:ksmM2} and \ref{eq:ksmM3} are: the mode-coupling matrixes $M^{\rm EE/BB,EE}_{\ell \ell'}$ (e.g., \cite{Brown2005,Hikage2011}, or Appendix B of \G{}),  which take into account the effects of masking; the factor $f_{\ell} = [(\ell+2)(\ell-1)]/[\ell(\ell+1)]$, which accounts for the mode-coupling matrix being applied to the shear field rather than to the convergence field directly; the pixel window function $F_{\ell}$; the non linear power spectrum $P_{\rm NL} (\ell/\chi,\chi)$, modelled using \textsc{HALOFIT} as detailed in \cite{Takahashi2014}; and the reduced skewness parameter $S_3$. The full derivation of $S_3$ is provided in Appendix A of \G{}, where it is  evaluated to leading order in perturbation theory with the addition of a small-scale refinement (in the form of analytical fitting formulae) based on N-body simulations from \cite{Scoccimarro2001}. In \G{} we determined the range (i.e., angular scales and redshift interval) of validity of our $S_3$ model to ensure that modelling uncertainties will not affect our cosmological analysis. {Since \G{}, however, some of our analysis choices changed; in particular, we updated the redshift distributions, the catalogue shape noise, the measurement covariance, and the nuisance parameters priors, to reflect the updates in the DES Y3 data and modelling. Moreover, we include galaxy-galaxy lensing information from small scales in the form of shear ratios. Therefore, we repeated the modelling validation performed in \G{} in Appendix~\ref{sect:validation}, using our updated analysis choices.} Moreover, we validated our modelling on two different sets of N-body simulations (T17 and PKDGRAV). 

\subsection{Systematic effects}\label{systematics}

We model astrophysical and measurement systematic effects through nuisance parameters, over which we marginalise when estimating the cosmological parameters. Here is a short description of the nuisance parameters used in this work; priors are summarised in Table~\ref{table1}.

\begin{table}
%\tiny
\caption {Cosmological and nuisance parameters. The cosmological parameters considered are $\Omega_{\rm m}$, $\sigma_8$, $\Omega_{\rm b}$,  $n_{\rm s}$ and $h$. The `calibration' nuisance parameters are the multiplicative shear biases $m_i$ and the mean photometric uncertainties of the weak lensing samples $\Delta z_i$ , {where the index $i$ runs over the tomographic bins}. The `astrophysical' nuisance parameters $A_{\rm IA,0}$ and $\alpha_{\rm IA}$ describe the intrinsic alignment model. The parameters $\Delta z^{\rm lens}_{i}$, $\delta^{\rm lens}_{z, i}$ and $b_{g}^{i}$ describe the mean photometric uncertainty, the width {of} photometric uncertainty, and the galaxy-matter bias of the lens sample used for the shear ratio likelihood (\S~\ref{sect:like_tt}). Note that the fact that the lens mean photometric uncertainties priors are not centred at 0 is related to a different definition of the priors compared to the sources' ones. In the `Prior' column we report either lower and upper boundaries (for flat priors) or the mean and standard deviation (for Gaussian priors; note that in this case we sample over a range much broader than the one $\sigma$ width). {Priors are described in \S~\ref{systematics}.}}
\centering
%\begin{adjustbox}{width=0.4\textwidth}
\begin{tabular}{cc}
\hline 
Parameter & Prior\tabularnewline
\hline 
\hline
\bf{Cosmological Parameters} & \\
%\hline
$\Omega_{\rm m}$ & $\mathrm{U}[0.1, 0.9]$ \tabularnewline
%\hline 
$\sigma_8$ & $\mathrm{U}[0.5, 1.4]$ \tabularnewline
%\hline 
$\Omega_{\rm b}$ &$\mathrm{U} [0.03, 0.07]$ \tabularnewline
%\hline 
$n_{\rm s}$& $\mathrm{U}[0.87, 1.07]$   \tabularnewline
%\hline 
$h$ &   $\mathrm{U}[0.55, 0.91]$ \tabularnewline
%\hline 
\hline
\bf{Calibration Parameters} & \\
%\hline 
$m_{1}$ & $\mathcal{N}(-0.0063,0.0091)$ \tabularnewline
%\hline
$m_{2}$ & $\mathcal{N}( -0.0198,0.0078)$ \tabularnewline
%\hline
$m_{3}$ & $\mathcal{N}( -0.0241,0.0076)$ \tabularnewline
%\hline
$m_{4}$ & $\mathcal{N}(-0.0369, 0.0076)$ \tabularnewline
%\hline
$\Delta z_{1}$ & $\mathcal{N}(0.0,0.018)$ \tabularnewline
%\hline
$\Delta z_{2}$ & $\mathcal{N}(0.0,0.015)$ \tabularnewline
%\hline
$\Delta z_{3}$ & $\mathcal{N}(0.0,0.011)$ \tabularnewline
%\hline
$\Delta z_{4}$ & $\mathcal{N}(0.0, 0.017)$ \tabularnewline
%\hline
\hline
\bf{Intrinsic Alignment Parameters} & \\
%\hline
$A_{\rm IA,0}$ & $\mathrm{U}[-5, 5]$ \tabularnewline
%\hline 
$\alpha_{\rm IA}$ & $\mathrm{U}[-5, 5]$ \tabularnewline
%\hline
\hline
\bf{Shear Ratios Parameters} & \\
%\hline 
$\Delta z^{\rm lens}_{1}$ & $\mathcal{N}(-0.009,0.007)$
\tabularnewline
%\hline 
$\Delta z^{\rm lens}_{2}$ & $\mathcal{N}(-0.035,0.011)$
\tabularnewline
%\hline 
$\Delta z^{\rm lens}_{3}$ & $\mathcal{N}(-0.005,0.006)$
\tabularnewline
%\hline 
$\delta ^{\rm lens}_{z,1}$ & $\mathcal{N}(0.975,0.062)$
\tabularnewline
%\hline 
$\delta ^{\rm lens}_{z,2}$ & $\mathcal{N}(1.306,0.093)$
\tabularnewline
%\hline 
$\delta ^{\rm lens}_{z,3}$ & $\mathcal{N}(0.870,0.054)$
\tabularnewline
%\hline 
$b^{1}_{g}$ & $\mathrm{U}[0, 3]$\\
$b^{2}_{g}$ & $\mathrm{U}[0, 3]$\\
$b^{3}_{g}$ & $\mathrm{U}[0, 3]$\\
\hline 
\end{tabular}
\label{table1}
\end{table}

\textit{Photometric redshift uncertainties}. The first type of nuisance parameters are `calibration' parameters that model uncertainties in the photometric redshift estimates from the SOMPZ method. Such uncertainties are parameterised through a shift $\Delta z$ in the mean of the redshift distributions:
\begin{equation}
n^i(z) = \hat{n}^i(z+\Delta z ),
\end{equation}
where $\hat{n}^i$ is the original estimate of the redshift distribution  for  bin $i$. We assume DES Y3 priors for the shift parameters. The priors also include the additional photo-$z$ uncertainty due to blending \citep{y3-imagesims}. This parameterisation of the redshift uncertainties was shown to be adequate for the DES Y3 2-point analysis \citep{y3-hyperrank,y3-cosmicshear1}; we none the less explore in \S~\ref{sect:results} a more complex parameterisation of redshift uncertainties that also accounts for uncertainties in the shape of the redshift distributions. 

\textit{Multiplicative shear biases}. {Biases coming from the shear measurement pipeline are modelled through an average multiplicative parameter $1+m^i$ for each tomographic bin}. The effect of multiplicative shear biases on the measured moments can be modelled via:

\begin{equation}
\langle \kappa^2_{\theta_0}\rangle^{i,j} \rightarrow (1+m^i) (1+m^j)\langle \kappa^2_{\theta_0}\rangle^{i,j},
\end{equation}
\begin{equation}
\langle \kappa^3_{\theta_0}\rangle^{i,j,k} \rightarrow (1+m^i) (1+m^j) (1+m^k)\langle \kappa^3_{\theta_0}\rangle^{i,j,k}.
\end{equation}
We assume Gaussian priors for each of the $m^i$, estimated following \cite{y3-imagesims}.

\textit{Intrinsic galaxy alignments {(IA)}}. We model IA following the non-linear alignment (NLA) model \citep{Hirata2004,Bridle2007,Pyne2022}. It can be included in our modelling introducing  $\delta_{\rm I} = A(z) \delta$, which is the density contrast responsible for the intrinsic alignment, related to the matter density contrast $\delta$. In the NLA model, the IA amplitude can be written as a power law:
\begin{equation}
A(z) = - A_{\rm IA,0} \left(\frac{1+z}{1+z_0} \right)^{\alpha_{\rm IA}} \frac{c_1 \rho_{m,0}}{D(z)},
\end{equation}
with $z_0= 0.62$, $c_1\rho_{\rm crit}=0.0134$, wih $\rho_{\rm crit}=\rho_{m,0}/ \Omega_{\rm m}$ \citep{Bridle2007} and $D(z)$ the linear growth factor \citep{Hamilton2001}.
For second moments, the NLA model can be incorporated in our theoretical predictions by modifying the lensing kernel:
\begin{equation}
q^i(\chi) \rightarrow q^i(\chi) + A(z(\chi)) \frac{n^i(z(\chi))}{\avg{n^i}}\frac{dz}{d\chi}.
\end{equation}
{For third moments, we make the assumption that the NLA contribution follows the perturbation theory relation for the actual signal. \citep{Pyne2022} have shown this is in reasonable agreement with measurements from hydrodynamical simulations, so we follow them and modify Eq. \ref{eq:ksmM3} as follows:}
\begin{multline}
{q^i q^jq^k}  \rightarrow  {q^i q^jq^k} + \frac{A^2+2A}{3} \left( q^i q^j n^k + {\rm cycl.} \right) + \\ \frac{A^2+2A^3}{3} \left( q^i n^j n^k + {\rm cycl.} \right) + {A^4}  \left( n^i n^j n^k  \right) ,
\end{multline} 

where in the above equation we dropped the redshift dependence for sake of simplicity; moreover, we used $n = \frac{n^i(z(\chi))}{\avg{n^i}}\frac{dz}{d\chi}$, and cycl. refers to the cyclic permutation of the indexes $i, j,
k$ for the terms in parenthesis. We marginalise over $A_{\rm IA,0}$ and $\alpha_{\rm IA}$ assuming flat priors. The fiducial DES Y3 3x2pt analysis adopted a different, more general model for the intrinsic galaxy alignment, called `TATT' (Tidal Alignment and Tidal Torquing; \cite{blazek19}), that can capture the `tidal torquing' relevant for determining the angular momentum of spiral galaxies. Tidal torquing is ignored in the NLA model, which can account only for the tidal alignment of galaxies. We did not implement such a general model here; the DES Y3 cosmic shear analysis \citep{y3-cosmicshear2} found a weak preference for simpler IA modelling (i.e., for NLA rather than TATT), obtaining consistent cosmological constraints when different IA prescriptions were assumed. For this reason we use the NLA model as our fiducial choice.

%The NLA model is usually used in the context of 2-point %correlation statistics, but the above equation generalises %it to third moments as well. The amplitude of the IA %contribution can be written as a power law:
%%
%\begin{equation}
%A(z) = A_{\rm IA,0} \left(\frac{1+z}{1+z_0} %\right)^{\alpha_{\rm IA}} \frac{c_1 \rho_{m,0}}{D(z)},
%\end{equation}
%%
%with $z_0= 0.62$, $c_1\rho_{\rm crit}=0.0134$ %\citep{Bridle2007} and $D(z)$ the linear growth factor %\citep{Hamilton2001}. W
%
%

%The NLA model can be incorporated in our theoretical %predictions by modifying the lensing kernel:
%%
%\begin{equation}
%q^i(\chi) \rightarrow q^i(\chi) - A(z(\chi)) %\frac{n^i(z(\chi))}{\avg{n^i}}\frac{dz}{d\chi}.
%\end{equation}
%%
%The NLA model is usually used in the context of 2-point %correlation statistics, but the above equation generalises %it to third moments as well. The amplitude of the IA %contribution can be written as a power law:
%%
%\begin{equation}
%A(z) = A_{\rm IA,0} \left(\frac{1+z}{1+z_0} %\right)^{\alpha_{\rm IA}} \frac{c_1 \rho_{m,0}}{D(z)},
%\end{equation}
%%
%with $z_0= 0.62$, $c_1\rho_{\rm crit}=0.0134$ %\citep{Bridle2007} and $D(z)$ the linear growth factor %\citep{Hamilton2001}. 

\textit{Shear ratio parameters}.
{We include  in the analysis galaxy-galaxy lensing small scale information in the form of ratios of galaxy-galaxy lensing measurements \citep{y3-shearratio}. These measurements use as lenses the first three tomographic redshift bins of the \textsc{MagLim} lens galaxy sample \citep{y3-2x2maglimforecast}. When modelling the shear ratio measurements, we marginalise over the uncertainties in the photo-$z$ estimates of the lens samples through a shift $\Delta z^{\rm lens}$ in the mean of the redshift distributions and a stretch $\delta^{\rm lens}$ in their widths:}
\begin{equation}
n^{{\rm lens},i}(z) = \delta^{\rm lens}\hat{n}^{{\rm lens},i}(\delta^{\rm lens}\left[z-\avg{z}\right]+\Delta z^{\rm lens} ),
\end{equation}
{where $\avg{z}$ is the mean redshift of the lens sample. Priors on  $\Delta z^{\rm lens}$ and $\delta^{\rm lens}$  are provided in \cite{y3-2x2ptaltlensresults}. We also marginalise over the galaxy-matter bias $b_g^i$ of the three lens samples using broad flat priors.}

\subsection{Map making and moments estimator}\label{sect:map_making}

We describe here how we measure the second and third moments of the convergence field starting from a weak lensing catalogue. The following applies to both data and simulated catalogues, as they come in the same format.

Starting from the catalogue, we first generate convergence maps for each tomographic bin. The convergence maps used in this work are estimated using a full-sky generalisation of the \cite{KaiserSquires} algorithm, first developed by \cite{Wallis2017}. The map-making process for the DES Y3 convergence maps is explained in full detail in \cite{y3-massmapping}, together with a thorough validation of the maps. Here, we briefly summarise the procedure. 

{We use the weak lensing catalogue shear estimates to create pixelized maps for the two components of the shear field. The maps are constructed using  \textsc{HEALPIX} with \textsc{NSIDE} = 1024 (corresponding to a pixel size of $3.44~\mathrm{arcmin}$). The estimated value of the complex shear per pixel is given by:}
\begin{equation}
\label{eq:pixelvalue}
\gamma_{\rm obs}^{\nu} = \frac{\sum_{j=1}^{n}\epsilon_j^{\nu}w_{j}}{\bar{R}\sum_{j=1}^{n}w_j}, \,\, \nu=1,2,
\end{equation}
 {where $\epsilon_j$ is the per-galaxy observed ellipticity, $\nu$ refers to the two shear field components, $n$ is the total number of galaxies in the pixel, $\bar{R}$ is the average \textsc{METACALIBRATION} response of the sample ($\bar{R}=1$ for simulated catalogues), and $w_j$ is the per-galaxy inverse variance weight. The sum runs over all the galaxies in the pixel. Shear maps for each tomographic bins are created. As specified in \S~\ref{systematics}, we do not explicitly correct for the multiplicative shear bias when making the maps, but rather we account for it during the cosmological inference. Any non-zero mean shear is subtracted from the catalogue before creating the maps.}

We then convert the shear maps into a curl-free E-mode convergence map $\hat{\kappa}_E$ and a divergence-free B-mode convergence map $\hat{\kappa}_B$ using a spin transformation. This is achieved by using the \textsc{HEALPIX} function \textsc{MAP2ALM} to decompose the shear maps in spherical harmonic space obtaining the coefficients $\hat{\gamma}_{E,\ell m}$, $\hat{\gamma}_{B,\ell m}$, and then calculating $\hat{\kappa}_{E,\ell m}$, $\hat{\kappa}_{B,\ell m}$  as:
\begin{equation}
\label{eq:KS}
{\gamma}_{\ell m} = \hat{\gamma}_{E,\ell m} + i {\gamma}_{B,\ell m} =  -\sqrt{\frac{(\ell +2)(\ell -1)}{\ell (\ell +1)}} ({\kappa}_{E,\ell m} + i {\kappa}_{B,\ell m}).
\end{equation}
Next we use the \textsc{HEALPIX} function \textsc{ALM2MAP} to convert these coefficients back to real space $\kappa_E$ and  $\kappa_B$ maps.
The maps are smoothed using a top-hat filter and different smoothing scales $\theta_0$. {In practice, this is achieved by multiplying the coefficients of the harmonic decompositions of the $\kappa_E$ and $\kappa_B$ maps by Eq. \ref{eq:filter}, prior to the conversion to real space.} Simple estimators then give the moments of a smoothed map:
\begin{equation}
\avg{\hat{\kappa}^2_{\theta_0}}^{i,j} = \frac{1}{N_{\mathrm{tot}}}\sum_{\mathrm{pix}}^{N_{\rm tot}} \kappa_{\theta_0,\mathrm{pix}}^i \kappa_{\theta_0,\mathrm{pix}}^j,
\end{equation}
\begin{equation}
\avg{\hat{\kappa}^3_{\theta_0}}^{i,j,k} = \frac{1}{N_{\mathrm{tot}}}\sum_{\mathrm{pix}}^{N_{\rm tot}} \kappa_{\theta_0,\mathrm{pix}}^i \kappa_{\theta_0,\mathrm{pix}}^j \kappa_{\theta_0, \mathrm{pix}}^k,
\end{equation}
where $i,j,k$ refers to different tomographic bins. {We estimate the moments for both the E- and B-mode convergence maps, although only the E-modes moments are used for the cosmological analysis.} The sum runs over all the pixels on the sky (thus including regions outside the footprint). {This is needed for two reasons: first, the transformation from the shear field to the convergence field is non-local and some power is transferred outside the footprint during the transformation; second, the smoothing of the maps also transfers some of the power from the pixels close to the edge to pixels outside the footprint. We have shown in \G{} that our modelling, together with the use of mode-coupling matrices, is able to take into account these effects (also including the lack of shear data outside the footprint, since the shear field is not defined there).}

Due to the presence of shape noise, the measurement of galaxy shapes is only a noisy estimate of the shear field $\gamma$. This also means that our estimate of the convergence field is noisy:
\begin{align}
\kappa_{E,{\rm obs}} = \kappa_{E, {\rm true}} + \kappa_{E,{\rm noise}},\\
\kappa_{B,{\rm obs}} = \kappa_{B,{\rm true}} + \kappa_{B,{\rm noise}}.
\end{align}
In the above equations, we omitted the smoothing angle $\theta_0$. The contribution of the noise to the convergence field can be estimated by randomly rotating the shapes of the galaxies and applying the full-sky spherical harmonics approach to obtain the convergence \citep{VanWaerbeke2013,Chang2018}. As the random rotation should completely erase the cosmological contribution, the resulting convergence signal just contains noise and averages to zero (but with a non-negligible variance). 

It follows that when estimating second and third moments from noisy convergence maps, it is necessary to properly de-noise the measured moments. Following \cite{VanWaerbeke2013}:

\begin{multline}
\label{eq:deno2}
\langle{\kappa}^2\rangle^{i,j} \rightarrow \langle  \kappa^2\rangle^{i,j} -  \langle\kappa \kappa_{{\rm rand}}\rangle^{i,j} -  \langle  \kappa_{\rm rand} \kappa \rangle^{i,j} -
 \langle \kappa^2_{{\rm rand}}\rangle^{i,j},
\end{multline}
\begin{multline}
\label{eq:deno3}
\langle{\kappa}^3\rangle^{i,j,k} \rightarrow \langle \kappa^3\rangle^{i,j,k} -  \langle \kappa^3_{{\rm rand}}\rangle^{i,j,k} - \\ \left[\langle  \kappa^2_{{\rm rand}}\kappa  \rangle^{i,j,k} - \langle  \kappa_{{\rm rand}}\kappa^2  \rangle^{i,j,k} + {\rm cycl.} \right],
\end{multline}
where $\rm{cycl.}$ refers to the cyclic permutation of the indexes $i$, $j$, $k$ for the terms in parenthesis. In the above equations, the term $\langle \kappa^2_{{\rm rand}}\rangle^{i,j}$  ($\langle \kappa^3_{{\rm rand}}\rangle^{i,j,k}$) is the noise-only contribution to the second (third) moments of the tomographic bins $i,j$(,$k$). Under certain conditions, most of these terms vanish; those terms that do not vanish need to be subtracted from the measured moments. We verified which terms vanish in Appendix~\ref{sect:noise_terms}.

%is the noise-only contribution to the second (third) moment of the tomographic bins $i,j$($k$); for $i \neq j$ it vanishes. The map $\kappa_{{\rm rand}}$ represents the estimate of the shape noise contribution to the convergence map; it is estimated by randomly rotating the galaxy shapes. The intrinsic ellipticity distribution of observed galaxies is not expected to be perfectly Gaussian, but by the central limit theorem, it would be the correct distribution in the limit of large numbers of galaxies averaged in the pixels of the convergence map \citep{Jeffrey2018}. If this holds, also the term $\langle \kappa^3_{{\rm rand}}\rangle^{i,j,k}$ (which is the noise-only contribution to the third moment of the tomographic bin  $i,j,k$) would vanish. Additional checks will need to be performed on DES Y3 data, as we do not include potential sources of noise inhomogeneities (e.g. astrophysical or observational systematics) in this work. Finally, we note that if the convergence field and the shape noise term in a given map pixel are uncorrelated, mixed terms should be consistent with zero. 

\section{Likelihood and Covariance}\label{sect:like_tt}

This section provides details about our data vector, likelihood and covariance. Our data vector consists of all the possible combinations of second and third moments involving the four weak lensing tomographic bins. This adds up to a total of 10 combinations of second moments and 20 combinations of third moments. For each of these second and third moments, we consider 10 equally (logarithmic) spaced smoothing scales $\theta_0 \in [3.2,200]$ arcmin. We then remove scales following \G{}, i.e. we remove angular scales smaller than a corresponding comoving scale $R_0$ given by $\theta_0 = R_0/ \chi(\avg{z})$, where $\chi(\avg{z})$ is the comoving distance of the mean redshift of a given tomographic bin. In the case of moments from different tomographic bins, we took the average of the mean $\avg{z}$ of the two bins. This scale cut is designed to remove scales significantly affected by modelling uncertainties that could contaminate the cosmological analysis, with the dominant uncertainty being contamination due to baryonic effects. \G{} determined the fiducial scale cut to be $24 h^{-1}$ Mpc when combining second and third moments. We adopt here a scale cut of $28 h^{-1}$ Mpc. This change is necessary because the simulated analysis in \G{} did not use the final setup for the analysis (e.g., inclusion of the shear-ratio likelihood, final values for redshift distributions, shape noise, effective number densities, covariance, etc.); we therefore repeated the scale cut analysis with  all the analysis ingredients updated, and determined $28 h^{-1}$ Mpc to be the correct scale cut to be used in this analysis (see Appendix~\ref{sect:scale_cuts} for more details).

We then compress our data vector using the Massively Optimised Parameter Estimation and Data compression (MOPED) algorithm \citep{Tegmark1997,Heavens2000,Gualdi2018} based on the  Karhunen-Lo\`eve algorithm, which allows us to reduce the dimensionality of our data vector to the number of model parameters considered. In our case, the number of parameters used to model the moments data vector is 15; therefore, the size of the compressed moments data vector is 15. The compression allows us to reduce the enlargement of the parameters posterior due to noise in the precision matrix estimate, as the covariance matrix is estimated from a limited number of simulations \citep{Hartlap2007}. The final enlargement depends on the size of the compressed data vector rather than on the size of the uncompressed data vector, which makes having an efficient compression scheme desirable. In particular: 
\begin{equation}
\label{eq:data_compression}
    d^{\rm compr}_{i} = \avg{d}^{T}_{,i} \hat{{C}}^{-1} d \equiv b_i d ,
\end{equation}
where $d$ is the full-length data vector, $\hat{{C}}$ is the measurement covariance, and $d^{\rm compr}_{i}$ is the $i$-th element of the compressed data vector. The index $i$ refers to the $i$-th model parameter $p$ considered, and $\avg{d}^{T}_{,i}$ is the derivative of the model data vector with respect to that parameter.

We evaluate the posterior of the parameters conditional on the data by assuming a Gaussian likelihood for the data, i.e.
\begin{equation}
\label{eq:ll}
- 2 \ln \mathcal{L} = f_2 f_1 [\hat{d}-M(p)] \hat{{C}}^{-1}[\hat{d}-M(p)]^T.
\end{equation}
Here $M(p)$ is our theoretical model, $\hat{d}$ is the data vector, and $\hat C^{-1}$ is the inverse of our covariance estimate. The posterior is then the product of the likelihood and the priors. 
Note that the quantities $M(p)$, $\hat{d}$ and $\hat C^{-1}$ in Eq.~\ref{eq:ll} are to be considered compressed quantities. The terms $f_1$ and $f_2$ account for noise introduced when the covariance matrix is estimated from random realisations of the data \citep{Hartlap2007,Dodelson2013,Friedrich2018b} and are given by:
\begin{equation}
\label{eq:hartlap}
f_1 = \frac{N_{\rm sims}-N_{\rm data}-2}{N_{\rm sims} -1} , 
\end{equation}
 \begin{equation}
\label{eq:DS}
f_2 = \left[1 + \frac{(N_{\rm data}-N_{\rm par})(N_{\rm sims}-N_{\rm data}-2)}{(N_{\rm sims}-N_{\rm data}-1)(N_{\rm sims}-N_{\rm data}-4)}\right]^{-1}\ ,
\end{equation}
where in our case the number of independent realisations used to estimate the covariance is $N_{\rm sims}$ (i.e. the number of independent simulations) and $N_{\rm data}$ is the length of the data vector. In the case of compressed quantities, $f_1,f_2\sim1$ as $N_{\rm sims}>> N_{\rm data}$.

\begin{figure}
\begin{center}
\includegraphics[width=0.45 \textwidth]{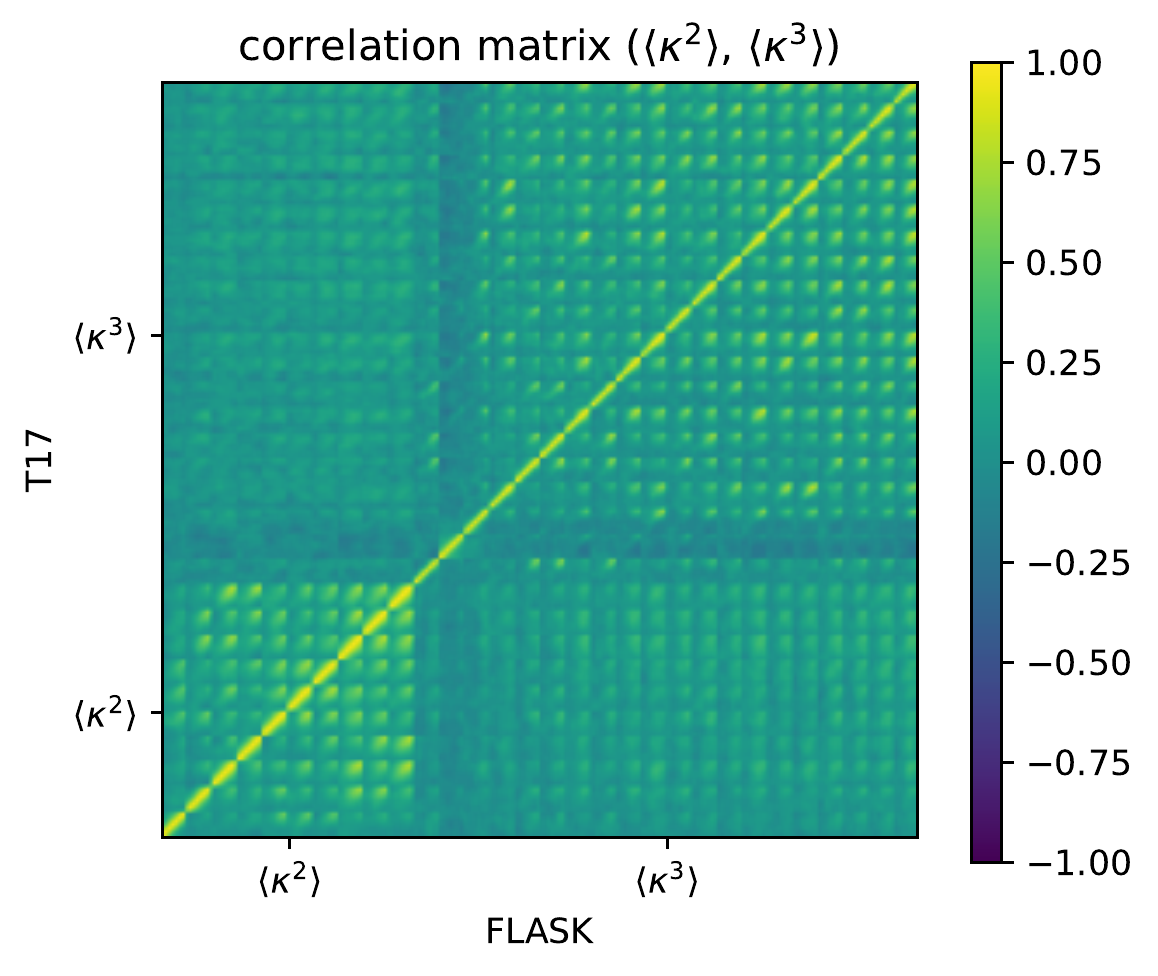}
\includegraphics[width=0.45 \textwidth]{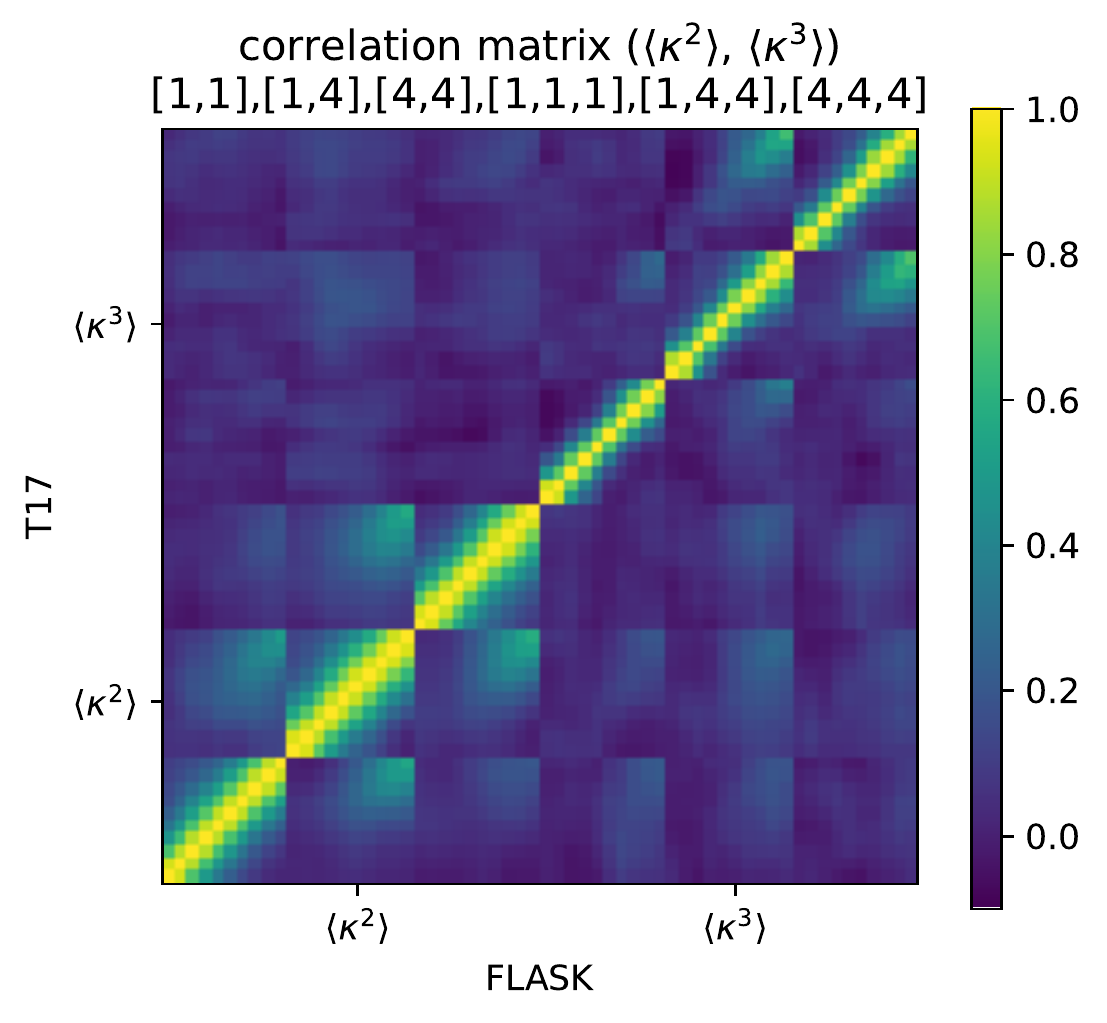}

\end{center}
\caption{\textit{Top}: measured correlation matrix of second and third moments from FLASK simulations (lower right triangle) and from T17 simulations (upper left triangle). No scale cut has been applied. {From bottom left to top right, we show the bins: [1,1], [2,2], [3,3], [4,4], [1,2], [1,3], [1,4], [2,3], [2,4], [3,4], [1,1,1], [2,2,2], [3,3,3], [4,4,4], [2,1,1], [3,1,1], [3,2,2], [4,1,1],[ 4,2,2], [4,3,3], [1,2,2], [1,3,3], [1,4,4], [2,3,3], [2,4,4], [3,4,4], [1,2,3], [1,2,4], [1,3,4], [2,3,4].} {The main difference between the two covariances is that the FLASK one has on average 5-10 per cent larger amplitude at large scales. The difference vanishes at small scales since those are dominated by shape noise}. \textit{Bottom}: same as the top image, but focusing on a few elements ({from bottom left to top right, we show }[1,1], [1,4], [4,4] for second moments and [1,1,1], [1,4,4] and [4,4,4] for third moments), and showing only the dynamical range [-0.1,1.0] effectively spanned by the elements of the correlation matrix. The diagonal blocks of the correlation matrix contain essentially all the non-negligible contributions. On large scales, where the cosmic variance contribution to the covariance dominates, the window function of the moments generates off-diagonal terms (within the block diagonal matrix) while on small scales these are due to the pixel window function. Note that in the absence of shape noise we also expect a contribution from non-linear evolution on small scales.The third moments correlation matrices are more diagonal than second moments ones owing to a larger shape noise contribution. Non-negligible cross-covariance between second and third moments is limited to very large scales, and is generally small (10-15 per cent at most).}
\label{fig:covariance}
\end{figure}
To correctly infer cosmological parameters from our data, we need an accurate estimate of the measurement uncertainty. Our fiducial method to estimate the covariance uses 1000 independent realisations of the convergence maps generated from the \textsc{FLASK} simulations. As an additional check, we also estimate the covariance using the PKDGRAV and T17 simulations. The PKDGRAV and T17 simulations (Fig. \ref{fig:covariance}) have been produced at cosmologies different to that of the \textsc{FLASK} simulations; hence, these alternative covariances provide extra validation against the dependency of our covariance on the value of cosmological parameters. More details are given in Appendix~\ref{sect:alternative_cov}. %Our fiducial method to estimate the covariance uses 432 independent realisations of the convergence maps generated from the T17 simulations. As additional check, we also estimate a second covariance using the PKDGRAV simulations. Note that the PKDGRAV simulations have been produced at a different cosmology than the T17 simulations; hence, this alternative covariance provides extra validation against the dependency of our covariance on the value of cosmological parameters.
Given a set of N-body simulations, for each realisation we measure the second and third moments of the smoothed convergence field and build the covariance matrix as:
\begin{equation}
\label{eq:cov_def}
\hat{{C}} = \frac{1}{\nu}{\sum_{i=1}^{\rm N_s}} (\hat{{d}}_i-\hat{{d}}){(\hat{{d}}_i-\hat{{d}})}^{T},
\end{equation}
where $\nu = {\rm N_s}-1$ with ${\rm N_s}$ the number of realisations, $\hat{{d}}_i$ the data vector measured in the $i$-th simulation, and $\hat{{d}}$ the sample mean. The data vector is made of a combination of second and third moments as measured at different smoothing scales. We also add to our covariance a `modelling uncertainty' related to the analytical fitting formulae describing the third moments at small scales (see \G{} for more details). We then compress the covariance following:
\begin{equation}
 \hat{{C}}_{ij}^{\rm compr} = b_i^T \hat{{C}} b_j.
\end{equation}
We tested that using the \textsc{FLASK} covariance we were able to correctly recover the input cosmology in simulations (Appendix~\ref{sect:validation}).  

In the inference, we also add an independent `shear ratio' likelihood \citep{y3-shearratio}. The shear ratio likelihood uses small scale information from the ratio of galaxy-galaxy lensing measurements (the mean tangential shear around lens galaxies) between two weak lensing source tomographic bins and a shared lens sample. Its inclusion improves the constraints on the redshift distributions and on other nuisance parameters of our model. The shear ratio data vector consists of nine scale-averaged  ratios. We use as a lens the first three tomographic redshift bins of the \textsc{MagLim} lens galaxy sample \citep{y3-2x2maglimforecast}. The shear ratios likelihood is modelled as an independent Gaussian likelihood, and uses an analytical covariance matrix. The assumption of independency is justified by the smallness of the scales involved in the shear ratio measurements (less than 6  $h^{ - 1}$ Mpc). We note that the scale cut for this work is 28 $h^{ - 1}$ Mpc, although the two scale cut limits cannot be directly compared since the mass map smoothing function and the galaxy-galaxy lensing angular bin kernels weight scales slightly differently. None the less, the independency of the shear ratio likelihood has been proven in the context of the DES Y3 3x2pt analysis \citep{y3-shearratio}. Because we adopt the same scale cut criteria as the DES 3x2pt analysis, we assume independency holds here as well. Lastly, since the shear ratio  covariance is analytical, we do not compress the shear ratio data vector. 

Having defined the likelihood, we sample the posteriors of our parameters using \textsc{Polychord} \citep{poly1,poly2}; this is a nested sampler that uses slice sampling within the nested iso-likelihood contours. For the cosmological parameters, we assume a flat $\Lambda$CDM cosmology and vary five parameters: $\Omega_{\rm m}$ (the density of the total matter today), $\sigma_{8}$ ({the amplitude of structure fluctuations in the present day Universe, parameterised as the standard deviation of the linear overdensity fluctuations on a 8$h^{ - 1}$ Mpc scale}), $\Omega_{\rm b}$ (the baryonic density in units of the critical density),  $n_{\rm s}$ (the spectral index of primordial density fluctuations), and $h$ (the dimensionless Hubble parameter). We assume wide flat priors on $\Omega_{\rm m}$ and $\sigma_8$ and adopt the informative priors on $h$, $n_{\rm s}$ and $\Omega_{\rm b}$ that were used in the DES Y3 2-point function {3$\times$2pt} analysis (see Table \ref{table1}). When constraining cosmological parameters, we marginalise over nuisance parameters describing mean photo-$z$ uncertainties, multiplicative shear biases and IA effects in our measurements. The modelling of our nuisance parameters is described in $\S$~\ref{systematics}. Photo-$z$ uncertainties are parametrised by a shift in the mean of the distribution (one for each tomographic bin). Priors for the shifts come from \citet*{y3-sompz}. Multiplicative shear bias priors are described in \citet{y3-imagesims}. We also assume wide flat priors for intrinsic alignment amplitudes. The addition of the shear-ratio likelihood to the analysis necessitates additional modelling parameters, summarised in Table~\ref{table1}. These are lens redshift parameters (modifying the mean redshift and the width of the lens sample redshift distributions) and one free (linear) galaxy bias parameter per lens bin.%, which is almost entirely unconstrained, and has almost no impact on the final likelihood %These describe the uncertainties in the redshift distribution of lens galaxies, as well as the relation between galaxies and dark matter, parametrized using a per-bin linear galaxy bias. 

Last, we note that since the theory predictions described in \S~\ref{sect:model} are time-consuming to compute due to the large number of cross-correlations and integrations involved, we implemented an emulator \citep{Heitmann2006,Habib2007} to speed up the calculations. In our implementation, the emulator provides fast theoretical predictions by interpolating over a number of predictions computed at a set of training points spanning the parameter space of interest (in our case, the 5 cosmological parameters). In particular, the quantities emulated are the terms
\begin{linenomath} % See https://tex.stackexchange.com/questions/461186
\begin{multline}
\langle \delta^2_{\theta_0}\rangle^{\rm EE/BB}(\chi)  \equiv  \sum_{\ell} \frac{2\ell+1}{4 \pi} f_{\ell}^{-1} W_{\ell}(\theta_0)^2 \times\\\sum_{\ell'}  M_{\ell \ell'}^{\rm EE/BB ,EE}P_{\mathrm{NL}} (\ell'/\chi,\chi) F_{\ell'}^2 f_{\ell'},
\end{multline}
\end{linenomath}
\begin{linenomath} % See https://tex.stackexchange.com/questions/461186
\begin{multline}
\langle \delta^3_{\theta_0}\rangle^{\rm  EE/BB} (\chi) \equiv
S_3\times\left[ \langle \delta^2_{\theta_0}\rangle^{\rm EE/BB}(\chi)\right]^2,
\end{multline}
\end{linenomath}
 which enter in the modelling of Eq. \ref{eq:ksmM2} and Eq. \ref{eq:ksmM3}. The accuracy of the emulator is sufficient to not bias the cosmological analysis, as demonstrated in \G{}.

\section{Pre-unblinding tests}\label{sect:unblinding}

%{add a comment on projection effects. Add a comment on degeneracy breaking based on fig 3. and best-fitting DV from second, third and second + third moments}.

\begin{figure*}
\begin{center}
\includegraphics[width=0.8 \textwidth]{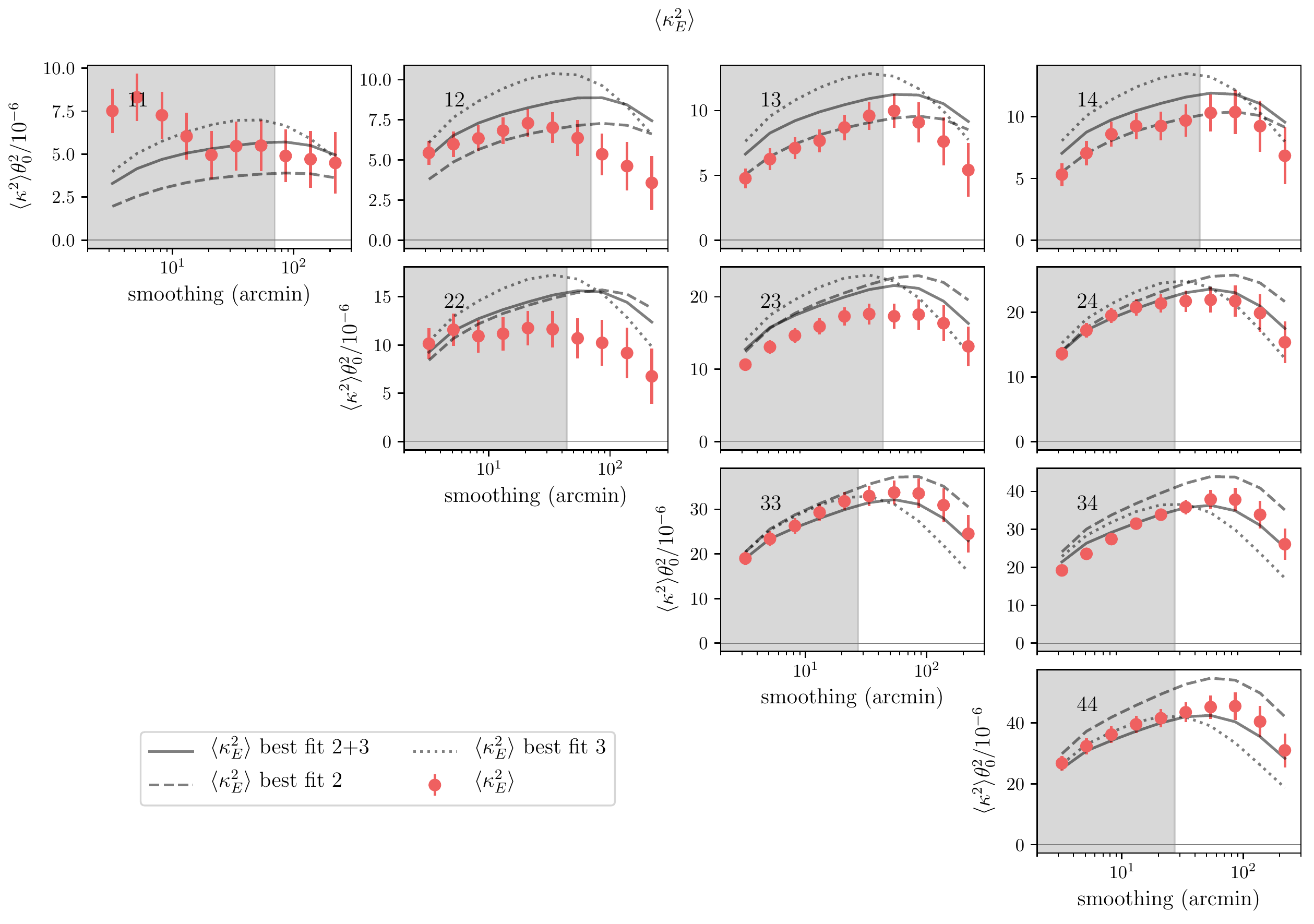}
\includegraphics[width=0.8 \textwidth]{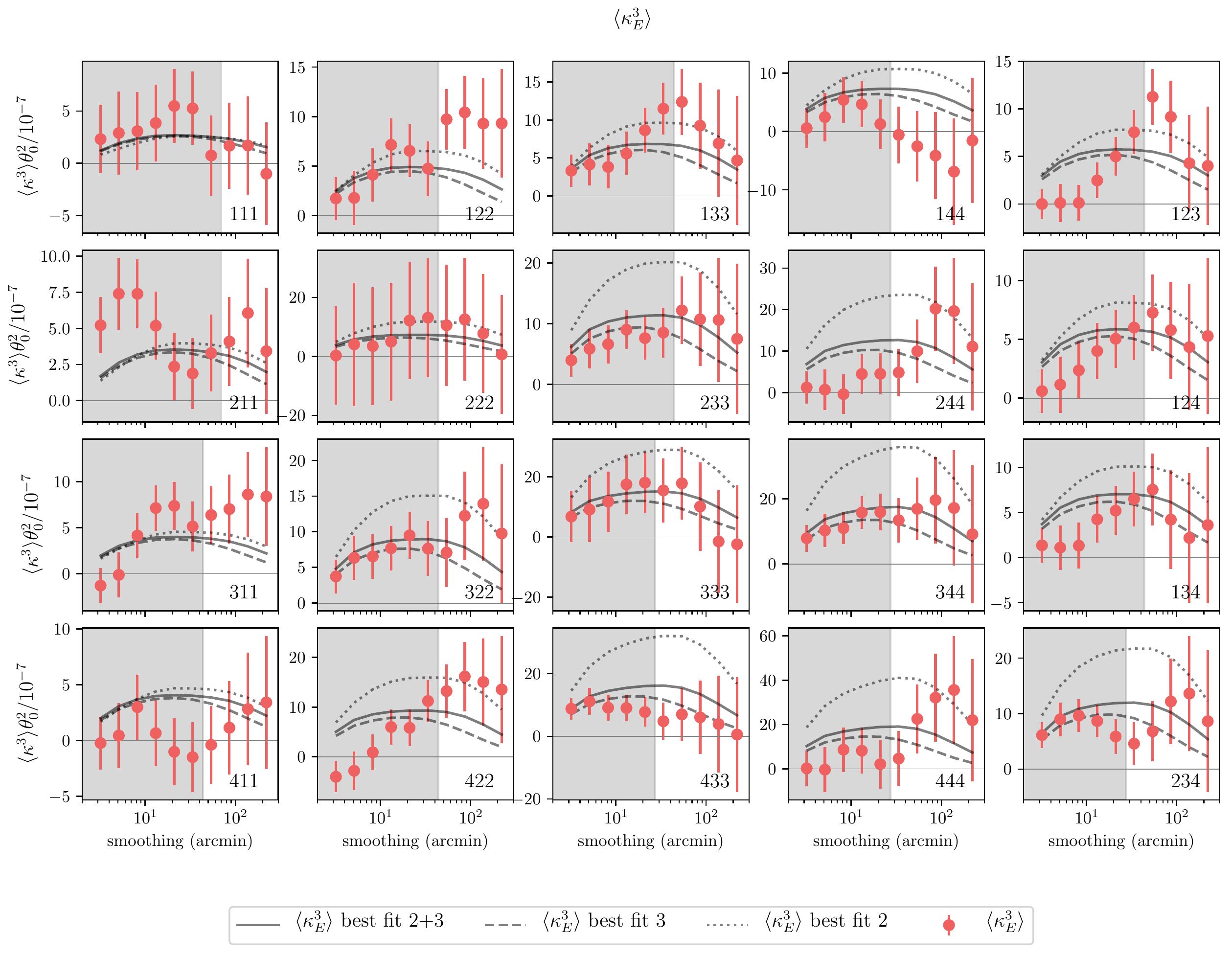}
\end{center}
\caption{Measured second moments (upper plots) and third moments (lower plots). Red points represent the measurement. Grey shaded regions highlight the scales removed by the analysis.  The conservative scale cut implemented in this analysis removes a large part of our data vector. Solid, dotted, and dashed lines represent the predictions obtained using the best-fitting cosmology of second and third moments analysis (either considered in combinations or alone). Data points are very correlated (Fig. \ref{fig:covariance}), so we caution the reader from any $\chi^2$-by-eye estimation. }
\label{fig:final_dv}
\end{figure*}

\begin{figure*}
\begin{center}
\includegraphics[width=0.7 \textwidth]{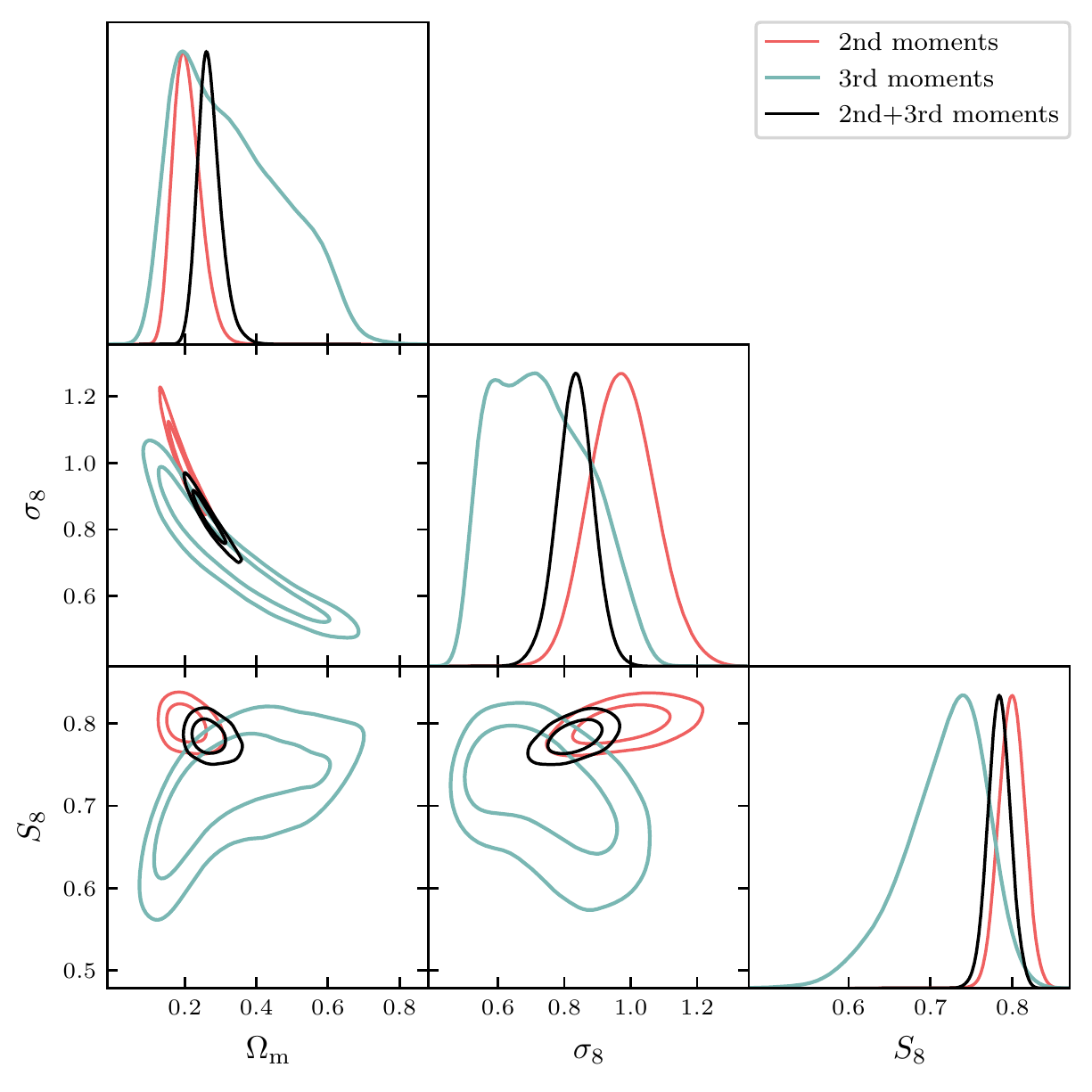}
\end{center}
\caption{Posterior distributions of the cosmological parameters $\Omega_{\rm m}$, $\sigma_8$, and $S_8$ for the second moments, third moments, and the combination of second and third moments. We note that our fiducial analysis include small-scale galaxy-galaxy lensing ratios (a.k.a. shear ratios, \S \ref{sect:like_tt}). The 2D marginalised contours in these figures show the 68 per cent and 95 per cent confidence levels. }
\label{fig:final_results}
\end{figure*}

\begin{figure*}
\begin{center}
\includegraphics[width=0.8\textwidth]{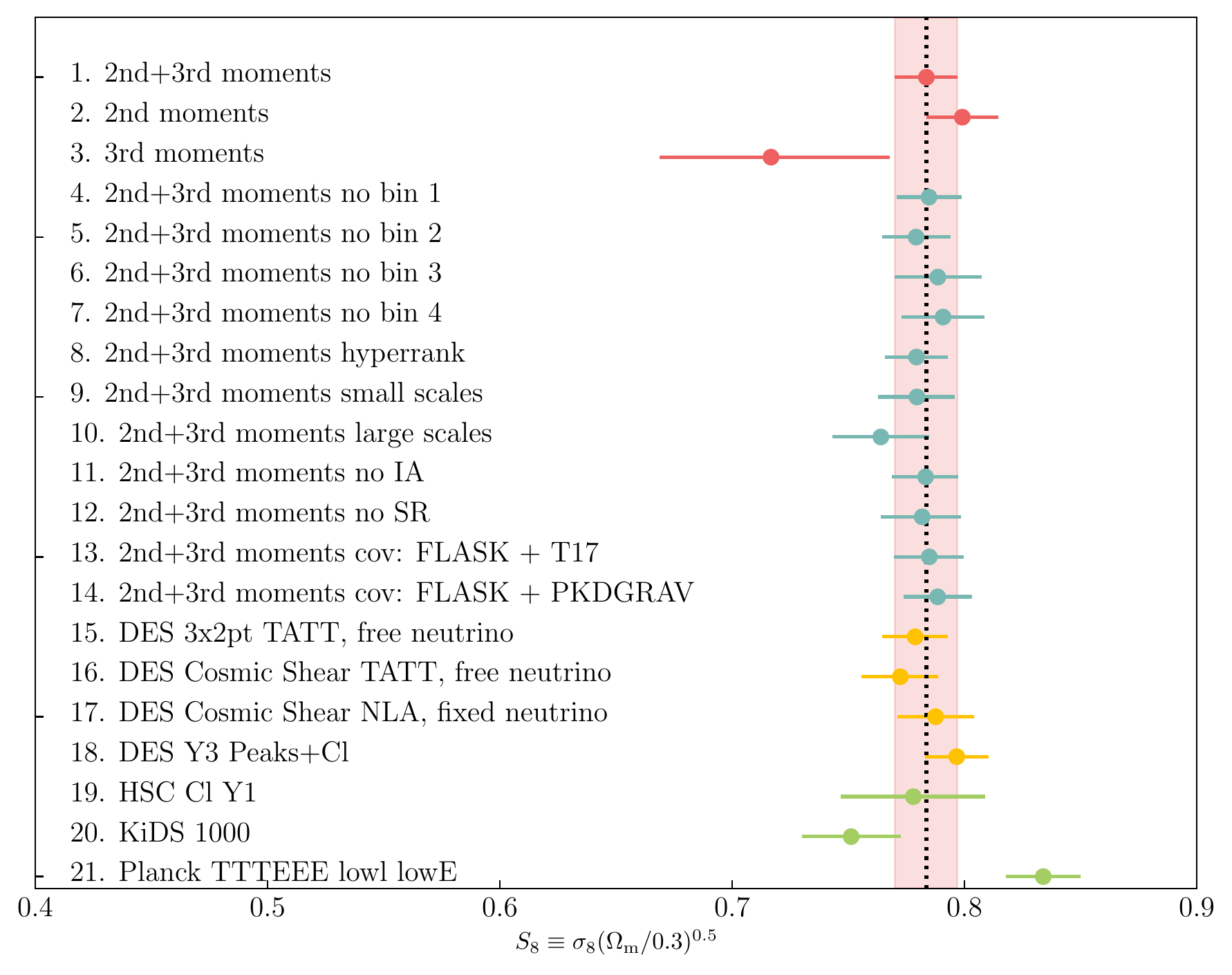}
\end{center}
\caption{Constraints on the cosmological parameter $S_8$; we report the mean of the posterior and the 68 per cent confidence interval. The first three lines are the fiducial results from this analysis. Following that are the $S_8$ values for a number of variations and systematic tests: removing one redshift bin at a time, using a different parameterization for the redshift distribution called  ``hyperrank'', considering only small or large scales, considering the case with no intrinsic alignement or no shear ratio (SR), and using different covariance matrixes (FLASK + T17 or FLASK + PKDGRAV), as explained in Appendixes \ref{sect:internal} and \ref{sect:alternative_cov}. Lastly, we compare with a number of results from other works, either with DES or external data. }
\label{fig:all}
\end{figure*}

\begin{table*}
%\tiny
\caption {Constraints on the cosmological parameters  $S_8$, $\Omega_{\rm m}$, and $\sigma_8$. For each parameter we report the mean of the posterior and the 68 per cent confidence interval. For the fiducial results (second moments, third moments, and the combination of the two) we also report the PPD goodness-of-fit $p$-value.}
\centering
\begin{tabular}{clccccc}

\hline
& & $S_8$ & $\Omega_{\mathrm{m}}$ & $\sigma_8$ & $p$-value \\ [0.2cm]

\hline
%\rule{{0pt}}{{4ex}} \multirow{4.5}{*}{\rotatebox{90}{\textbf{Fiducial}}}  & {2nd moments} & $ 0.799\pm0.015$  & $0.21\pm0.04$  & $0.98\pm0.10$ & 14.0/9.0  & 0.12   \\ 
%\rule{{0pt}}{{4ex}} & {3rd  moments} & $0.72\pm0.05$  & $0.33\pm0.16$  & $0.73\pm0.16$  & 12.7/10.3  & 0.26    \\ 
%\rule{{0pt}}{{4ex}}  & {2nd + 3rd moments} & $0.784\pm0.013$  & $0.27\pm0.03$  & $0.83\pm0.05$ & 10.9/9.1  & 0.29  \\ 
\rule{{0pt}}{{4ex}} \multirow{4.5}{*}{\rotatebox{90}{\textbf{Fiducial}}}  & {2nd moments} & $ 0.799\pm0.015$  & $0.21\pm0.04$  & $0.98\pm0.10$ & 0.21  \\ 
\rule{{0pt}}{{4ex}} & {3rd  moments} & $0.72\pm0.05$  & $0.33\pm0.16$  & $0.73\pm0.16$   & 0.63    \\ 
\rule{{0pt}}{{4ex}}  & {2nd + 3rd moments} & $0.784\pm0.013$  & $0.27\pm0.03$  & $0.83\pm0.05$ & 0.26  \\ 

\hline
\rule{{0pt}}{{4ex}} \multirow{16}{*}{\rotatebox{90}{\textbf{Variations}}}  &  {2nd + 3rd moments, no bin 1} & $0.785\pm0.014$  & $0.30\pm0.04$         & $0.79\pm0.06$  &  & \\ 
\rule{{0pt}}{{4ex}} &  {2nd + 3rd moments, no bin 2} & $0.779\pm0.015$  & $0.27\pm0.04$         & $0.83\pm0.06$  &  & \\ 
\rule{{0pt}}{{4ex}}  & {2nd + 3rd moments, no bin 3} & $0.789\pm0.019$  & $0.27\pm0.05$         & $0.83\pm0.08$  &  & \\ 
\rule{{0pt}}{{4ex}}  & {2nd + 3rd moments, no bin 4} & $0.791\pm0.018$  & $0.23\pm0.04$          & $0.92\pm0.08$  &  & \\ 
\rule{{0pt}}{{4ex}}  & {2nd + 3rd moments, hyperrank} & $0.779\pm0.014$  & $0.26\pm0.03$        & $0.83\pm0.05$  &  & \\ 
\rule{{0pt}}{{4ex}}  & {2nd + 3rd moments, small scales} & $0.780\pm0.017$  & $ 0.32\pm0.05$     & $0.76\pm0.07$  & &  \\ 
\rule{{0pt}}{{4ex}}  & {2nd + 3rd moments, large scales} & $0.76\pm0.02$  & $ 0.28\pm0.04$       & $0.79\pm0.07$  & &  \\ 
\rule{{0pt}}{{4ex}}  & {2nd + 3rd moments, no shear ratio} & $0.782\pm0.017$  & $ 0.27\pm0.04$            & $0.83\pm0.06$  & &  \\ 
\rule{{0pt}}{{4ex}}  & {2nd + 3rd moments, FLASK + T17} & $0.785\pm0.015$  & $ 0.27\pm0.03$      & $0.82\pm0.06$  & &  \\ 
\rule{{0pt}}{{4ex}}  & {2nd + 3rd moments, FLASK + PKDGRAV} & $ 0.788\pm0.015$  & $ 0.28\pm0.03$  & $0.82\pm0.06$  & &   \\ 
\hline
\rule{{0pt}}{{4ex}} \multirow{13}{*}{\rotatebox{90}{\textbf{Other works}}}  & DES Y3 Cosmic Shear,  &$ 0.772\pm 0.016$ &  $0.29\pm 0.05$  &  $ 0.79\pm 0.08$    &  &   \\ 
& TATT free neutrino \citep{y3-cosmicshear1,y3-cosmicshear2} & &   &    &  &   \\ 
\rule{{0pt}}{{4ex}}  & DES Y3 Cosmic Shear,  &$ 0.788\pm0.016$ &  $0.28\pm0.04$  &  $ 0.82\pm0.08$    &  &   \\ 
&   NLA fixed neutrino \citep{y3-cosmicshear1,y3-cosmicshear2}  & &   &    &  &   \\ 
\rule{{0pt}}{{4ex}}  & DES Y3 3x2pt, & $ 0.779\pm 0.014$ &   $ 0.33\pm 0.03$  &  $ 0.74\pm0.04$   &  &   \\ 
& TATT free neutrino \citep{y3-3x2ptkp} & &   &  &  &   \\ 
%\rule{{0pt}}{{4ex}}  & {DES Y3 3x2pt NLA fixed neutrino **} & $ 0.792\pm0.014$ &   $0.32\pm0.02$  &  $ 0.77\pm0.04$  &  &   \\
\rule{{0pt}}{{4ex}}  & DES Y3 Peaks + Cls \citep{Zuercher2021b}& $ 0.797\pm0.014$ &   $ 0.28\pm0.07$ &  $ 0.85\pm0.11$  &  &   \\ 
\rule{{0pt}}{{4ex}}  & KIDS-1000 \citep{Heymans2021} &$ 0.751\pm 0.021$  &  $ 0.29\pm0.08$   & $ 0.79\pm 0.13$  &  &   \\ 
\rule{{0pt}}{{4ex}}  &HSC Y1 CLs \citep{Hamana2019} & $ 0.778\pm 0.031$ &    $ 0.18\pm0.07$  &   $ 1.05\pm 0.16$   &  &   \\ 
\rule{{0pt}}{{4ex}}  & Planck 2018 TT,TE,EE + lowl + lowE \citep{aghanim2020planck} & $ 0.834\pm 0.016$ &    $ 0.316\pm 0.008$   &  $ 0.812\pm 0.007$   &  &   \\

%\rule{{0pt}}{{4ex}}  & Planck 2018 TT,TE,EE + lowE \citep{aghanim2020planck} & $ 0.832\pm 0.013$ &    $ 0.315\pm 0.007$   &  $ 0.811\pm 0.006$   &  &   \\ 

\hline
\end{tabular}
\label{table_results}
\end{table*}

Before proceeding to unblind the data vector and analyse the results of the unblinded analysis, we performed a number of tests. These tests complement the ones performed at the catalogue and map level presented in \cite{y3-shapecatalog,y3-massmapping}. We remind the reader that when this analysis was performed, the shape catalogue was already deemed science-ready and unblinded, and only the data vector level of blinding was enforced. The whole cosmological pipeline had already been demonstrated in \G{} to recover the true cosmology using realistic simulations. We none the less repeated the validation in simulations with the updated analysis choices (e.g., redshift distributions, shape noise, priors, etc.) in Appendix \ref{sect:validation}, using both T17 and PKDGRAV simulations. We also slightly changed the scale cut decided in \G{}, due to updates in the analysis choices. More details concerning the scale cuts are given in Appendix~\ref{sect:scale_cuts}. 

We first performed two tests at the data vector level:
\begin{itemize}
    \item We checked that additive biases due to PSF modelling errors were negligible at the data vector level, i.e., if neglected they would not bias our cosmological analysis. This test is similar to the test performed for the DES Y3 cosmic shear analysis \citep{y3-cosmicshear1}; more details are given in Appendix~\ref{sect:PSF_modelling_errors}.
    \item We tested that mixed moments between convergence maps E-mode and noise (e.g., $\langle \kappa_{{\rm N}}\rangle^{i,j}$) are consistent with expectations based on tests on N-body simulations; more details are given in Appendix~\ref{sect:noise_terms}.
\end{itemize}
We then ran our analysis on blinded data vectors, and checked that:
\begin{itemize}
    \item Cosmological constraints obtained using (blinded) second and third moments were consistent with each other. To this aim, we used posterior predictive distributions (PPD, \cite{Doux2021}); see Appendix~\ref{sect:internal}.
    \item The (blinded) posteriors of the systematic parameters did not concentrate at the edge of the prior. The level of agreement/disagreement with the prior was tested using a Gaussian estimator called the `update difference-in-mean' (UDM) statistic \citep{Raveri2019} (Appendix~\ref{sect:nuisance}).
\end{itemize}

{We then unblinded the data vectors and ran the fiducial analysis; before looking at the unblinded posteriors, we further checked that:}

\begin{itemize}
    \item The goodness-of-fit $p-$value on unblinded data vectors was larger than 1 per cent; see \S~\ref{sect:results}.
    \item The best-fitting cosmology provided a good description to second and third moments B-modes (which are not included in the data vector), see Appendix~\ref{sect:B-modes}. This was done in an automated fashion such that we did not look at the actual best-fitting values.
\end{itemize}

In order to quantify goodness-of-fit and internal consistency among different parts of our data vector, we use the PPD methodology developed by \cite{Doux2021} and adopted in the main DES Y3 3x2pt analysis. The PPD methodology derives a calibrated probability-to-exceed $p$; in the case of goodness-of-fit tests, this is achieved by drawing realisations of the data vector for parameters drawn from the posterior under study; for consistency tests (e.g., second moments vs. third moments), the realisations are drawn from disjoint subsets of the data vector. These realisations are then compared to actual observations and a distance metric ($\chi^2$) is computed in data space, which is then used to compute the $p$-value.

Once all these tests were passed, we looked at the unblinded posteriors of our analysis.

\section{Cosmological Constraints}\label{sect:results}

We present here the cosmological constraints obtained assuming the $\Lambda$CDM model, varying 5 cosmological parameters and 19 nuisance parameters (10 for the moments likelihood and 9 additional ones for the shear ratio likelihood), as summarised in Table~\ref{table1}. In addition to these parameters, we will also quote results in terms of the $S_8$ parameter, defined as
\begin{equation}
S_8 \equiv \sigma_8(\Omega_{\rm m}/0.3)^{\alpha} \,.
\end{equation} 
The value of $\alpha$ can be chosen such that $S_8$ best constrains the degeneracy between $\Omega_{\rm m}$ and $\sigma_8$. However, the second and third moments have a slightly different degeneracy direction and so there is no value of $\alpha$ that simultaneously optimises both. For sake of simplicity we adopt $\alpha=0.5$.

Fig.~\ref{fig:final_results} shows the posteriors for $S_8$, $\Omega_{\rm m}$, and $\sigma_8$ from the second and third moments individually, and from the combinations of the two. Third moments are much less constraining than second moments alone, but they are characterised by a slightly different degeneracy tilt in the $\sigma_8$-$\Omega_{\rm m}$ plane compared to second moments. The marginalised mean values of $S_8$, $\Omega_{\rm m}$, and $\sigma_8$ for the combination of second and third moments, along with the 68\% confidence intervals, are: 
\begin{linenomath} % See https://tex.stackexchange.com/questions/461186
\begin{align}
\Omega_{\rm m} &{} =   0.27\pm0.03  \\%0.27^{0.05}_{-0.04} ()\\
\sigma_8 &{} = 0.83\pm0.05 \\
S_8 &{} =  0.784\pm0.013  
\end{align}
\end{linenomath}
\noindent 
We report the constraints from the analysis of second and third moments individually in Table~\ref{table_results}, and for $S_8$ we aditionally provide a visual comparison in Fig. \ref{fig:all}. The combined moments analysis places a 1.7 per cent constraint on $S_8$ and a 10 per cent constraint on $\Omega_{\rm m}$, improving by $\sim 15$ and  $\sim 25$ per cent over constraints from second moments only. This level of improvement is expected (\G{}), and is due to the additional non-Gaussian information probed by third moments and the degeneracy breaking when second and third moments are combined. Table~\ref{table_results} also reports the $p$-values for the goodness-of-fit tests; these are well above the $p\textrm{-value}=0.01$ threshold. The unblinded data vectors, along with the best-fitting models from our posteriors, are shown in Fig.~\ref{fig:final_dv}. We caution the reader from any $\chi^2$-by-eye estimate, as the different scales are highly correlated (especially for second moments, where adjacent scales have a correlations higher than 90 per cent). Constraints from second and third moments are consistent with each other, although it is evident from Fig.~\ref{fig:final_results} that they probe different parts of the parameter space in the $\sigma_8$-$\Omega_{\rm m}$ plane. 

In Appendix \ref{sect:internal} we use PPD to quantify the internal consistency of our data sets. In particular, we tested the compatibility between second and third moments constraints, between small and large scales, and between parts of the data vector using different redshift bins. These tests were performed prior to unblinding, using blinded data vectors, and were repeated after unblinding (although only the compatibility of second and third moments was considered as an unblinding criterion). {In Appendix \ref{sect:internal} we also perform a test analysing the data vector using a different parameterisation of the redshift uncertainties, called hyperrank \citep{y3-hyperrank}.}

The results reported here were obtained using the \texttt{FLASK} covariance; in addition, we tested that our results do not change significantly when using the covariances estimated using the T17 or PKDGRAV simulations (Appendix~\ref{sect:alternative_cov}).

\begin{figure}
\begin{center}
\includegraphics[width=0.5\textwidth]{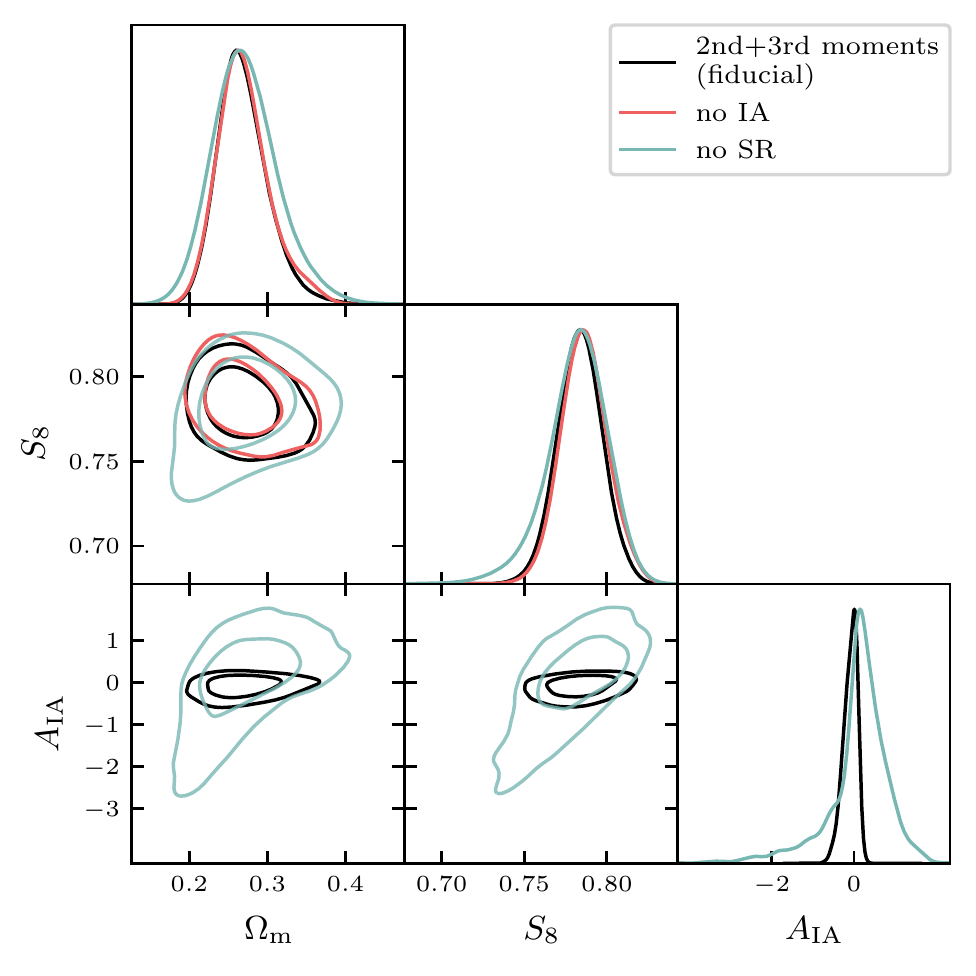}
\end{center}
\caption{Posterior distributions of the cosmological parameters $\Omega_{\rm m}$ and $S_8$, and the IA amplitude parameter $A_{\rm IA}$, for the combination of second and third moments. ``SR'' stands for shear ratio. The 2D marginalised contours in these figures show the 68 per cent and 95 per cent confidence levels.}
\label{fig:IA}
\end{figure}

\begin{figure}
\begin{center}
\includegraphics[width=0.4 \textwidth]{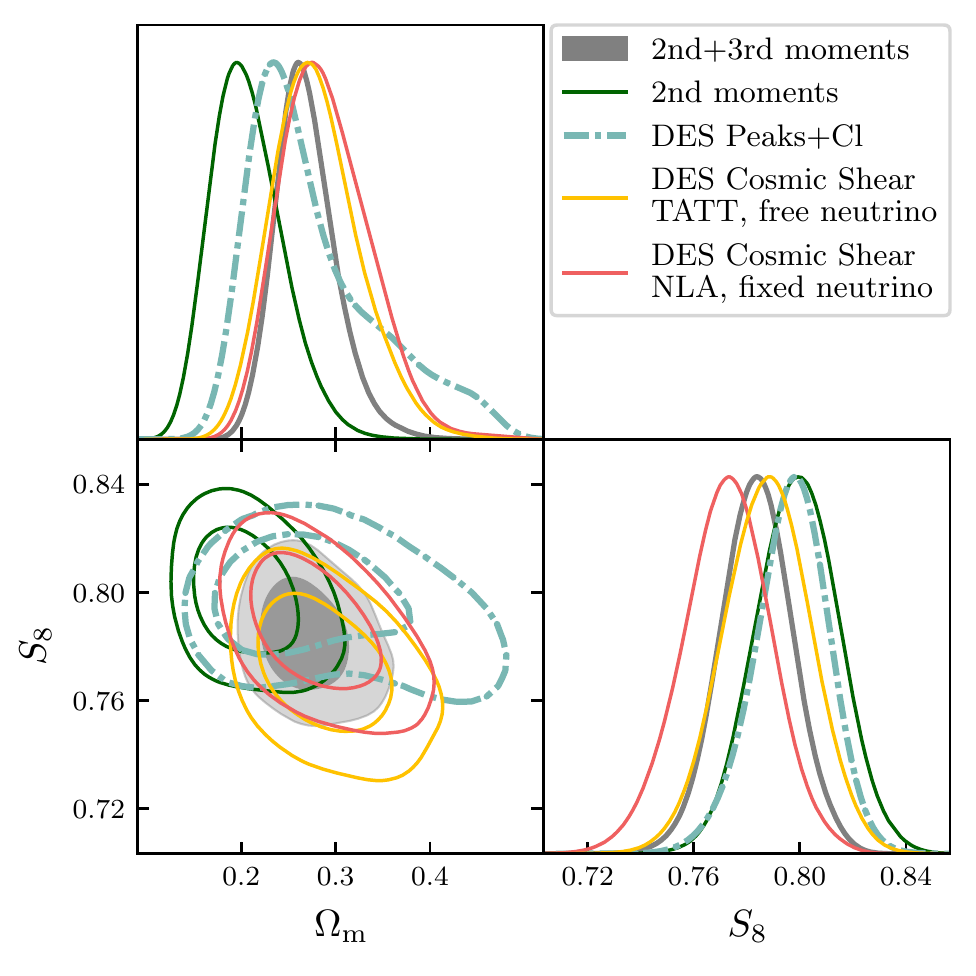}
\includegraphics[width=0.4 \textwidth]{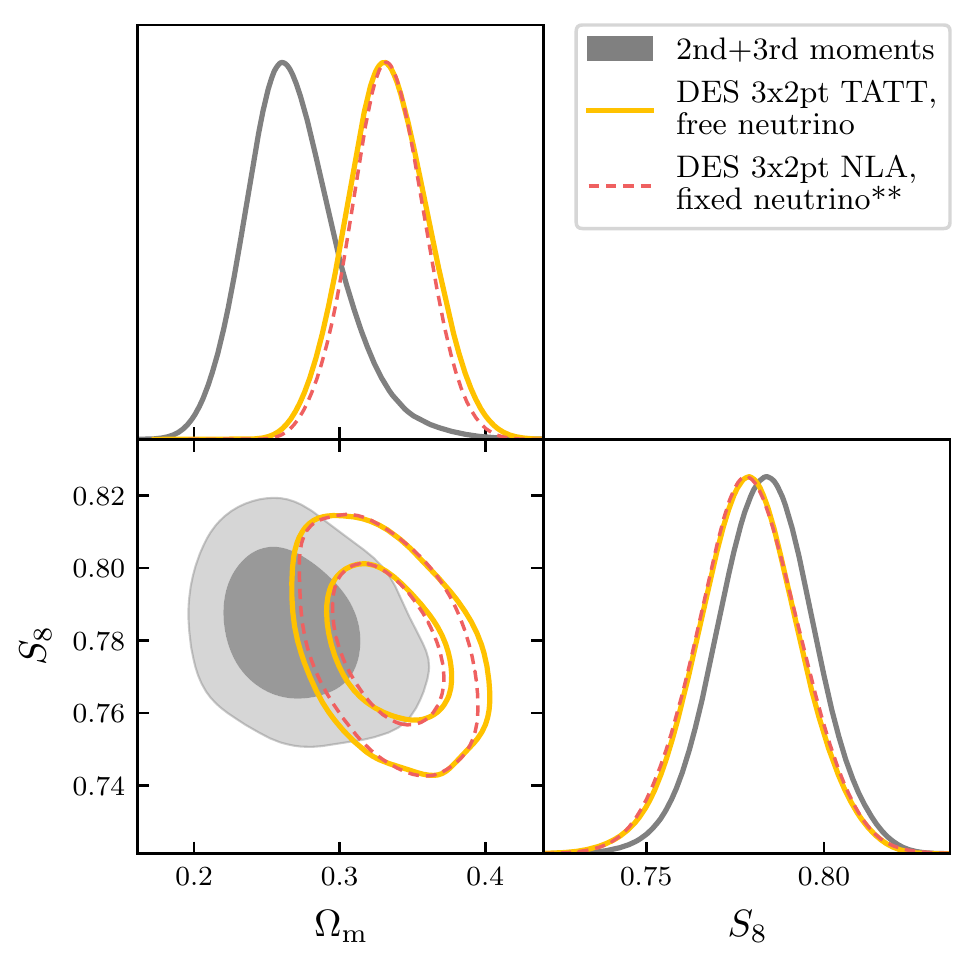}
\end{center}
\caption{Posterior distributions of the cosmological parameters $\Omega_{\rm m}$ and $S_8$. \textit{Top panel:} we show the posteriors for the moments analysis, two versions of the DES Y3 cosmic shear analysis, and the DES Y3 Peaks + Power spectrum analysis. For readability we do not show the third moments constraints separately. \textit{Bottom panel:} we show the posteriors for the moments analysis and for two versions of the DES Y3 3x2pt analysis. \textit{**: the DES 3x2pt NLA + fixed neutrino analysis is unlikely to pass our scale cut criteria, see \S \ref{sect:des_comparison} for more details. For this reason, we shifted the contours on top of the DES 3x2 TATT posterior, so as to not unveil the exact location of the (potentially biased) posterior.} The 2D marginalised contours in these figures show the 68 per cent and 95 per cent confidence levels.}
\label{fig:des}
\end{figure}

\begin{figure}
\begin{center}
\includegraphics[width=0.45 \textwidth]{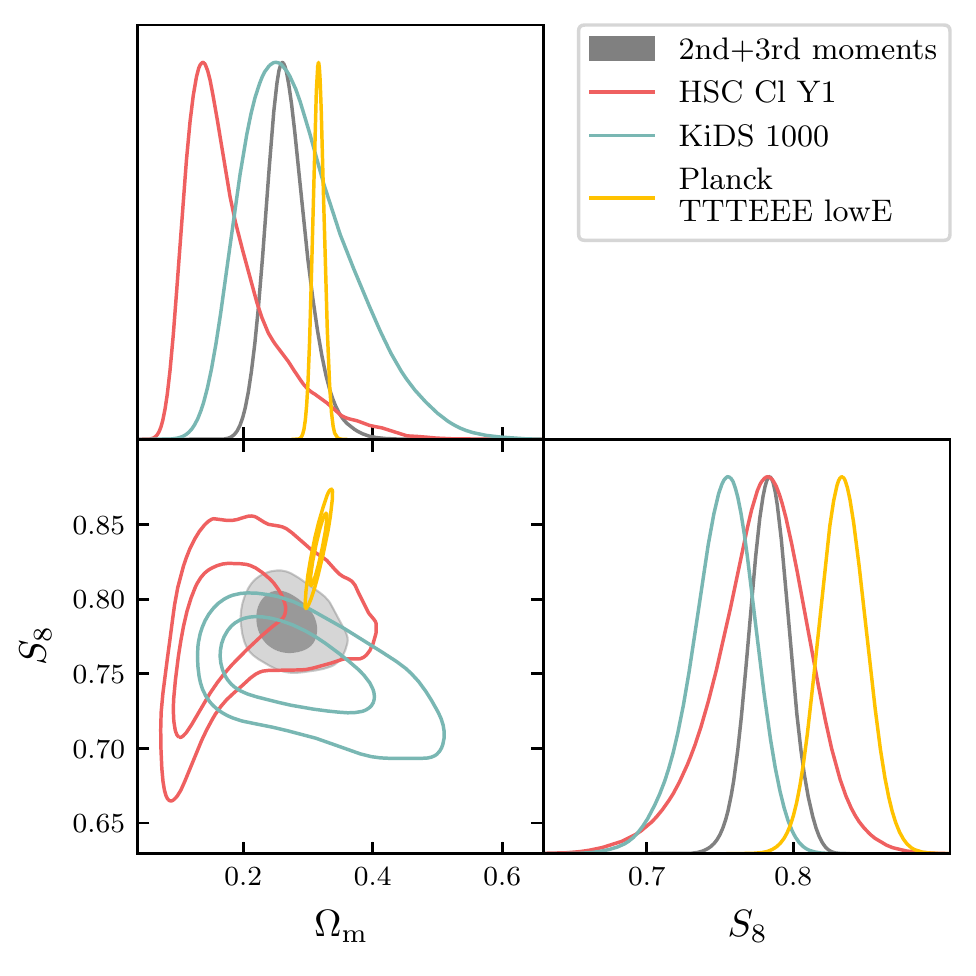}
\includegraphics[width=0.45 \textwidth]{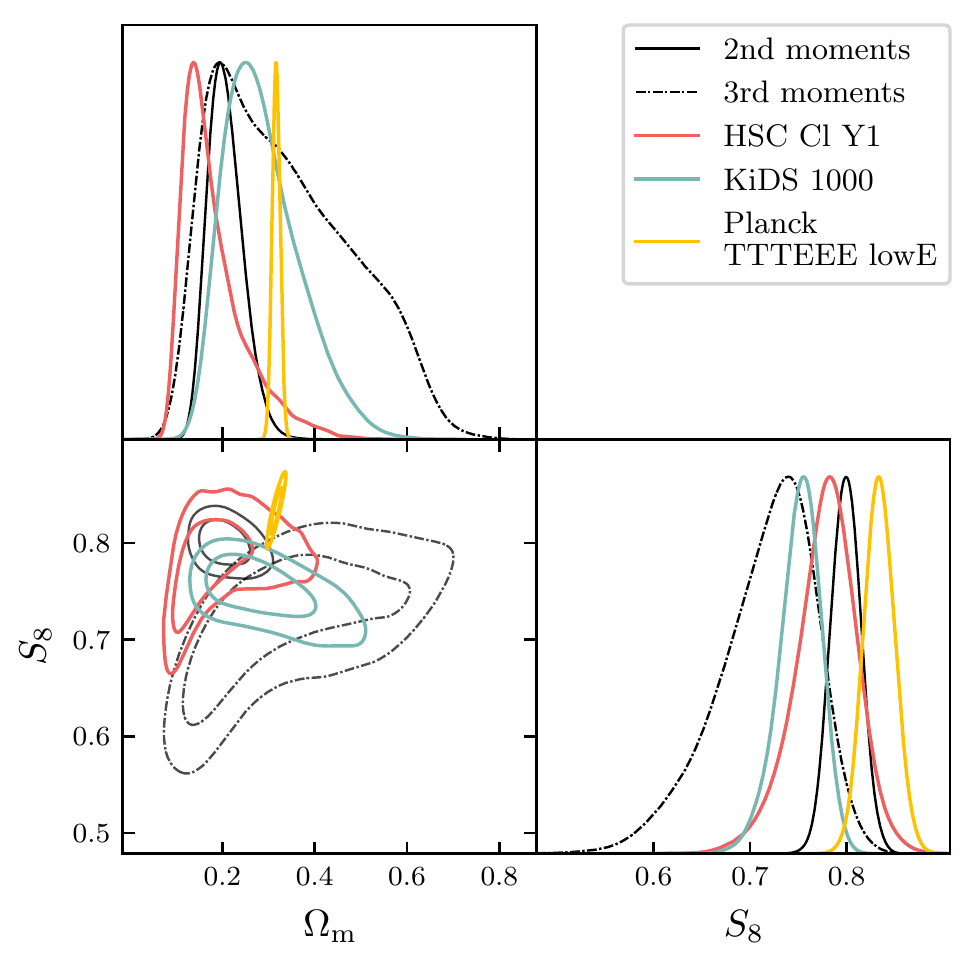}
\end{center}
\caption{Posterior distributions of the cosmological parameters $\Omega_{\rm m}$ and $S_8$, for the moments analysis and for the recent results of the KIDS-1000 survey \citep{Asgari2021}, HSC \citep{Hikage2019}, and \textit{Planck} \citep{aghanim2020planck}. Table \ref{tension} shows the tension between the DES moments and other analyses. The upper panel only shows the combination of second and third moments, whereas the lower panel shows second and third moments constraints individually.  }
\label{fig:planck}
\end{figure}

\subsection{Intrinsic alignment constraints and impact of the shear ratio likelihood}

Intrinsic alignment (IA) is a potentially important contribution to the shear signal. We show in Fig. \ref{fig:IA} the posterior of the IA amplitude parameter $A_{\rm IA}$ for the combination of second and third moments. Our results are compatible with a null IA signal, as the amplitude of the IA signal is constrained to $A_{\rm IA} = -0.09 \pm 0.17$. Most of the constraint on IA comes from the shear ratio likelihood (Fig.\ref{fig:IA}), although when performing the analysis without shear ratio we also obtain a null IA signal of $A_{\rm IA} = 0.09\pm0.6$. The improvement in the IA constraints due to the inclusion of shear ratio is expected \citep{y3-shearratio}; moreover, because of the slight degeneracy between the IA amplitude parameter and $S_8$, shear ratio also improves the $S_8$ constraints ($\sim 25$ per cent). The constraints obtained analysing second and third moments only are also very similar: $-0.08\pm0.17$ and $-0.10\pm0.15$ for second and third moments, respectively. The tighter constrain on $A_{\rm IA}$ from third moments is due to a projection effect related to the broader constraints on $\Omega_{\rm m}$. These results are compatible with the DES Y3 cosmic shear and 3x2pt analyses results \citep{y3-cosmicshear2,y3-cosmicshear1,y3-3x2ptkp}, which also find an IA amplitude consistent with zero. Lastly, we ran an additional test analysing our data vector assuming no IA ($A_{\rm IA}=0$); the results are shown in Fig. \ref{fig:IA} and are almost identical to the fiducial results. The only difference between the no IA model and the fiducial analysis is that the former strongly constrains the nuisance parameter $\Delta z_1$ (redshift uncertainty of the first redshift bin). In particular, the posterior of that parameter is shrunk by half, although it is still consistent with zero. The IA model (NLA) used in this work is simpler than the fiducial model (TATT) adopted by the DES Y3 3x2pt analysis \citep{blazek19}. However, \cite{y3-cosmicshear2} finds that simpler IA models such as NLA are sufficient for modeling the DES Y3 data, so we do not think any of the conclusions in this work are affected by our (simpler) IA modelling choice.

\subsection{Comparison with DES constraints}\label{sect:des_comparison}

We discuss here how the parameter constraints obtained from this work compare with the ones obtained by other cosmological analyses using DES Y3 data (cosmic shear, 3x2pt, and lensing peaks). Marginalised posteriors for $S_8$ and $\Omega_{\rm m}$ are shown in Fig. \ref{fig:des} and (for $S_8$ only) in Fig.\ref{fig:all}; we also report the numerical values in Table \ref{table_results}. While the level of agreement can be noted in these figures, we cannot quantify it using the PPD metric, as we do not have the cross-covariance of moments with the other data vectors (a requirement of the PPD method).

The comparison that is probably the most relevant is with cosmic shear, which is a two-point correlation of the same lensing field. Our second moment should be consistent with it, although as discussed below the weighting of different scales (in particular in Fourier space) differs. The peaks statistic uses different non-Gaussian information from the third moment, so that is an interesting comparison as well. For completeness we include the 3x2pt results although these use the clustering of lens galaxies (a different probe of the mass distribution). However within the context of $\Lambda$CDM, the results should agree provided the theoretical predictions are accurate and the mitigation of systematic errors in each analysis is reliable.  

\paragraph*{DES Y3 cosmic shear.} The first comparison with other DES Y3 constraints is with the cosmic shear analysis \citep{y3-cosmicshear2,y3-cosmicshear1}. We compare with the constraints from two slightly different cosmic shear analyses: the first one is a $\Lambda$CDM  analysis {which assumes a more complex IA model (the TATT model), and marginalises over the neutrino mass, whereas the second one, which better matches the analysis choices adopted in this work, assumes NLA as IA model and fixes the neutrino mass to zero.\footnote{We remind the reader that neutrinos are not included in the modelling of moments, so their mass is automatically fixed to zero.}}  Both adopt the DES Y3 $\Lambda$CDM optimised scale cut\footnote{The DES Y3 $\Lambda$CDM optimised scale cuts are similar to the ones adopted in this work. In particular, they have been chosen so as to have the DES Y3 3x2pt $S_8$-$\Omega_{\rm m}$ constraints unbiased (i.e., $<0.3\sigma$) for a $\Lambda$CDM cosmology, with respect to potential baryonic contamination. The scale cuts adopted for the fiducial DES Y3 3x2pt results are more conservative because they also consider a $w$CDM cosmology.}. The constraints from the combination of second and third moments are in good agreement with the constraints from both cosmic shear analyses. 

In terms of constraining power on the $S_8$ and $\Omega_{\rm m}$ parameters, the NLA + fixed neutrino cosmic shear analysis is similar to the analysis using second moments only. The combined moments analysis is more constraining, due to the additional non-Gaussian information and the degeneracy breaking of the third moments. Although both cosmic shear and second moments are Gaussian statistics and they both probe the shear power spectrum, their posteriors do not have to perfectly overlap, as they weight power spectrum multipoles differently (Appendix \ref{sect:comp_wf}). In particular, our scale cuts exclude some of the higher wavenumber contributions to $\xi_{+-}$. None the less, the peaks of the second moments and the combination of second and third moments posteriors are consistent with the peak of the NLA + fixed neutrino cosmic shear posterior in the $S_8$-$\Omega_{\rm m}$ plane (1$\sigma$ and 0.15$\sigma$, respectively).

\paragraph*{DES Y3 3x2pt.} Similar to the DES Y3 cosmic shear analysis, we compare to two different versions of the DES Y3 3x2pt analysis \citep{y3-3x2ptkp}: a first one that assumes $\Lambda$CDM, the TATT model and marginalises over the neutrino mass, and a second one which better matches the analysis choices adopted in this work, assuming NLA as IA model and fixing the neutrino mass to zero. {We report the latter analysis for a visual comparison of the constraining power, but we caution the reader that it is unlikely to pass our scale cuts criteria, which impose a maximum bias of 0.3 $\sigma$ in the $S_8$-$\Omega_{\rm m}$ plane in case of baryonic contamination. {This analysis was not presented in \citep{y3-3x2ptkp}, and no adequate scale cut was determined. For sake of simplicity, we decided to use the same scale cut adopted in the 3x2pt TATT + free neutrino mass analysis, which is likely too aggressive.} %, which having less constraining power than the NLA + fixed neutrino analysis, 
%The TATT + free neutrino mass analysis is more conservative than the NLA + fixed neutrino analysis and it this scale cut passes the criteria with exactly a 0.3 bias \citep{y3-3x2ptkp} for the 
%, the NLA + fixed neutrino analysis is likely to fail the criteria, isnce i
This is because we know that the 3x2pt TATT + free neutrino mass analysis passes the scale cuts criteria with exactly a 0.3$\sigma$ bias \citep{y3-3x2ptkp}; the NLA + fixed neutrino analysis, having slightly more constraining power, is likely to fail those criteria. To avoid misinterpreting these results, we decided to shift the contours to lie on top of the DES 3x2pt TATT posterior, such that the real position is unknown and the posterior can only be used to get a sense of the effect of different analysis choices on the constraining power of the 3x2pt analysis.} The DES 3x2pt analysis relies on three different probes: cosmic shear, galaxy-galaxy lensing, and galaxy clustering. Remarkably, the $S_8$ constraining power from the moments analysis is 10 per cent better than that from the DES Y3 3x2pt analysis, despite not relying on a lens sample or the 2x2pt part of the data vector. The DES Y3 3x2pt constraints are, however, slightly more stringent in terms of $\Omega_{\rm m}$ (by 10 per cent), due to the significant contribution from the galaxy-galaxy lensing and galaxy clustering part of the analysis. The posteriors show good overlap, with the moments peak being $\sim 1.1 \sigma$ away from the DES Y3 3x2pt TATT + free neutrino analysis peak in the $S_8$-$\Omega_{\rm m}$ plane. Given that the constraints come from different probes we can consider the posteriors to be in reasonably good agreement.

\paragraph*{DES Y3 Peaks + Power spectrum analysis.}  \cite{Zuercher2021b} use peak counts to extract non-Gaussian information from the convergence field, and combine this with constraints from the power spectrum of convergence maps. The comparison of our analysis with theirs is interesting for two reasons: 1) similar to this analysis, it exploits some non-Gaussian information of the convergence field to constrain cosmological parameters; 2) the Peaks + Power spectrum analysis uses an independent, completely different framework to provide theory predictions for the observables -- they forward model the measurements using a Gaussian process emulator built using N-body simulations of different cosmologies. The analysis choices of the Peaks + Power spectrum analysis and our moments analysis are very similar, the main difference being that the former does not use the shear ratio likelihood and uses somewhat tighter priors for the $n_{\rm s}$, $h_{100}$, and $\Omega_{\rm b}$ parameters. The results from these two analyses are in agreement (Fig. \ref{fig:des}), with the peaks of their posterior within $1 \sigma$ of ours in the $S_8$-$\Omega_{\rm m}$ plane. Similar to the moments analysis,  the Peaks + Power spectrum analysis finds an IA amplitude consistent with zero.

\subsection{Comparison with external data sets}
We compare here our parameter constraints with the results obtained using external data sets. In particular, we compare with the recent results of the KIDS-1000 survey \citep{Asgari2021}, HSC \citep{Hikage2019}, and \textit{Planck} \citep{aghanim2020planck}. In order to estimate the tension between different analyses, we calculate a Monte Carlo estimate of the probability of a parameter difference \citep{Raveri2019,Raveri:2019gdp}, using the \texttt{tensiometer} software. In the case of uncorrelated data sets, the probability of the parameter difference reads:
\begin{equation} \label{Eq:ParameterDifferencePDF}
\mathcal{P}(\Delta \theta) = \int_{V_p} \mathcal{P}_A(\theta) \mathcal{P}_B(\theta-\Delta \theta) d\theta,
\end{equation}
where $V_p$ is the prior support and $\mathcal{P}_A$ and $\mathcal{P}_B$ are the two posterior distributions of the parameters. 
The probability of an actual shift in parameter space is obtained from the density of parameter shifts:
\begin{equation} \label{Eq:ParamShiftProbability}
\Delta = \int_{\mathcal{P}(\Delta\theta)>\mathcal{P}(0)} \mathcal{P}(\Delta\theta) \, d\Delta\theta,
\end{equation}
which is the posterior mass above the contour of constant probability for no shift, $\Delta\theta=0$. Due to the discrete nature of our posterior samples, the integral in Eq. (\ref{Eq:ParamShiftProbability}) is evaluated using a Monte Carlo approach \citep{Raveri:2019gdp}.

A visual comparison between the results of the moments analysis and the results obtained from external data sets is provided in Fig. \ref{fig:planck} for the $S_8$ and $\Omega_{\rm m}$ parameters and in Fig.\ref{fig:all} for $S_8$ only; additionally, the probability of the parameter difference is reported in Table \ref{tension}. The moments analysis is in good agreement with the other weak lensing analyses considered here ($\lesssim1\sigma$), and it is the most constraining one (owing both to the larger data set and to the extra non-Gaussian information probed by the moments).

When comparing with the results from the \textit{Planck} analysis, however, we measure a larger tension, at the level of $2.2-2.8 \sigma$, depending on the combination of moments considered (see Fig. \ref{fig:planck}). The third moment independently is in $2.8\sigma$ tension with {\it Planck}, which provides a cross-check on the other analyses of 2-point correlations. Note that the joint constraint, though tighter, is in slightly lower tension. Interestingly, the moments analysis is significantly more constraining than \textit{Planck} for the $S_8$ parameter.

When comparing results from different analyses, we did not try to unify different analysis choices (e.g, priors, scale cuts, etc.); this complicates the comparison \citep{Chang2019}. Nevertheless, the moments analysis, in line with other weak lensing analyses, favours lower $S_8$ values than \textit{Planck}.

%{x Bhuv: expand Planck paragraph; mention S8 constraints are similar;mention .}

\begin{table}
%\tiny
\caption {Probability of the parameter difference (computed over the full parameter space) between the DES Y3 moments analysis and three analyses using external data sets: KIDS-1000 survey \citep{Asgari2021}, HSC \citep{Hikage2019}, and \textit{Planck} \citep{aghanim2020planck}. }
\centering
%\begin{adjustbox}{width=0.7\textwidth}
\begin{tabular}{|c|c|c|c|}
 \hline
 & \textbf{\textit{Planck}} TTTEEE & HSC Y1& KIDS-1000\\

 &  lowl lowE & Power spectrum&\\
 \hline
2nd moments      & 2.7$\sigma$ & 0.3$\sigma$& 0.9$\sigma$\\
3rd moments      & 2.8$\sigma$ & 1.2$\sigma$ & 0.2$\sigma$\\
2nd+3rd moments  & 2.2$\sigma$ & 0.9$\sigma$ & 0.6$\sigma$\\
\hline
\end{tabular}
%\end{adjustbox}
\label{tension}
\end{table}

\section{Summary}\label{sect:summary}
We presented a cosmological analysis of the second and third moments of  weak lensing mass (convergence) maps from the third year (Y3) data of the Dark Energy Survey (DES). The second moment of the convergence as a function of smoothing scale contains information similar to standard shear 2-point statistics, whereas the third moment, or skewness, contains additional non-Gaussian information. 
Several theoretical studies have explored the use of statistics beyond 2-point correlations to extract additional non-Gaussian information from lensing data. The 3-point function is the lowest order statistic in perturbation theory and is the simplest to model and interpret. Its  signal-to-noise is significantly smaller than for 2-point correlations, but its dependence on the key cosmological parameters ($S_8$ and $\Omega_{\rm m}$) differs, enabling partial degeneracy breaking and improved  constraints on  cosmological parameters. Our study is the first to test these theoretical expectations with data in a comprehensive way, following  an end-to-end analysis of mock catalogues that included the expected leading sources of systematic uncertainty (see \G{}). We note that the counts of peaks in the lensing field are analysed in a separate DES paper \citep{Zuercher2021b} and other non-Gaussian statistics such as the topological Minkowski functionals as well as deep learning approaches have been proposed as well (see \S 1 for a review).

Our analysis relies on 100 million galaxy shapes measured over 4139 square degrees, which have been used to reconstruct the convergence field in four source redshift bins. The data has been analysed in the context of the $\Lambda$CDM model, varying 5 cosmological parameters ($\Omega_{\rm m}$, $\sigma_8$, $n_{\rm s}$, $\Omega_{\rm b}$, and $h_{\rm 100}$) and 19 nuisance parameters (modelling astrophysical and measurement uncertainties). One of our goals is to quantify the tension between CMB and late time estimates of $S_8$ and other relevant parameters. In view of several recent measurements reporting tension between the amplitude of mass fluctuations in the late times vs. early universe (as probed by the CMB), we have carried out measurements and consistency tests of $\Lambda$CDM rather than pursue extended cosmological models. The modelling used to describe the second and third moments measured in data is analytical: as described in \G{} we have built an emulator to obtain rapid predictions from perturbation theory calculations well tested with N-body simulations. Thus the cosmological analysis here does not rely on large suites of N-body simulations to forward model the signal.

The combined analysis of second and third moments was able to constrain $S_8 \equiv \sigma_8(\Omega_{\rm m}/0.3)^{0.5}$ with 1.7 percent uncertainty and $\Omega_{\rm m}$ with 10 percent: in particular, we obtained $S_8 =  0.784\pm0.013$ and $\Omega_{\rm m} =   0.27\pm0.03 $. The third moments improved the constraints on $S_8$ and $\Omega_{\rm m}$ by $\sim 15$ and $\sim 25$ per cent, respectively, in line with the expectation based on simulations (\G{}). The improvement is due to the degeneracy breaking and the non-Gaussian information probed by the third moments. The goodness-of-fit $p-$value of the data vectors (second, third, and the combination of second and third moments) was found to be way larger than 1 per cent, which is our criterion for a reasonable goodness-of-fit.

We performed our analysis following the blinding scheme proposed by \cite{y3-blinding}. Before unblinding the analysis, we performed a number of systematic tests which had been defined as unblinding criteria: we checked that additive biases due to PSF modelling errors were small enough to not bias the cosmological analysis; that mixed moments between convergence map E-modes and noise were consistent with expectations based on tests on N-body simulations;  that cosmological constraints obtained using second and third moments were consistent with each other using posterior predictive distributions (PPD, \cite{Doux2021});  that the best-fitting cosmology provided a good description of the B-modes of the second and third moments as well (the B-modes  were not included in the data vector used for the cosmological analysis); that the posteriors of the nuisance parameters did not concentrate at the edge of the prior, tested using the Gaussian estimator update difference-in-mean (UDM) statistic \citep{Raveri2019}. All these tests were successfully passed. After unblinding, we further used PPD to assess the internal consistency of other subsets of the data vector (small vs. large scales, or across redshift bins); we also tested that our results were robust against different modelling choices for the covariance matrix used in the analysis, or the inclusion of small-scale galaxy-galaxy lensing ratios (a.k.a. shear ratios, \cite{y3-shearratio}). All tests performed after unblinding validated the robustness of our results.

Constraints from the combination of second and third moments were found to be compatible with constraints from the DES Y3 cosmic shear analysis \citep{y3-cosmicshear1,y3-cosmicshear2}, the DES Y3 3x2pt analysis \citep{y3-3x2ptkp}, and the DES Peaks + Power spectrum analysis \citep{Zuercher2021b}. In terms of constraining power, the addition of non-Gaussian information via the third moments in the analysis may be regarded as  successful -- the  constraints on $S_8$ and $\Omega_{\rm m}$ were shown to be tighter than from DES cosmic shear, and, for $S_8$, similar to the DES 3x2pt constraint.  

We compared our constraints to two contemporaneous lensing surveys:  the KIDS-1000 survey \citep{Asgari2021} and  the HSC Y1 data \citep{Hikage2019}, finding  agreement ($\lesssim 1 \sigma$). 

When compared to predictions based on CMB data from the \textit{Planck} satellite \citep{aghanim2020planck}, we estimate a $2.2-2.8\sigma$ tension in the full parameter space, depending on the combination of moments considered (see Table \ref{tension}). The moments analysis  favours lower $S_8$ values compared to \textit{Planck}, in line with other weak lensing analyses. Interestingly the third moment by itself is in tension with Planck at the 2.8$\sigma$ level: this is significant since additive lensing systematics are more likely to impact the second moment than the third. So the third moment provides a useful check on the `low $S_8$' cosmic tension between the late time and early universe.

{We expect to improve the analysis presented in this work and apply it to future data, such at the final DES Y6 data. Based on the investigation performed in \G\, we expect to further improve our constraining power on $S_8$ by roughly 20 per cent, if we take into account the expected increase in the source number density. We plan to be able to model baryonic effects, which should allow us to push our analysis to smaller scales, improving constraints (up to 20 per cent, Appendix \ref{sect:scale_cuts}) and learning about baryonic physics. We are also planning to expand our modelling to include massive neutrinos and the full $w$CDM parameter space.}

\section*{Acknowledgements}
Funding for the DES Projects has been provided by the U.S. Department of Energy, the U.S. National Science Foundation, the Ministry of Science and Education of Spain, 
the Science and Technology Facilities Council of the United Kingdom, the Higher Education Funding Council for England, the National Center for Supercomputing 
Applications at the University of Illinois at Urbana-Champaign, the Kavli Institute of Cosmological Physics at the University of Chicago, 
the Center for Cosmology and Astro-Particle Physics at the Ohio State University,
the Mitchell Institute for Fundamental Physics and Astronomy at Texas A\&M University, Financiadora de Estudos e Projetos, 
Funda{\c c}{\~a}o Carlos Chagas Filho de Amparo {\`a} Pesquisa do Estado do Rio de Janeiro, Conselho Nacional de Desenvolvimento Cient{\'i}fico e Tecnol{\'o}gico and 
the Minist{\'e}rio da Ci{\^e}ncia, Tecnologia e Inova{\c c}{\~a}o, the Deutsche Forschungsgemeinschaft and the Collaborating Institutions in the Dark Energy Survey. 

The Collaborating Institutions are Argonne National Laboratory, the University of California at Santa Cruz, the University of Cambridge, Centro de Investigaciones Energ{\'e}ticas, 
Medioambientales y Tecnol{\'o}gicas-Madrid, the University of Chicago, University College London, the DES-Brazil Consortium, the University of Edinburgh, 
the Eidgen{\"o}ssische Technische Hochschule (ETH) Z{\"u}rich, 
Fermi National Accelerator Laboratory, the University of Illinois at Urbana-Champaign, the Institut de Ci{\`e}ncies de l'Espai (IEEC/CSIC), 
the Institut de F{\'i}sica d'Altes Energies, Lawrence Berkeley National Laboratory, the Ludwig-Maximilians Universit{\"a}t M{\"u}nchen and the associated Excellence Cluster Universe, 
the University of Michigan, NSF's NOIRLab, the University of Nottingham, The Ohio State University, the University of Pennsylvania, the University of Portsmouth, 
SLAC National Accelerator Laboratory, Stanford University, the University of Sussex, Texas A\&M University, and the OzDES Membership Consortium.

Based in part on observations at Cerro Tololo Inter-American Observatory at NSF's NOIRLab (NOIRLab Prop. ID 2012B-0001; PI: J. Frieman), which is managed by the Association of Universities for Research in Astronomy (AURA) under a cooperative agreement with the National Science Foundation.

The DES data management system is supported by the National Science Foundation under Grant Numbers AST-1138766 and AST-1536171.
The DES participants from Spanish institutions are partially supported by MICINN under grants ESP2017-89838, PGC2018-094773, PGC2018-102021, SEV-2016-0588, SEV-2016-0597, and MDM-2015-0509, some of which include ERDF funds from the European Union. IFAE is partially funded by the CERCA program of the Generalitat de Catalunya.
Research leading to these results has received funding from the European Research
Council under the European Union's Seventh Framework Program (FP7/2007-2013) including ERC grant agreements 240672, 291329, and 306478.
We  acknowledge support from the Brazilian Instituto Nacional de Ci\^encia
e Tecnologia (INCT) do e-Universo (CNPq grant 465376/2014-2).

This manuscript has been authored by Fermi Research Alliance, LLC under Contract No. DE-AC02-07CH11359 with the U.S. Department of Energy, Office of Science, Office of High Energy Physics.

\section*{Data availability}
The simulated data used in this work has been generated using the public code \texttt{FLASK} (\url{http://www.astro.iag.usp.br/~flask/}), the public T17 simulations (\url{http://cosmo.phys.hirosaki-u.ac.jp/takahasi/allsky_raytracing/}), and the public code \textsc{PKDGRAV} \citep{potter2017pkdgrav3}. The full \mcal\ catalogue will be made publicly available following publication, at the URL \url{https://des.ncsa.illinois.edu/releases}. The code used in this article will be shared on request to the corresponding author.

%%%%%%%%%%%%%%%%%%%%%%%%%%%%%%%%%%%%%%%%%%%%%%%%%%

%%%%%%%%%%%%%%%%% APPENDICES %%%%%%%%%%%%%%%%%%%%%

\appendix
\section{Scale cuts}\label{sect:scale_cuts}
\begin{figure*}
\begin{center}
\includegraphics[width=0.4 \textwidth]{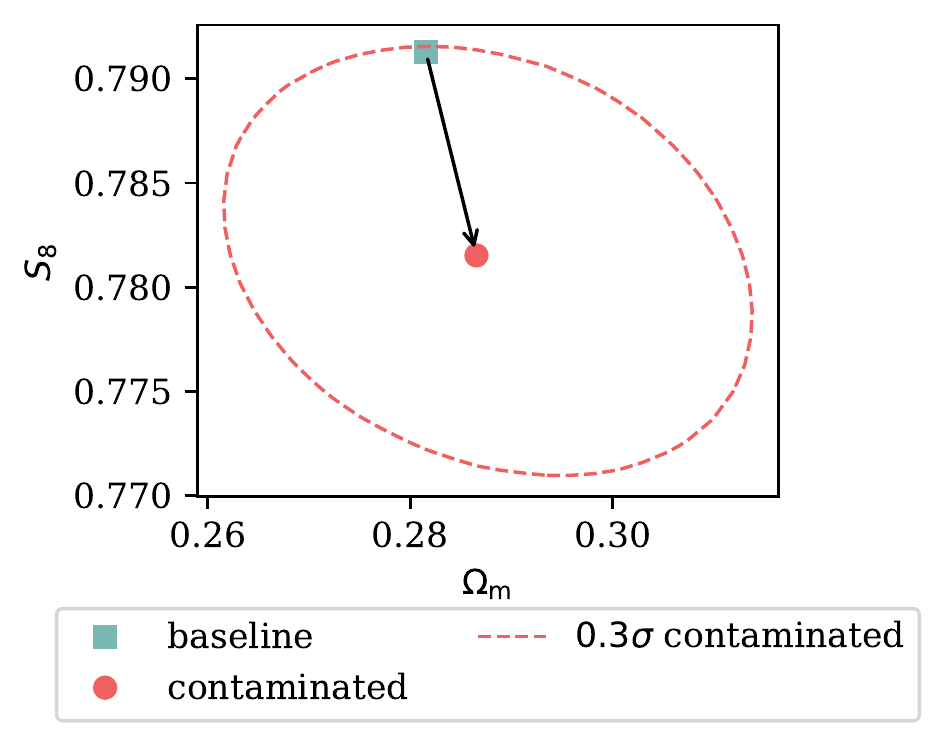}
\includegraphics[width=0.4 \textwidth]{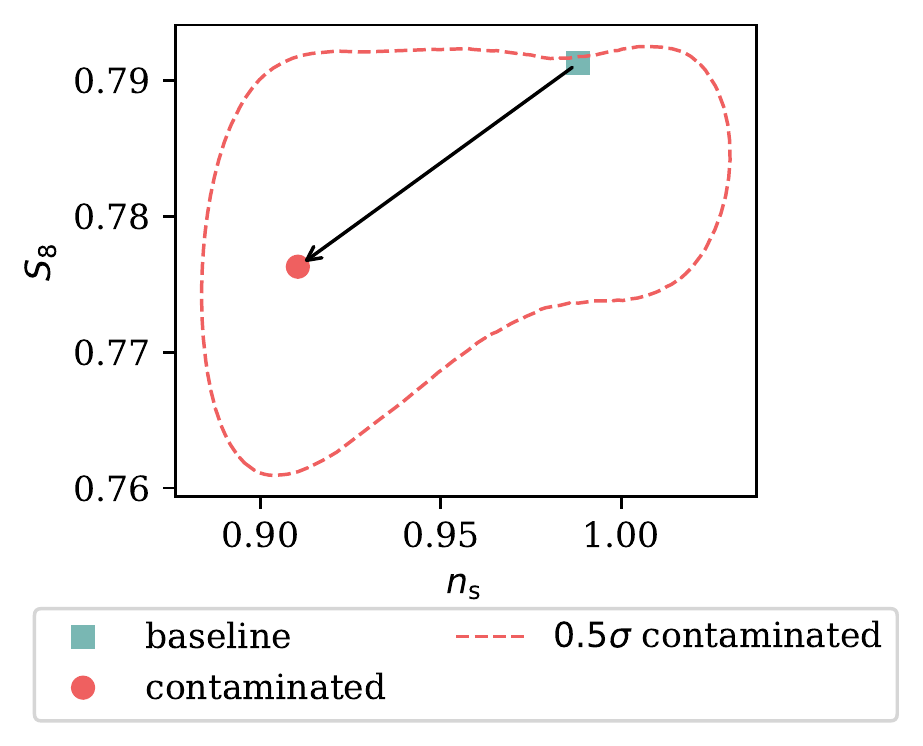}
\end{center}
\caption{Parameter posteriors used to determine the scale cuts for the cosmological analysis. Constraints from the combination of second and third moments are shown. `Baseline' refers to an analysis performed on a theory data vector, `Contaminated' refers to the analysis performed on a data vector contaminated by the impact of baryonic feedback (see Appendix~\ref{sect:scale_cuts}). The dashed lines demarcate the $0.3\sigma$ or $0.5\sigma$ contours for the 2D marginalised
constraints of the contaminated data vector; the filled square and circle show the peak of the posteriors for the Contaminated and Baseline data vectors, respectively.}
\label{fig:scale_cuts}
\end{figure*}

We repeat on DES Y3 data the scale cut test we performed on simulated data in \G{} in order to determine which part of the data vector can be used in the cosmological analysis. The reason the test is repeated is that some details of the analysis have been updated since \G{} (mostly the nuisance parameters priors and the redshift distributions). The scale cut test is performed by contaminating a theory data vector with the known dominant systematic effect that is not part of our model, namely baryonic feedback based on hydrdynamical simulations as described in \G{}. Then, we check that the cosmological parameters posterior obtained by analysing the contaminated data vector is not substantially biased with respect to the posterior with an uncontaminated data vector. 

We adopted the `optimised scale cut criteria' for the $\Lambda$CDM cosmology adopted by the main DES cosmological analysis \citep{y3-cosmicshear2,y3-cosmicshear1,y3-3x2ptkp}. The criterion requires the peak of the marginalised 2-D posterior of $\Omega_{\rm m}$ and $S_8 \equiv \sigma_8 (\Omega_{\rm m}/0.3)^{0.5}$ obtained by analysing the contaminated data vector to be within $0.3 \sigma$ of the values obtained with the uncontaminated one.  As we partially constrain $n_{\rm s}$, we also require the peak of the marginalised 2-D posterior of $n_{\rm s}$ and $S_8$ to be within $0.5 \sigma$ of the baseline value. {We arbitrarily chose a larger value for the $n_{\rm s}$ and $S_8$ criteria because $n_s$ is only partially constrained and the posterior might be artificially too sharp. We also note that the DES Y3 3x2pt analysis does not assume any scale cut criteria on $n_s$.} We also check that the $\chi^2$ of the best-fitting cosmology of the analysis of the contaminated data vector is within $0.3$ of the expected spread of the $\chi^2$ distribution. Since the length of the compressed data vector is 15, we require the best-fitting $\chi^2 < 1.6$\footnote{{Note that we are considering a $\chi^2$ statistic and not a reduced $\chi^2$ statistic. The reported $\chi^2$ might seem small due to the small number of d.o.f (15, due to data-compression) and due to the lack of measurement noise in the input data vectors. For negligible contamination we would expect a best-fit  $\chi^2 = 0$ (instead of $\chi^2 \sim $ d.o.f. for a noisy data vector).}}. This second criterion ensures that the best-fitting $\chi^2$ from the analysis on data is unbiased. We note that these last two checks have not been included in the scale cut criteria in the main DES cosmological analysis. 

In \G{}, we determined that a scale cut of $R_0 = 24 h^{-1}$ Mpc was sufficient (such that scales smaller than $\theta_0 = R_0/ \chi(\avg{z})$ were removed, where $\avg{z}$ is the average of the mean redshift of different tomographic bins). When repeating this test, we had to use slightly large scales ( $R_0 = 28 h^{-1}$ Mpc) to pass the scale cut criteria,  due to our updated analysis choices (e.g., inclusion of the shear-ratio likelihood, final values for redshift distributions, shape noise, effective number densities, covariance, etc.). Results are shown in Fig.~\ref{fig:scale_cuts}: the peak of the 2-D posterior of the contaminated data vector is $0.28\sigma$ off the baseline value in the $\Omega_{\rm m}$ - $S_8$ plane, and $0.48\sigma$ in the $n_{\rm s}$ - $S_8$ plane; we also obtain a best-fitting $\chi^2=0.91<1.6$ for the contaminated data vector. Therefore, the scale cut of $R_0 = 28 h^{-1}$ Mpc is deemed sufficient. 

{We note that our scales cut removes a significant number of data points from our measurement. This has a non negligible impact on our constraining power. Using a simulated data vector, we estimate that we would improve our constraints on $S_8$ and $\Omega_{\rm m}$ by a further 20 per cent if we could apply no scale cut. This assumes we had a perfect knowledge of the baryonic effects on our data vector, which, unfortunately, is not the case for this analysis. }

\section{Validation of the modelling on N-body simulations}\label{sect:validation}

\begin{figure*}
\begin{center}
\includegraphics[width=0.85 \textwidth]{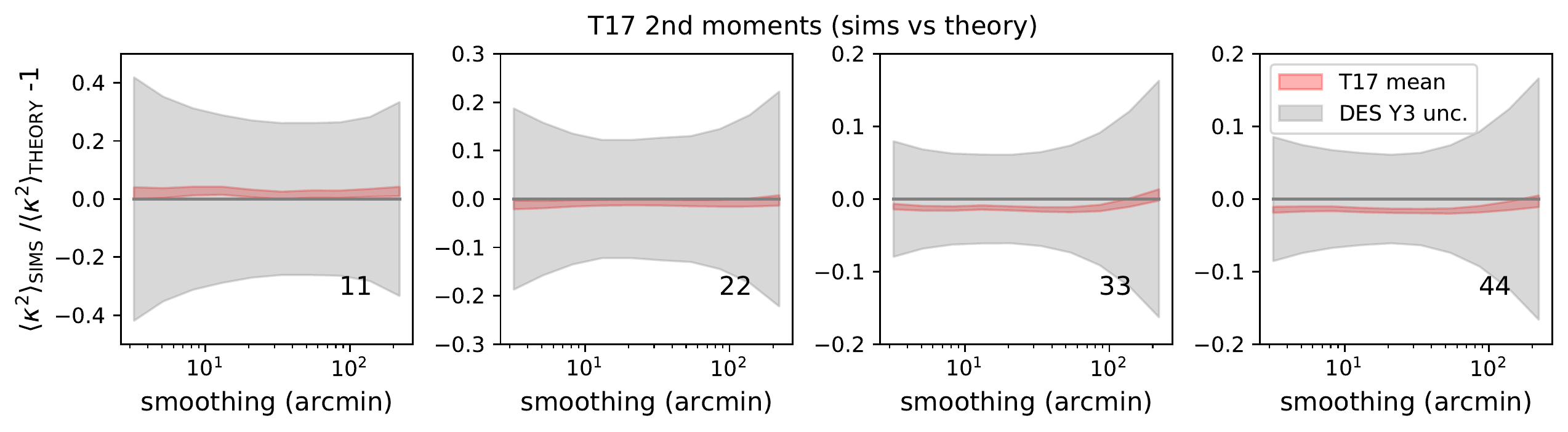}
\includegraphics[width=0.85 \textwidth]{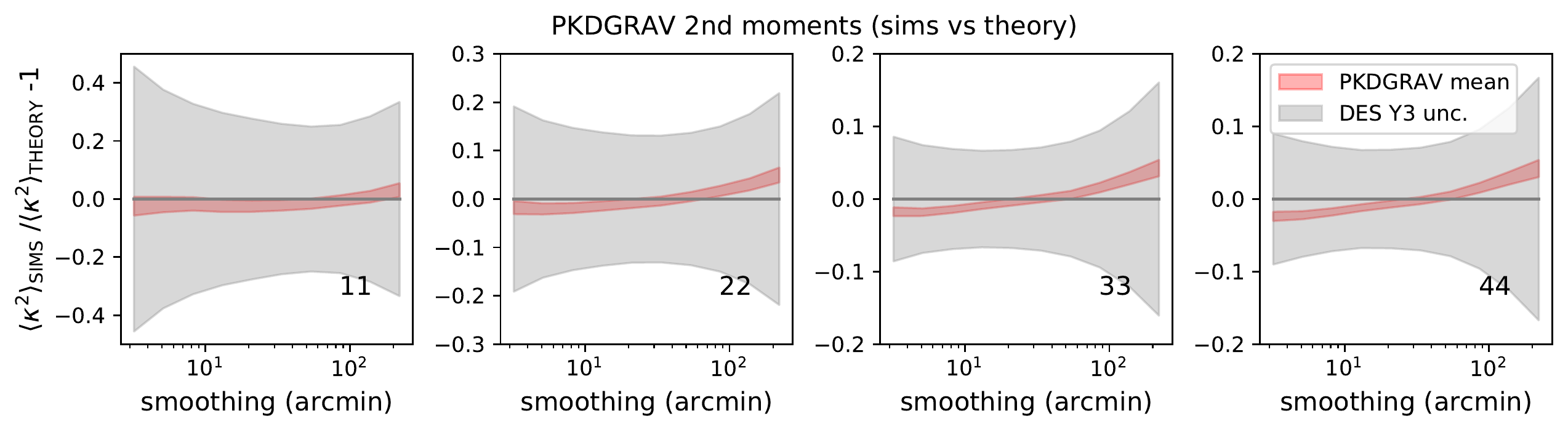}
\includegraphics[width=0.85 \textwidth]{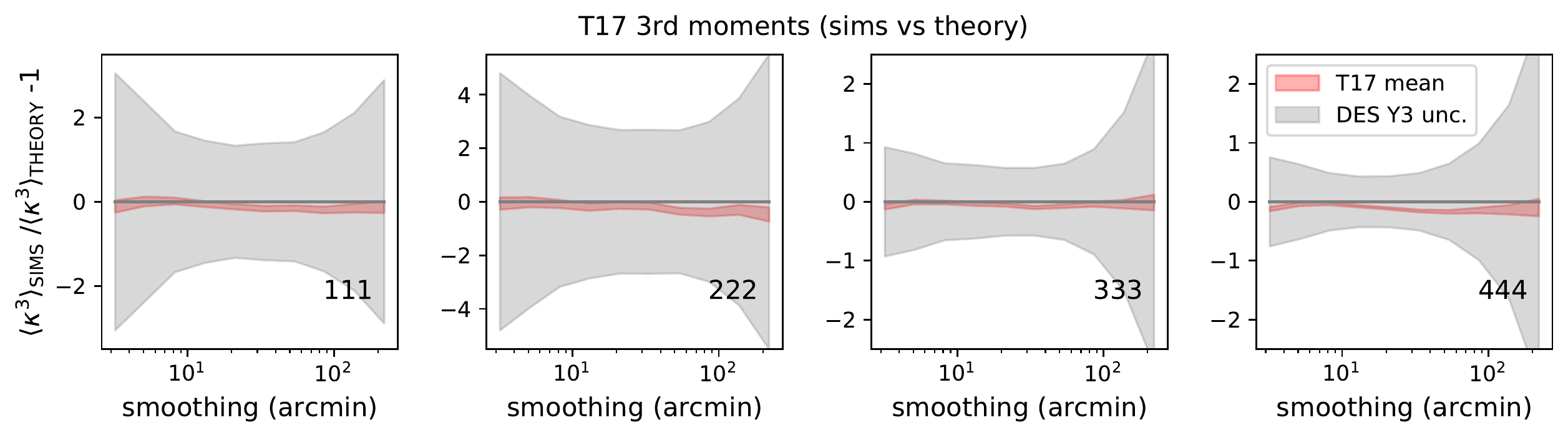}
\includegraphics[width=0.85 \textwidth]{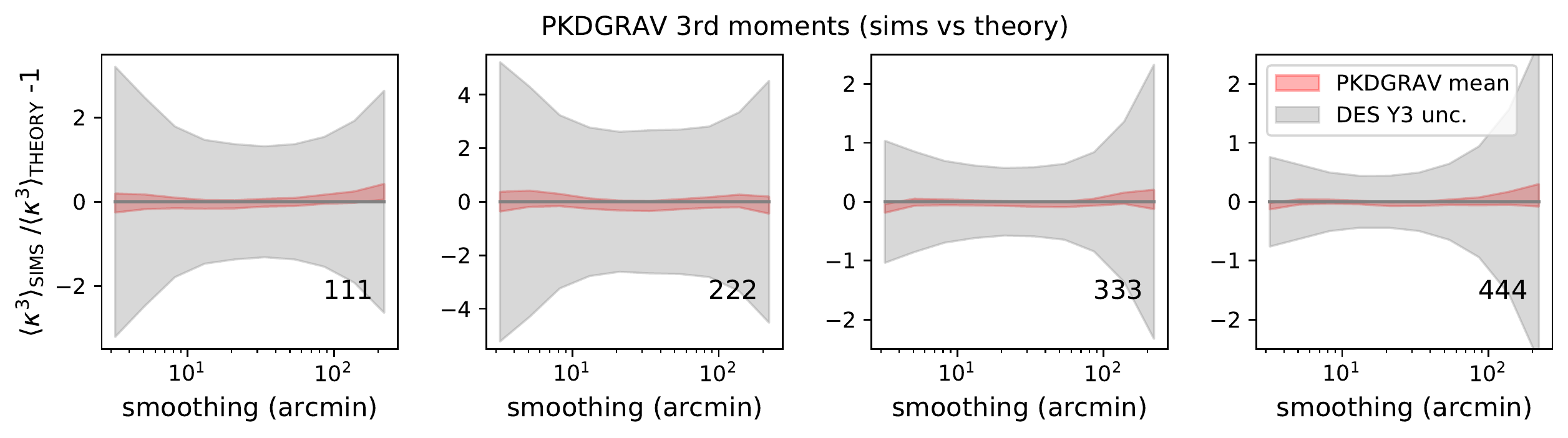}
\end{center}
\caption{Comparison between theory predictions and moments as measured in N-body simulations. The red bands encompass the 68 percentile of the moments as measured on all the realisations from the T17 or PKDGRAV simulations. The grey bands represent the expected measurement uncertainty for one individual realisation, which represents the DES Y3 survey. No scale cut is applied here. Only `auto' moments are shown. It is evident that the theoretical model agrees with N-body simulations to well within the statistical uncertainty of the survey. }
\label{fig:theory_sims_comparison}
\end{figure*}

\begin{figure*}
\begin{center}
\includegraphics[width=0.4 \textwidth]{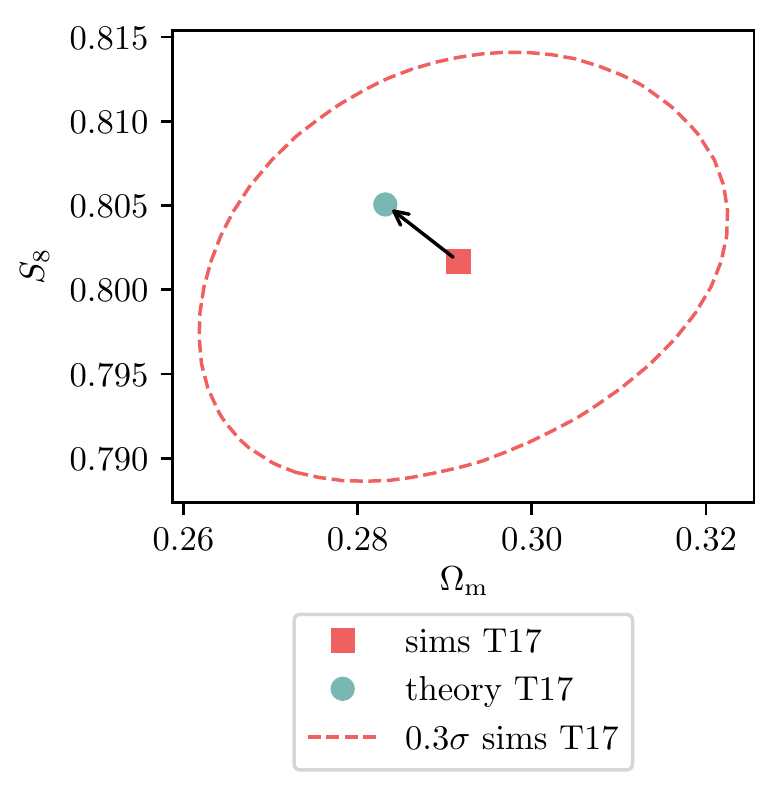}
\includegraphics[width=0.4 \textwidth]{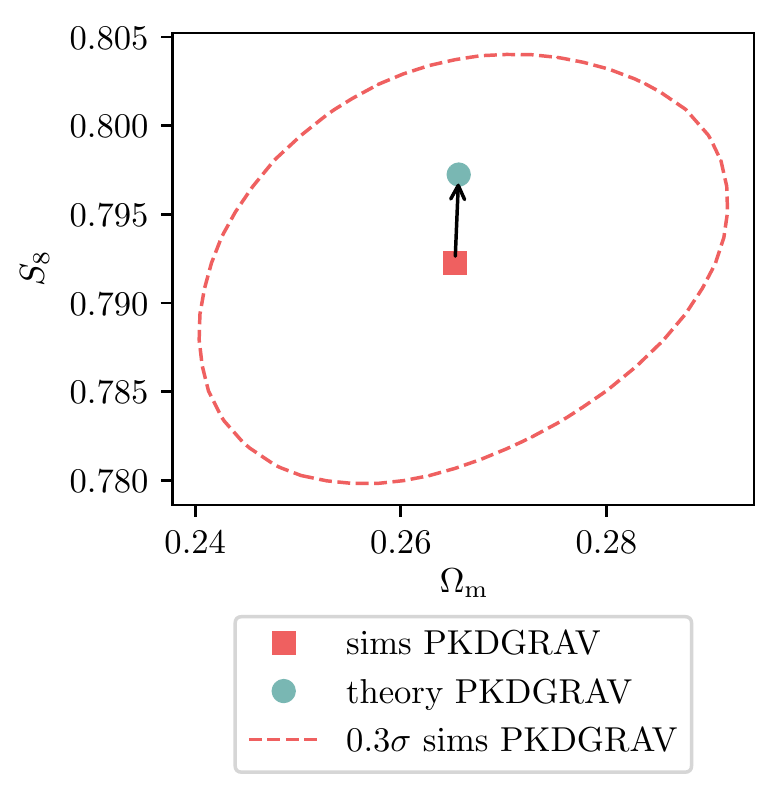}
\end{center}
\caption{Parameter posteriors used to validate our modelling of second and third moments with N-body simulations. Constraints from the combination of second and third moments are shown. `Theory' points refer to the peak of the posteriors of the synthetic data vectors computed at the cosmology of either the T17 or PKDGRAV simulation, `sims' squares refer to the peaks of the posteriors of the analyses run on the average data vectors from all the realisations of the N-body simulations. The lines demarcate the $0.3\sigma$ contours for the 2D marginalised
constraints of the contaminated data vector.}
\label{fig:cosmology_validation}
\end{figure*}

We repeat in this Appendix the validation of our theoretical modelling performed in \G{}. We repeat that validation  for two reasons: 1) some of our analysis choices have been updated (e.g., priors, redshift distributions, covariance, etc.); 2) we perform the validation on two independent N-body simulations (whereas in \G{} we compared only to one). 

We show first in Fig.~\ref{fig:theory_sims_comparison} the comparison between theory predictions and the data vector as measured in the two sets of N-body simulations. For the data vector, we take the average of the data vector measured in every realisation available. The mean offset between measurements and predictions is 0.5 and 8 per cent for second and third moments of the T17 simulations, and 0.005 per cent for both second and third moments of the PKDGRAV simulations (note that there are scale dependent residuals that are larger, but they average down when computing a mean offset). Note also that these numbers for the second moments are in agreement with the quoted uncertainties for the power spectrum for the two sets of simulations \citep{Takahashi2017,potter2017pkdgrav3}; in particular, the pattern seen for the second moments of the PKDGRAV simulations is similar to that shown in \cite{Zuercher2021b}.

We used the measured data vector (averaged over all the available realisations) and the scale cut determined in Appendix \ref{sect:scale_cuts} to run two simulated cosmological analyses, one for each set of simulations. We compared the posteriors of  $\Omega_{\rm m}$ and $S_8 \equiv \sigma_8 (\Omega_{\rm m}/0.3)^{0.5}$ with the posteriors obtained running the same cosmological analysis on a synthetic data vector at the `true' cosmology of the two simulations. Results are shown in Fig. \ref{fig:cosmology_validation}, showing a good recovery of the true cosmological parameters.
\section{Additive biases due to PSF error}\label{sect:PSF_modelling_errors}
\begin{figure*}
\begin{center}
\includegraphics[width=0.95 \textwidth]{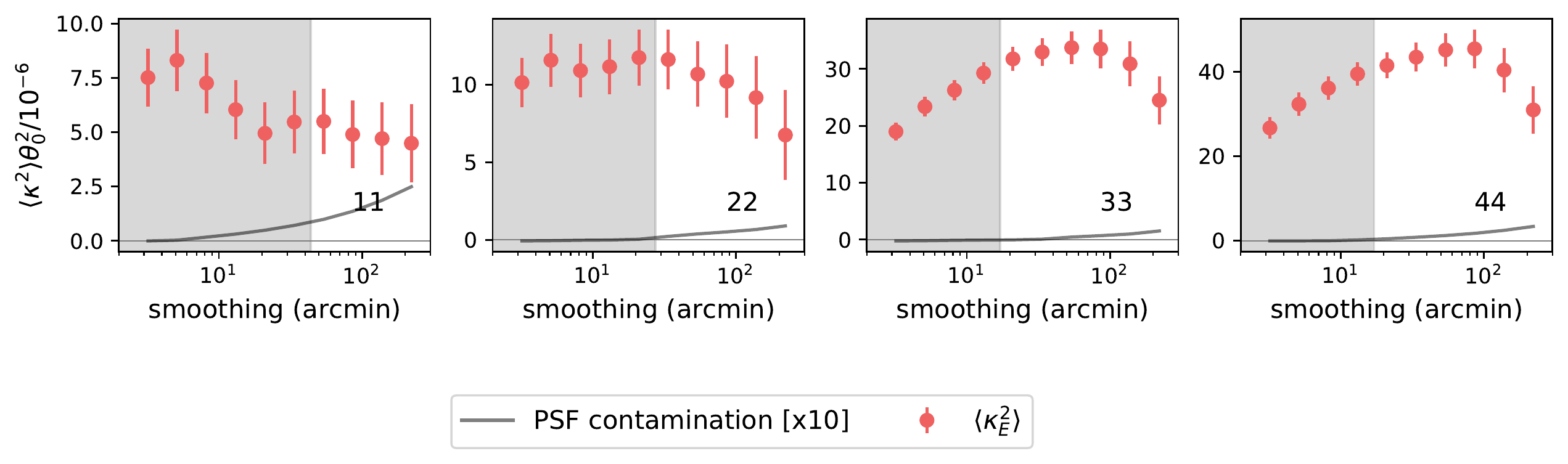}
\includegraphics[width=0.95 \textwidth]{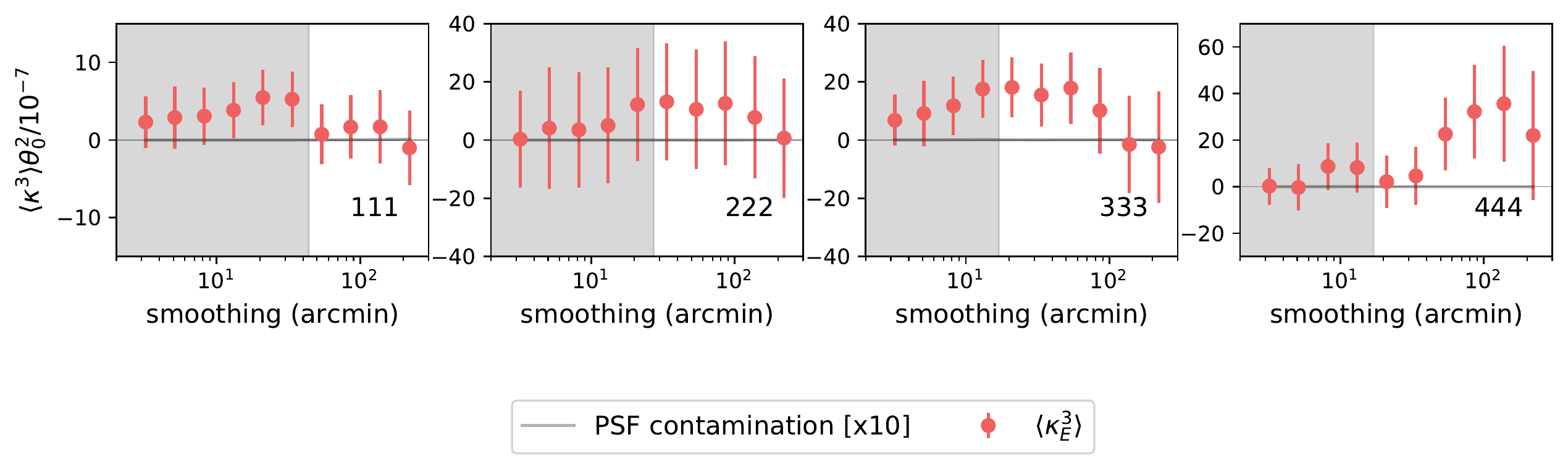}
\end{center}
\caption{Contribution of PSF modelling errors to the second and third moments of the convergence fields. The contribution, estimated as explained in Appendix~\ref{sect:PSF_modelling_errors}, is shown as the black lines. Due to its small amplitude, the signal has been multiplied by a factor of 10. The red points represent the measured moments of the convergence field. We only show `auto' moments here, although the pattern is very similar for all the other moments. Grey shaded regions highlight the scales removed by the scale cut used in the analysis.  }
\label{fig:EE_PSF_contamination}
\end{figure*}
\begin{figure}
\begin{center}
\includegraphics[width=0.45 \textwidth]{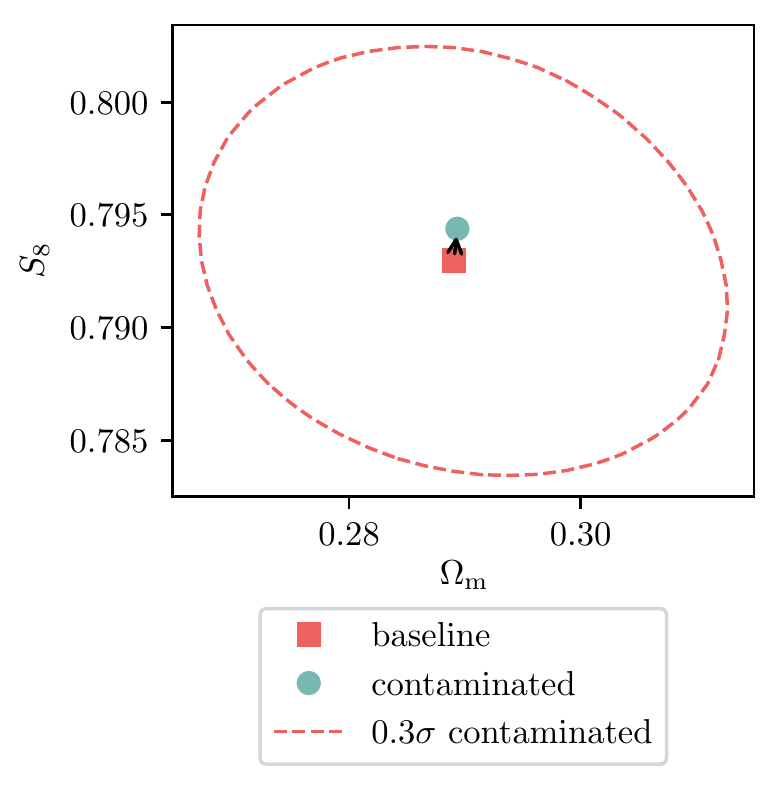}
\end{center}
\caption{Parameter posteriors used to determine the level of PSF additive bias contamination. Constraints from the combination of second and third moments are shown. `Baseline' refers to an analysis performed on a theory data vector, `contaminated' refers to the analysis performed on a data vector contaminated by the impact of PSF additive biases (see Appendix~\ref{sect:PSF_modelling_errors}). The dashed lines demarcate the $0.3\sigma$ contours  for the 2D marginalised
constraints of the contaminated data vector; the filled square and circle show the peak of the posteriors. }
\label{fig:EE_PSF_contamination_contours}
\end{figure}
We quantify in this Appendix the level of contamination of our data vector due to additive biases related to PSF misestimation. PSF misestimation can cause additive biases in the  measured galaxy shapes such that $\vecgest=\vecg +\delta \vest^{\textrm{sys}}_{\textrm{PSF}}+\delta\vest^{\textrm{noise}}$. These spurious contributions can be characterised assuming a model for the PSF modelling errors and using a catalogue of `reserved' stars that have not been used to train the PSF model. In what follows, we parameterise additive biases due to PSF misestimation following \cite{Jarvis2016}, \cite{y3-shapecatalog} (other modelling choices also exist in literature, e.g, \cite{Giblin2020}). In particular, we assume that:
\begin{equation}
\label{eq:new}
\delta \vest^{\textrm{sys}}_{\textrm{PSF}}=\alpha \textbf{e}_{\rm model}+\beta\left(\textbf{e}_{\rm *}-\textbf{e}_{\rm model}\right)+\eta\left(\textbf{e}_{\rm *}\frac{T_{\textrm{\rm *}}-T_{\rm model}}{T_{\rm *}}\right),
\end{equation}
where $\alpha,$ $\beta$, and $\eta$ are coefficients estimated from data, $e_{\rm *}$ is the PSF ellipticity measured directly using the reserved stars catalogue, $T_{\rm model}$ is the modelled PSF size, and $T_{\rm *}$ is the PSF size measured from the reserved stars catalogue. The coefficients $\alpha,$ $\beta$, and $\eta$ for the DES Y3 shape catalogue have already been estimated in \cite{y3-shapecatalog} for the non tomographic case and in \cite{y3-cosmicshear1} for the tomographic case. In what follows, we will use the values from \cite{y3-cosmicshear1}, as we are interested in the contamination of our tomographic moments.

 An empirical method was used to estimate the contribution to the measured moments due to PSF additive biases. We first created maps of $\textbf{e}_{\rm model}$, $\textbf{e}_{\rm *}$, and $\frac{T_{\textrm{\rm *}}-T_{\rm model}}{T_{\rm *}}$ from the reserved stars catalogue. Using the estimated values for $\alpha,$ $\beta$, and $\eta$, we then created maps of $\delta \vest^{\textrm{sys}}_{\textrm{model}}$, one for each tomographic bin. Last, we computed the second and third moments of the smoothed version of the $\delta \vest^{\textrm{sys}}_{\textrm{model}}$ maps, in exactly the same way that we estimated the moments of the convergence maps (\S~\ref{sect:map_making}). In order to estimate the contribution due to noise (that has to be subtracted from the raw, measured moments), we adopted a different technique as the two components of the $\delta \vest^{\textrm{sys}}_{\textrm{model}}$ field cannot just be randomly rotated as in the case of galaxies. We created two additional versions of the $\delta \vest^{\textrm{sys}}_{\textrm{model}}$ maps, obtained by sampling two disjoint halves of the reserved stars catalogue. We made sure the two halves spanned the footprint uniformly. We then measured the moments of the difference of the two maps, $\delta \vest^{\textrm{sys}}_{\textrm{model,DIFF}}$. In this way, the true signal should cancel, leaving only a contribution due to noise. The noise contribution to the moments of the $\delta \vest^{\textrm{sys}}_{\textrm{model}}$ maps can be related to the signal measured from  $\delta \vest^{\textrm{sys}}_{\textrm{model,DIFF}}$ as follows:
\begin{equation}
 \langle (\delta \vest^{\textrm{sys}}_{\textrm{model,NOISE}})^2\rangle^{i,j} =  4 \langle (\delta \vest^{\textrm{sys}}_{\textrm{model,DIFF}})^2\rangle^{i,j},
\end{equation}
\begin{equation}
 \langle (\delta \vest^{\textrm{sys}}_{\textrm{model,NOISE}})^3\rangle^{i,j,k} =  8 \langle (\delta \vest^{\textrm{sys}}_{\textrm{model,DIFF}})^3\rangle^{i,j,k},
\end{equation}
for any combination of tomographic bins $i$, $j$, and $k$. The second and third moments contribution due to PSF biases, once the noise term has been subtracted, is shown in Fig.~\ref{fig:EE_PSF_contamination}. It can be seen clearly that such contribution is subdominant with respect to the moments of the convergence field, and that it mostly affects the large scales of the second moments. To further evaluate the impact of PSF modelling errors, we ran a cosmological analysis on a theory data vector contaminated by the measured moments of the PSF bias, and compared to the results obtained with a cosmological analysis performed on an uncontaminated theory data vector. The results are shown in Fig. \ref{fig:EE_PSF_contamination_contours}, demonstrating that PSF additive biases have a negligible impact on our analysis.

\section{Noise terms and source clustering}\label{sect:noise_terms}

\begin{figure*}
\begin{center}
\includegraphics[width=0.85 \textwidth]{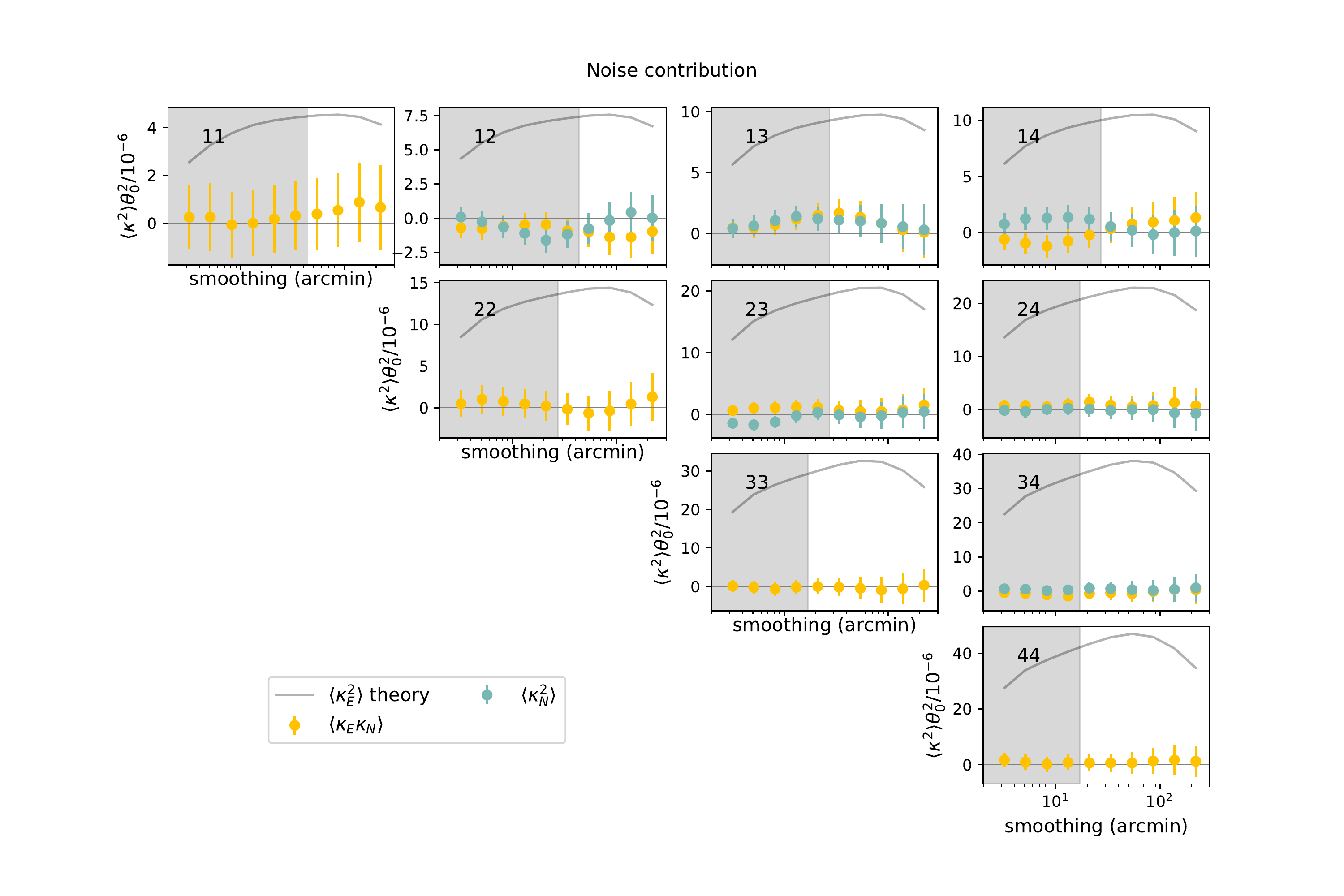}
\includegraphics[width=0.85 \textwidth]{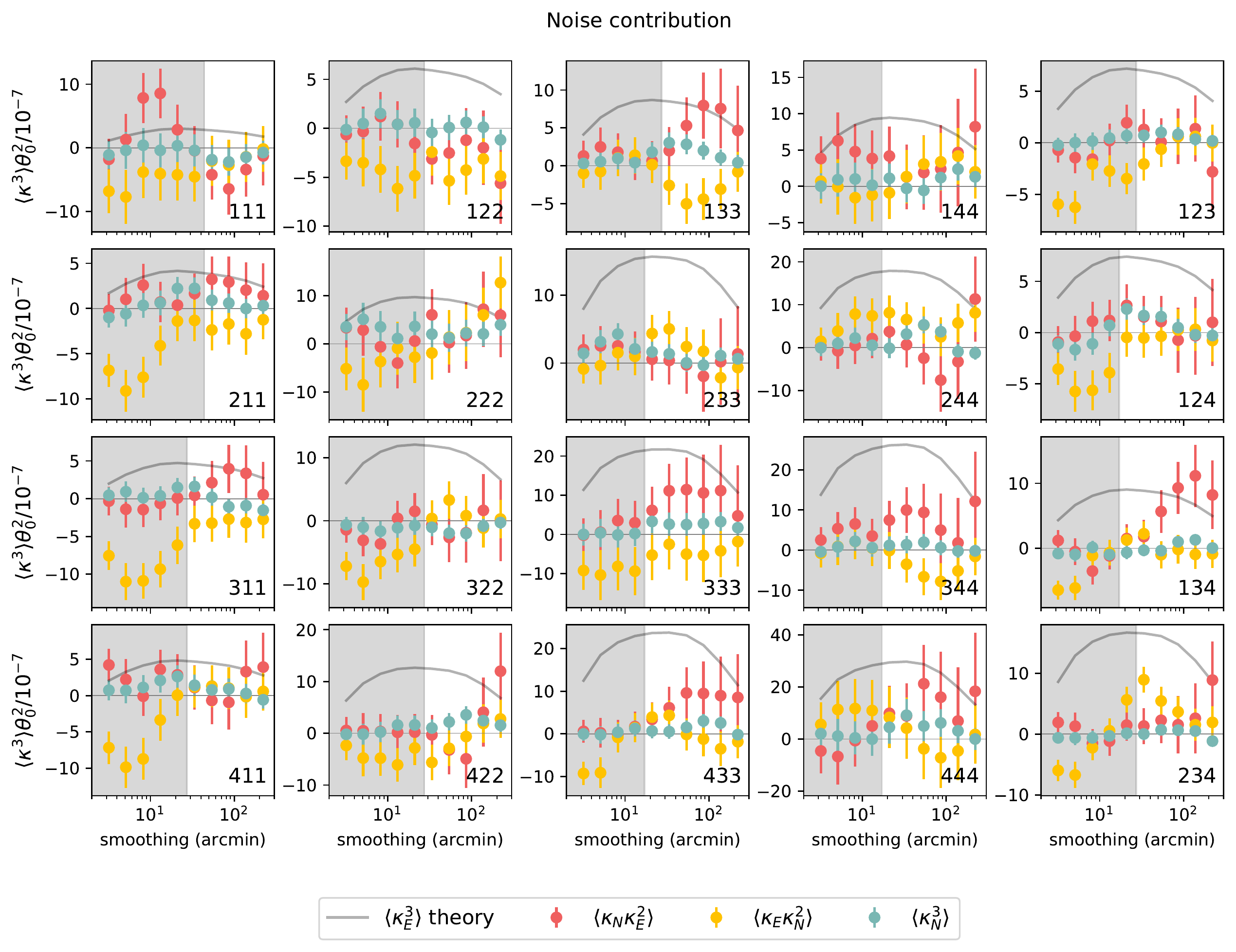}
\end{center}
\caption{Measured moments involving the noise contribution to the convergence map. We do not show $\langle\kappa_{{\rm N}}\rangle^{i,j}$ for $i=j$, as these moments are much larger than the signal $\langle\kappa_{{\rm E}}\rangle^{i,j}$ and are always subtracted. The grey line is shown for reference and represents the expected theoretical signal for E-mode second and third moments.  Grey shaded regions highlight the scales removed by the analysis.  }
\label{fig:noise}
\end{figure*}

We can only have a noisy estimate of the shear field and so when computing the moments of the convergence maps the contribution due to noise has to be properly subtracted, as explained in \S~\ref{sect:map_making} (Eqs. \ref{eq:deno2} and \ref{eq:deno3}). It is standard procedure to subtract only the contributions that are known to differ from zero, so as to not unnecessarily inflate the statistical uncertainty of our measurement. We show in Fig. \ref{fig:noise} the noise terms as measured from the data. The noise contribution to the convergence map $\kappa_{\rm N}$ has been obtained by randomly rotating the galaxy shapes and repeating the map-making procedure. For the second moments, we do not show the terms $\langle \kappa_{{\rm N}}\rangle^{i,j}$ when $i=j$, as they are much larger than the measurement $\langle \kappa_{{\rm E}}\rangle^{i,j}$ (at small scales, they are one order of magnitude larger) and are always subtracted. All the other terms are compatible with zero; the only exception concerns mixed terms of the form $\langle  \kappa \kappa^2_{\rm N}\rangle^{i,j,k}$, which presents some deviations from zero at small scales, especially in the moments involving the first tomographic bin (with a significance of $\chi^2\sim 20-25 / 10 \ \textrm{d.o.f.}$, depending on the bin combinations). This is in line with what we found in simulations in \G{}, {where such terms did not vanish due to correlations between the pixels' shape noise and the shear field value, induced by the intrinsic clustering of the sources \citep{y3-generalmethods}. These terms are subtracted from the measured moments before proceeding with the cosmological analysis; due to our scale cut this has a very small impact on the data vector used for the cosmological analysis. By using simulated data vectors with and without source clustering effects, we tested that this procedure is sufficient to remove the effect of source clustering and to have unbiased cosmological constraints (ignoring source clustering effects produces a shift in 2D $\Omega_{\rm m}$ - $S_8$ plane of only 0.08$\sigma$)}.
%source galaxy density--convergence field correlations. 

%These terms are subtracted from the measured moments before proceeding with the cosmological analysis; note that due to our scale cut, this has a very small impact on the data vector used for the cosmological analysis.

\section{B modes}\label{sect:B-modes}

\begin{figure*}
\begin{center}
\includegraphics[width=0.9 \textwidth]{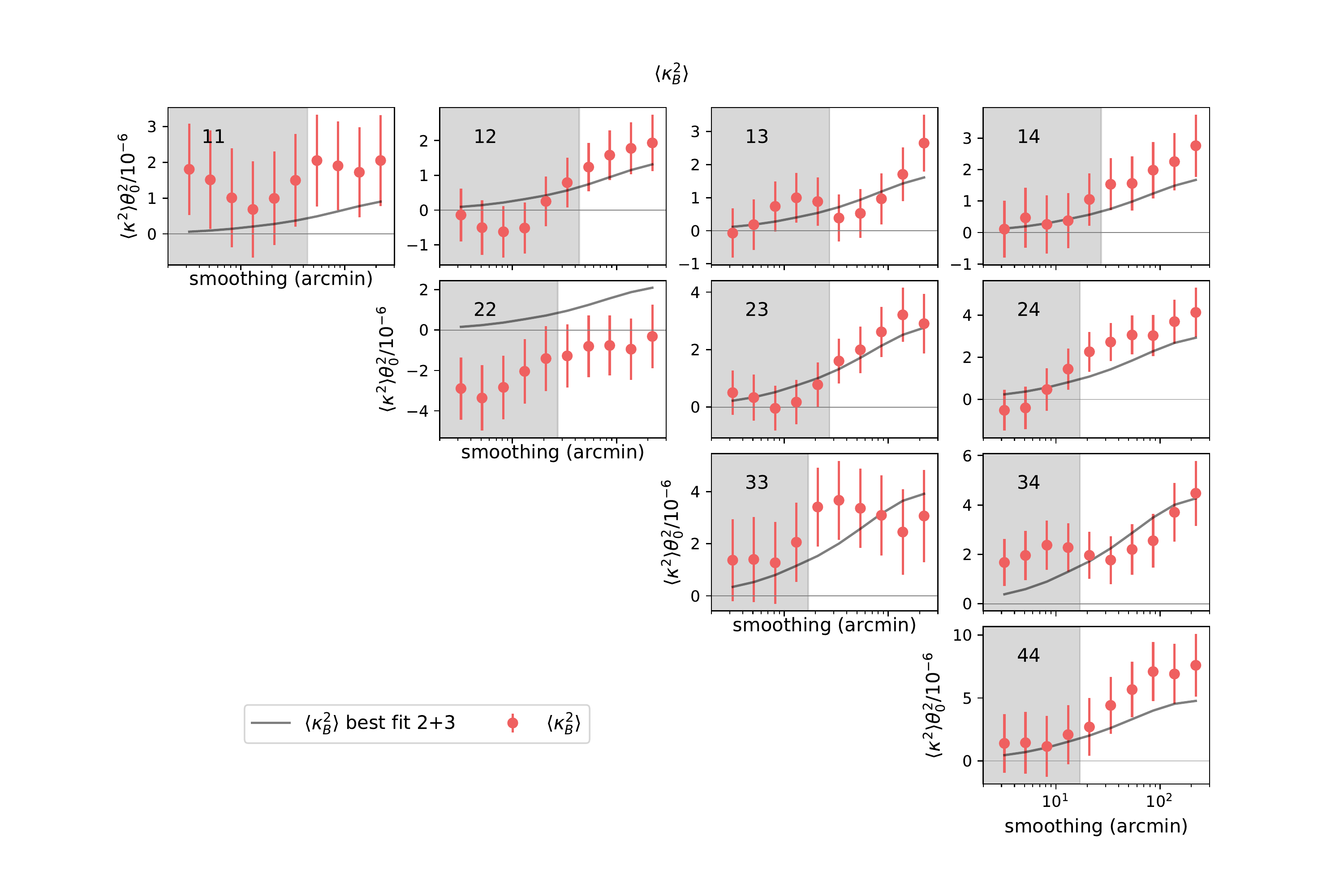}
\end{center}
\caption{Measured second and third moments of the B-modes convergence maps.   Grey shaded regions highlight the scales removed by the analysis.  The solid line represents the predicted B-modes at the best-fitting cosmology of E-modes second and third moments.}
\label{fig:B-modes}
\end{figure*}
We show in this Appendix the measured moments of the B-modes of the convergence  maps. As we used the Kaiser-Squires algorithm to obtain the mass maps, non-null B-modes are expected as a consequence of mask effects \citep{y3-massmapping}, and are not necessarily associated with any observational systematic. The measured second moments are shown in Fig.~\ref{fig:B-modes}. B-modes second moments are significantly non-zero; in the same figure, we also overplot the predicted B-modes given the best-fitting cosmology of the E-modes second moments, showing good agreement with the observed B-modes moments ($\chi^2=51/50 \ \textrm{d.o.f}$). We do not detect any B-modes third moments at a significant level ($\chi^2=127/108 \ \textrm{d.o.f}$); this is in line with the expected sensitivity of our data set and with the tests performed in \G{}.

\section{Internal consistency tests}\label{sect:internal}
\begin{table}
    \centering
    \begin{tabular}{l c c c c }
        \hline 
        PPD test \,\, & $p$-values & \\ %& $\Delta S_8$
         \hline 
        \textit{Data splits} \\  \\
        Bin 1 \textit{vs.} no bin 1 & 0.648   \\
        Bin 2 \textit{vs.} no bin 2 & 0.148   \\
        Bin 3 \textit{vs.} no bin 3 & 0.659  \\
        Bin 4 \textit{vs.} no bin 4 & 0.260   \\
        Large \textit{vs.} small scales & 0.391 \\ 
        Small \textit{vs.} large scales & 0.350  \\
        2nd \textit{vs.}
        3rd & 0.32 \\
        3rd \textit{vs.}
        2nd & 0.49 \\
        \hline 
    \end{tabular}

    \caption{\label{tab:IC} Summary of internal consistency test $p$-values. All internal consistency tests pass the pre-defined (arbitrary) threshold of 0.01. Besides the second vs third moments tests, all the other tests have been performed on the data vector including the combination of second and third moments. }
\label{table:PPD}
\end{table}

We quantify here the internal consistency of our data sets. Such tests, which rely on the PPD method, were performed prior to unblinding (using blinded data vectors) and were repeated after unblinding (although only the compatibility of the second and third moments was considered as an unblinding criterion).

\begin{figure}
\begin{center}
\includegraphics[width=0.45 \textwidth]{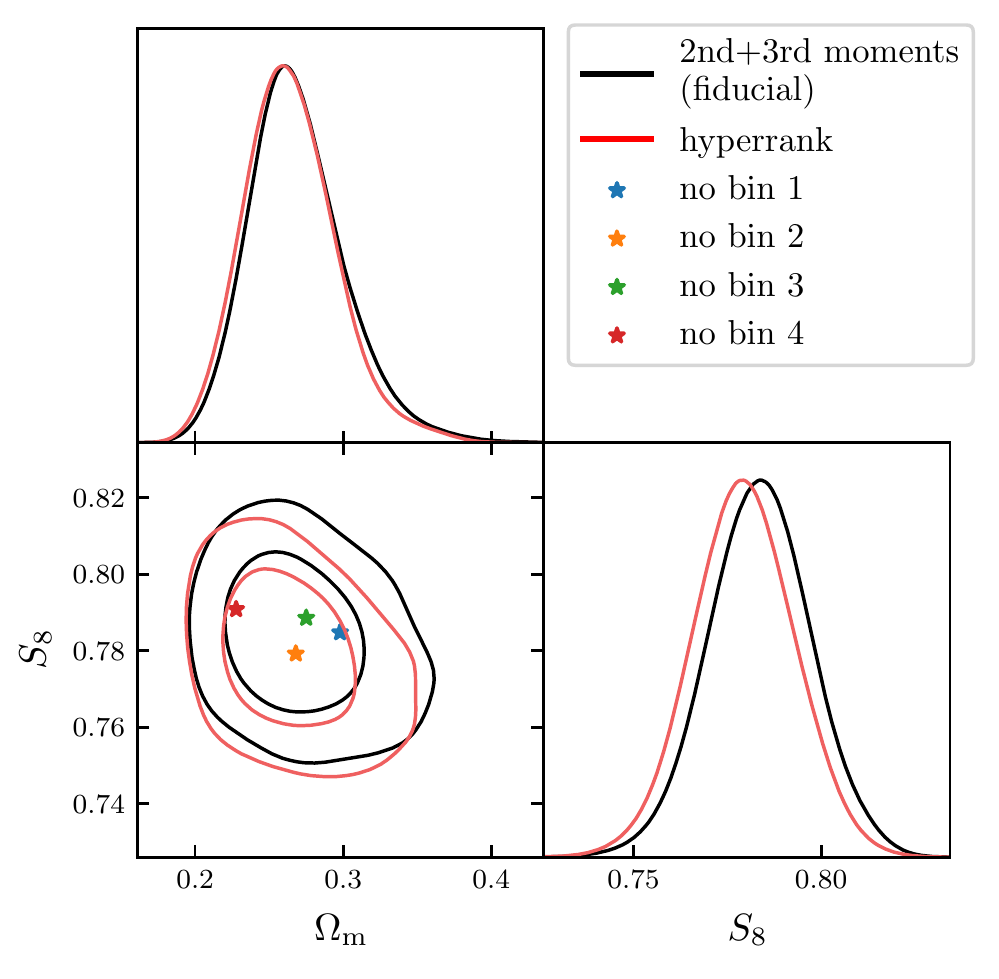}
\end{center}
\caption{Posterior distributions of the cosmological parameters $\Omega_{\rm m}$ and $S_8$ for the combination of second and third moments. We compare the 2D marginalised posterior obtained using hyperrank to model redshift uncertainties to the fiducial results. We also show the peak of the posteriors (the coloured stars in the plot) obtained removing one redshift bin at a time from the analysis (and using the fiducial redshift uncertainties model).}
\label{fig:redshift_tests}
\end{figure}
\begin{figure}
\begin{center}
\includegraphics[width=0.45 \textwidth]{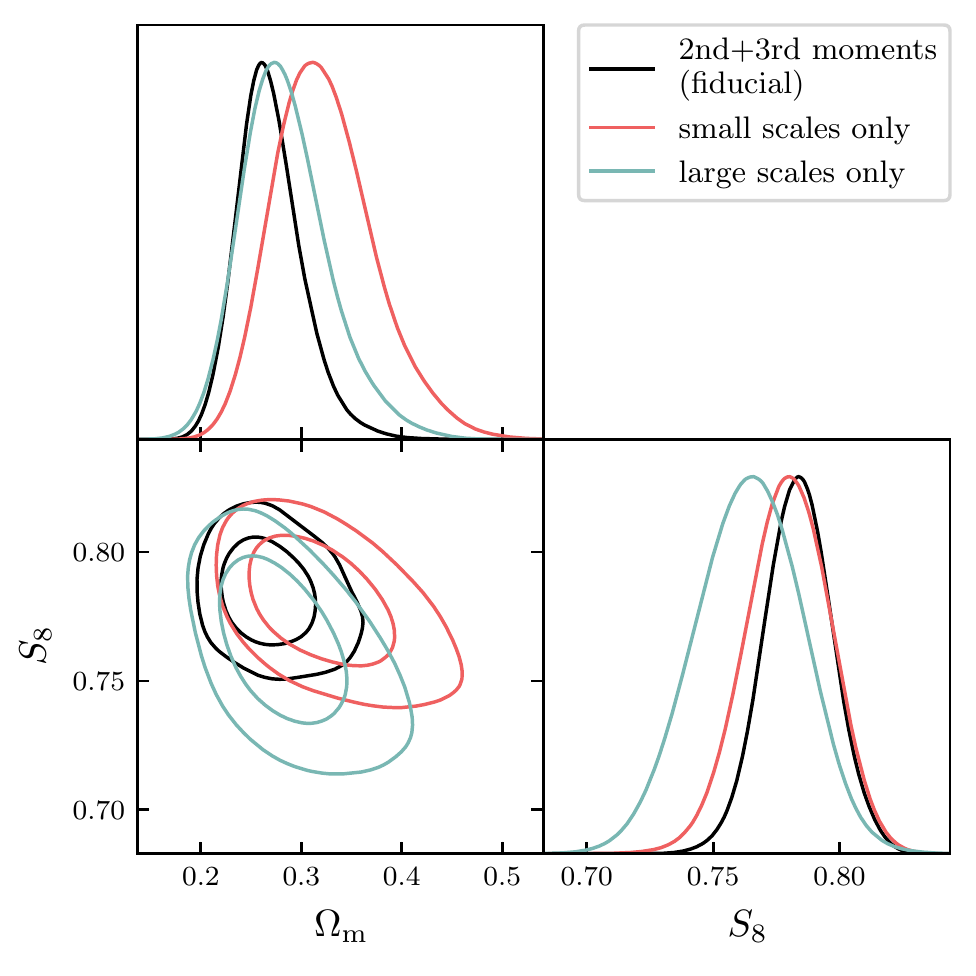}
\end{center}
\caption{Posterior distributions of the cosmological parameters $\Omega_{\rm m}$ and $S_8$ for the combination of second and third moments. The 2D marginalised contours in these figures show the 68 per cent and 95 per cent confidence levels. The figure shows the  posteriors obtained using only the small or the large scales of the data vector.}
\label{fig:scale_tests}
\end{figure}

\paragraph*{Compatibility between second and third moments.} This first test was one of the unblinding criteria.  Using PPD, we can check that second and third moments posteriors are consistent with each other, so that we can run the analysis using the combined data vector. The PPD $p$-values for $p(\langle \kappa^2\rangle|\langle \kappa^3\rangle)$ and $p(\langle \kappa^3\rangle|\langle \kappa^2\rangle)$ are reported in Table~\ref{table:PPD} and are well above the $p=0.01$ threshold. We note that these values need not be the same as the two PPDs are not symmetric. 

%We test first the compatibility between the posterior distributions of second and third moments. In particular, we are interested in comparing the PPD of second moments conditioned the observed third moments, and vice-versa. This was a required step prior to unblinding; therefore, it was first performed on blinded posteriors, then it was repeated on the unblinded analysis.

%Fig.~\ref{fig:PPD} shows the PPD realisations for second moments conditioned on the observed third moments, and vice-versa. We obtain $p-$value=0.32 and $p-$value=0.49 for the PPD of second moments conditioned on the observed third moments, and vice-versa, indicating compatibility between the probes. It can be noted from Fig.~\ref{fig:PPD} that the 68\% confidence interval of the PPD realisations of second moments conditioned on the observed third moments is larger  than the observed second moments: this is expected, as third moments are less constraining than second moments.

\paragraph*{Redshift tests.} Two types of internal consistency checks involving redshift distributions are performed. We performed these checks only using the combination of second and third moments; we did not perform them for second or third moments only.

The first check concerns the impact of removing individual redshift bins from the analysis. In order to perform this test, we again use the PPD. We first repeated our cosmological analysis removing all the second and third moments pairs and triples involving one particular redshift bin. We then sampled from those posteriors (one per bin), and compared using PPD to the observed second and third moments pairs and triples involving that particular redshift bin. This test is meant to highlight potential biases that might preferentially impact the low or high redshift end of our sample. The $p$-values from the PPD test, for each tomographic bin removed, are reported in Table \ref{table:PPD}: all the values are safely within our threshold. Fig.~\ref{fig:redshift_tests} shows the peaks of the posteriors in the $\Omega_{\rm m}$-$S_8$ plane of the analyses performed removing one bin at a time, and they are within the 1 $\sigma$ contour of the fiducial analysis. The biggest changes are obtained removing bin 4 (the posterior moves towards lower $\Omega_{\rm m}$ values) and removing bin 3 (the constraining power deteriorates more than with the other bins,  see Table~\ref{table_results}). This is not unexpected as bin 3 and 4 are the most constraining ones. 

The second test involves using a different parameterisation of the redshift uncertainties, called ``hyperrank'' \citep{y3-hyperrank}. With the hyperrank setup, a number of realisations of redshift distributions that encompass the redshift calibration uncertainties are provided. During the cosmological analysis, such realisations are marginalised over, instead of simply marginalising over the mean of the redshift distributions. Hyperrank is more complete as a method because it also accounts for uncertainties on the higher order moments of the redshift distributions. In the DES Y3 cosmic shear analysis, hyperrank has been proven to deliver very similar results compared to the simpler marginalisation over the the mean of the redshift distributions \citep{y3-cosmicshear1}. We perform here a similar test, analysing our data vector marginalising over the hyperrank realisations. The results of this alternative approach are shown in Fig. \ref{fig:redshift_tests}. We measure no significative difference with respect to our fiducial setup, demonstrating that for our analysis marginalising over the uncertainties of the mean of the redshift distribution is sufficient.

%Table~\ref{table_results}
%showing only small deviations from the fiducial posterior obtained from all the redshift bins. It can be noted that removing the first tomographic bin almost produces no change in the posterior, due to its limited constraining power. 

\paragraph*{Small scales vs. Large scales.} We check for internal consistency between the small and large scales of our data vector. In order to do so, we split our data vector in two halves that have similar constraining power and that retain only either small or large scales. {This is achieved by imposing a cut at the comoving scale of $56 h^{-1}$ Mpc, which we converted to an angular scale cut as explained in \S \ref{sect:like_tt}}. We use PPD to check for consistency between the two halves of the data vector. The PPD values are reported in Table \ref{table:PPD}; all the values are safely within our threshold. We note that for this test we considered the combination of second and third moments; we did not perform this test for second or third moments only. We show in Fig. \ref{fig:scale_tests} the posterior obtained using the two halves of the data vector; interestingly, the two halves of the data vector are associated to posteriors with a slightly different degeneracy direction in the $S_8$-$\Omega_{\rm m}$ plane. This is a similar behaviour to what has been found in the DES Y3 cosmic shear analysis \citep{y3-cosmicshear1}.

\section{Tests with alternative covariances}\label{sect:alternative_cov}

\begin{figure}
\begin{center}
\includegraphics[width=0.45 \textwidth]{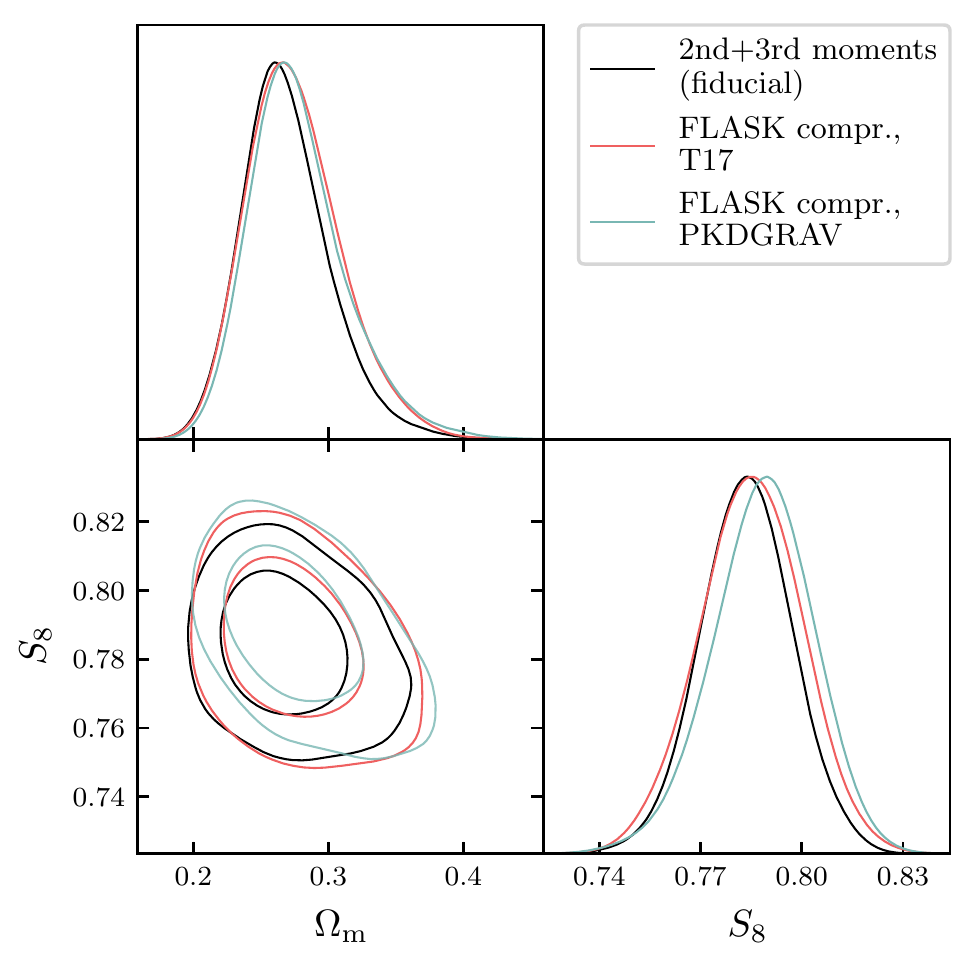}
\end{center}
\caption{Posterior distributions of the cosmological parameters $\Omega_{\rm m}$ and $S_8$ for the combination of second and third moments. The 2D marginalised contours in this figure show the 68 per cent and 95 per cent confidence levels. The figure shows the posteriors obtained using the T17 and PKDGRAV covariances, with the data vector compressed using the FLASK covariance.}
\label{fig:different_covariances}
\end{figure}

We explore in this Appendix the effect on our posteriors of using different covariances. We have at our disposal three covariances, obtained using \textsc{FLASK} realisations, T17 N-body simulations, and PKDGRAV N-body simulations. These covariances assume different cosmologies, and in particular, model the higher order moments of the convergence field slightly differently. This is easy to understand for \textsc{FLASK} covariance, since  \textsc{FLASK} assumes the convergence field to be lognormal, which is only an approximation. We should also expect some differences between the T17 and PKDGRAV simulations, based on the different agreement with theory predictions shown in Fig. \ref{fig:cosmology_validation}. Such differences probably stem from the different resolution settings of the two sets of N-body simulations.

When using the PKDGRAV and T17 covariances for the inference, we still use the FLASK covariance to compress the data vector since FLASK comes with the largest number of independent realisations. \footnote{We cannot use PKDGRAV realisations to do the compression, for instance, because the number of independent realisations is similar to (or, depending on the scale cut choice, smaller than) the length of the uncompressed data vector. This would imply that the covariance used for the compression is barely (or not) invertible, making the compression inaccurate (or impossible to be performed).}  According to \citep{Heavens2017}, this should not bias the inference, but (in the worst case) makes the compression sub-optimal. We show in Fig. \ref{fig:different_covariances} the posteriors obtained using different covariance matrices for the combined second and third moments on data, and report the values of $S_8$ in Table~\ref{table_results}; the posteriors and the values of the constraints are very similar in the three cases, implying that our modelling of the covariance matrix is adequate. 

%We ran multiple version of our analysis on data, using \textsc{FLASK}, T17 and PKDGRAV covariances; we also used the T17 and PKDGRAV covariances compressed using the \textsc{FLASK} covariance. For the compression stage, we note that when considering the combined second and third moments, we had to regularise the covariance used for the compression so as to have a good condition number and not be affected by numerical noise; this was achieved by adding a `floor' value to the smallest eigenvalues of the covariance. The regularised covariances had a slightly larger diagonal elements for the third moment at large scales. The compressed covariances did not need any regularisation. We noticed that introducing such a regularisation caused shifts of the order of $0.1-0.2$ $\sigma$, with the PKDGRAV covariance being the one affected the most.

\section{Parameter 1D posteriors and tension with the priors}\label{sect:nuisance}
\begin{figure*}
\begin{center}
\includegraphics[width=0.8 \textwidth]{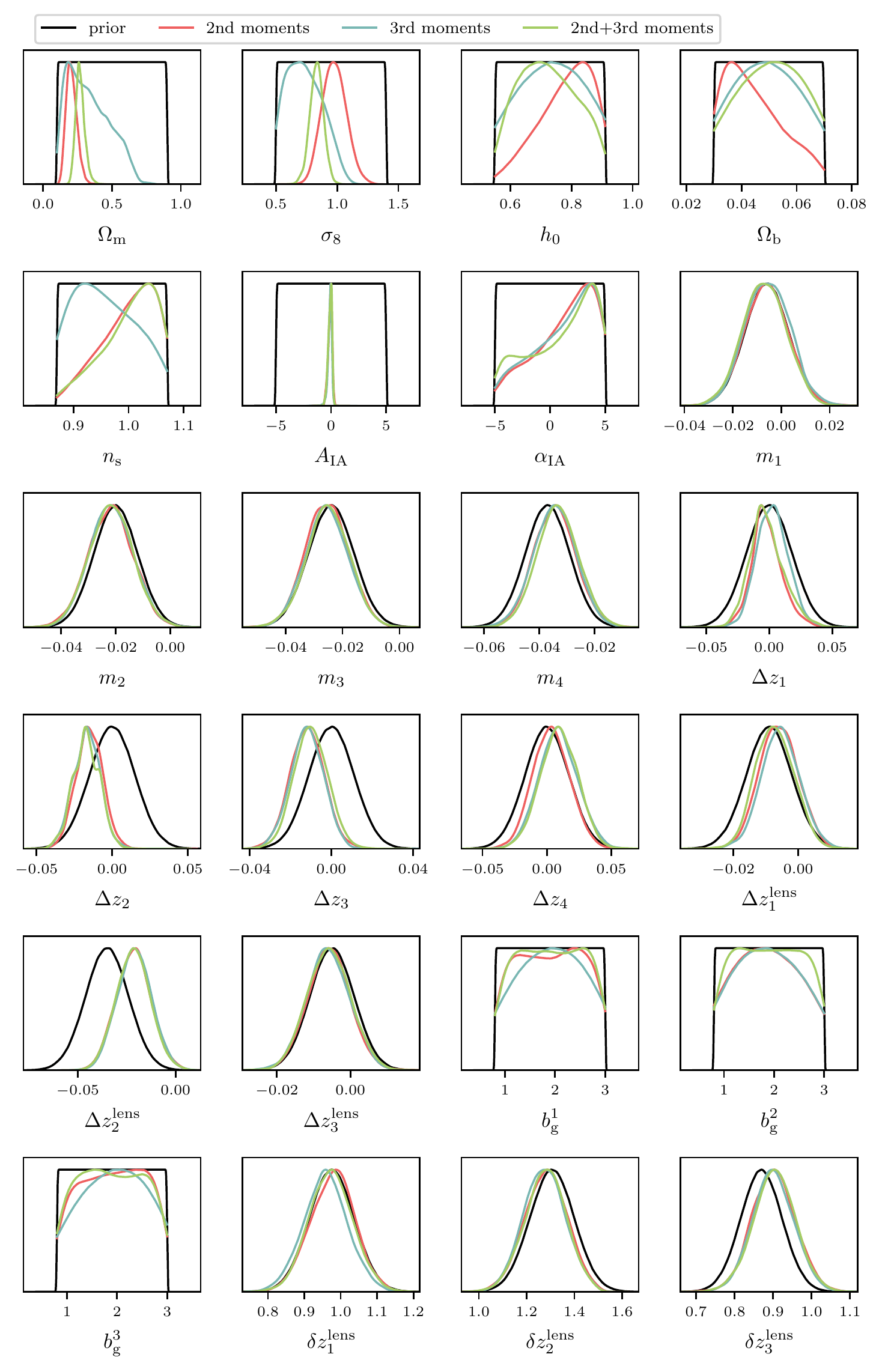}
\end{center}
\caption{1D marginalised posteriors for each parameter varied in the cosmological analysis. We show posteriors for second moments, third moments and the combination of second and third moments, and compare them with their prior. Note that in some cases the edges of the 1D posteriors might look to extend over the prior edges; however, it is only a visual effect due to the smoothing of the plotting script close to the edge of the prior.}
\label{fig:1d}
\end{figure*}

We test here whether the best-fitting models are in tension with their priors. This test was performed prior to unblinding. The 1D posteriors and their priors for all the parameters varied in this analysis are shown in Fig.~\ref{fig:1d}. In order to quantify the level of tension with the priors we use a Gaussian estimator called the `update difference-in-mean' (UDM) statistic \citep{Raveri2019}. The UDM statistic compares the mean parameters from the prior $\hat{\theta}^{\rm p}$ with the updated values $\hat{\theta}^{\rm p + d}$ obtained running the analysis on data. %{This statistic assumes either flat or Gaussian priors (which is satisfied for all the parameters considered in this analysis, see Table~\ref{table1}), and requires the posterior of the well-constrained parameters to be approximately Gaussian (which is also satisfied)}. 
In particular, we can define

%This is evident in Fig.~\ref{final_result} for the \textit{Planck} prior case: for the two highest redshift bins, all the best-fitting models are well below their priors. 
%
\begin{equation}
Q_{\rm UDM} = (\hat{\theta}^{\rm p + d }-\hat{\theta}^{\rm p})^T \left( \mathcal{C}^{\rm p} -\mathcal{C}^{\rm p + d } \right)^{-1}(\hat{\theta}^{\rm p + d }-\hat{\theta}^{\rm p}),
\end{equation}
where the difference in the mean of the parameters $(\hat{\theta}^{\rm p + d }-\hat{\theta}^{\rm p})$ is weighted by the inverse covariance of the parameters. If the parameters are Gaussian distributed then $Q_{\rm UDM}$ is chi-squared distributed with ${\rm rank}(\mathcal{C}^{\rm p + d } -\mathcal{C}^{\rm p})$ degrees of freedom. The UDM tension for second moments, third moments, and the combination of second and third moments is $0.6\sigma$, $1.2\sigma$, and $0.8\sigma$, respectively, indicating no tension. {We note that most of the parameter posteriors are actually prior dominated (without being in tension with the prior). This is fine as long as we trust our priors. {The main parameters constrained by the analysis ($\sigma_8$ and $\Omega_{\rm m}$ through $S_8$, and the intrinsic alignment amplitude $A_{\rm IA}$)}, however, are not prior dominated. The effective number of parameters $N_{\rm eff}$ constrained by the analysis can be computed as}
\begin{equation} \label{Eq:NumberEffectiveParameters}
N_{\rm eff} = N -{\rm tr}[(\mathcal{C}^{\rm p })^{-1}\mathcal{C}^{\rm p +d}],
\end{equation}
where N is the number of free parameters in the analysis. For instance, if we restrict to the five cosmological parameters, $N_{\rm eff}$ is only 2.6, 1.5, and 2.6 for second moments, third moments, and the combination of second and third moments, respectively.\newline

\section{Comparison between 2nd moments and cosmic shear window function}\label{sect:comp_wf}

\begin{figure}
\begin{center}
\includegraphics[width=0.45 \textwidth]{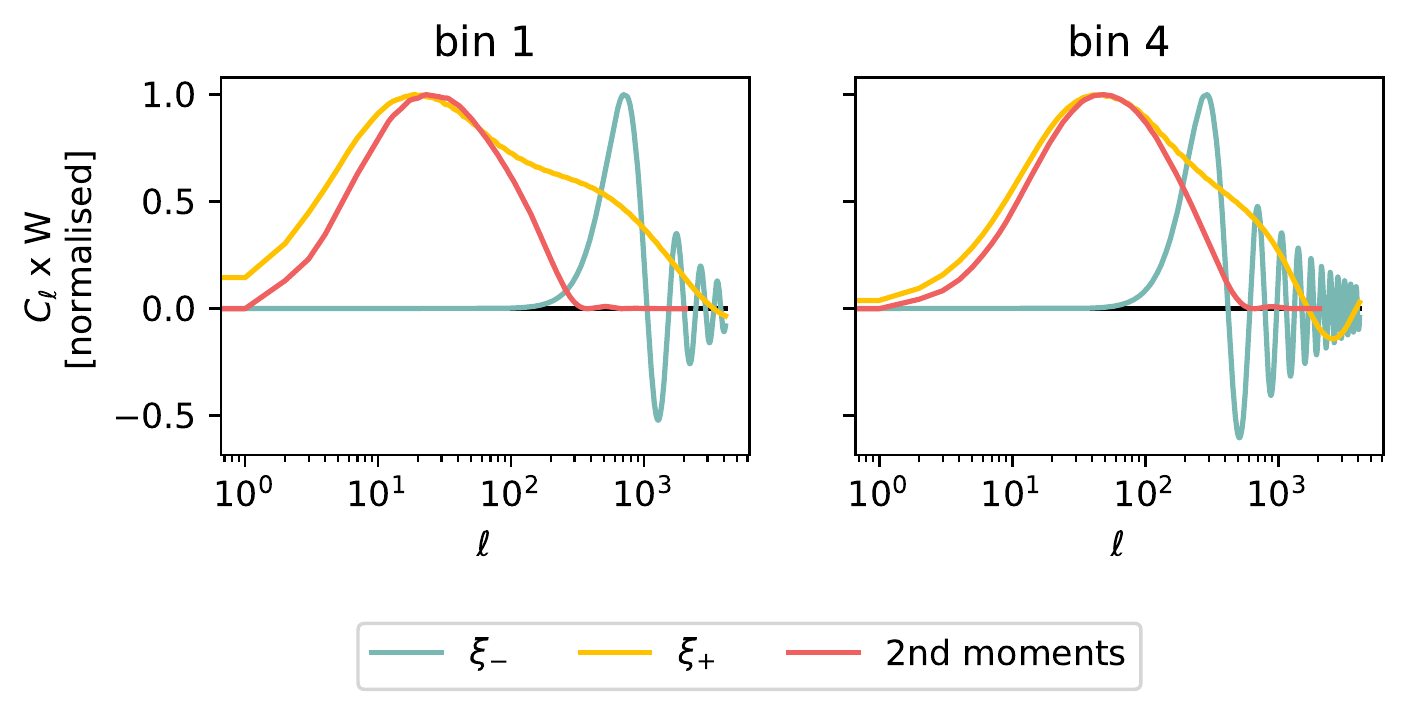}
\end{center}
\caption{Shear power spectrum for a given redshift bin multiplied by the window function $W$ of a given statistics (2nd moments, $\xi_{+-}$), normalised to their maximum value. For each statistics, the window function is computed at the minimum angular or smoothing scale allowed by the scale cut. The window functions for $\xi_{+-}$ are defined as $J_{0/4}(\ell \theta)$, whereas the second moments window function is $(2\ell+1)[W_{\ell}(\theta_0)]^2$, with $W_{\ell}(\theta_0)$ given by Eq. \ref{eq:filter}. For 2nd moments, we also convolved the power spectrum by the mode coupling matrix.}
\label{fig:comp_wf}
\end{figure}

We discuss here a potential explanation for the $\sim 1 \sigma$ difference between the peaks of the DES Y3 cosmic shear and the 2nd moments analysis in the $S_8$-$\Omega_{\rm m}$ plane (Fig. \ref{fig:des}). Both cosmic shear and second moments are Gaussian statistics and they both probe the shear power spectrum, but their posteriors do not have to perfectly overlap, as they weight power spectrum multipoles differently. Moreover, the process adopted to determine the scales that can be used for each tomographic bin is different for the two analyses: for the moments analysis we adopted a cut based on a well-determined physical scale $\theta_0 = R_0/ \chi(\avg{z})$ (Appendix \ref{sect:scale_cuts}), whereas the cosmic shear analysis adopted a criterion that evenly distributed a given $\Delta \chi^2$ among tomographic bins (where the $\Delta \chi^2$ is computed between a synthetic data vector contaminated with baryonic effects and an uncontaminated data vector). Although both criteria are designed to minimise the impact of baryons on the $S_8$-$\Omega_{\rm m}$ constraints, they can contribute to the different sensitivity of the statistics to the shear power spectrum multipoles.

Fig. \ref{fig:comp_wf} shows, for the first and the last redshift bins, how the different statistics weight the multipoles of the shear power spectrum at the minimum angular scale allowed by their scale cut. The cosmic shear scale cut allows $\xi_{+-}$ to probe significantly higher multipoles compared to the second moments, whose window function is more compact. When considering the redshift bin 1 (4), $\sim 30$ ($\sim 25$) per cent of the S/N of the $\xi_{+-}$ data vector passing the scale cut comes from $\ell>200$. For 2nd moments the contribution to the S/N from $\ell>200$ is significantly smaller: $\sim 1$ ($\sim 10$) per cent. Although a more quantitative assessment of the compatibility between the 2nd moments and cosmic shear constraints should be performed via PPD, we consider the pieces of evidence provided in this Appendix sufficient to justify the differences between the two analyses shown in Fig. \ref{fig:des}.

%%%%%%%%%%%%%%%%%%%%%%%%%%%%%%%%%%%%%%%%%%%%%%%%%%

% Don't change these lines

\label{lastpage}


%\bibliographystyle{mnras}		  %SWITCH TO MNRAS
\bibliography{bibliography}

\begin{thebibliography}{117}
\providecommand{\natexlab}[1]{#1}
\providecommand{\url}[1]{\texttt{#1}}
\providecommand{\urlprefix}{URL }
\providecommand{\eprint}[1][]{\url{#1}}

\bibitem[{{Abbott} et~al.(2018){Abbott} \& {Abdalla} et~al.}]{desy1_results}
{Abbott}, T.~M.~C., {Abdalla}, F.~B., {Alarcon}, A., et~al., 2018, \prd, 98, 4,
  043526, \eprint arXiv:{1708.01530}

\bibitem[{Aghanim et~al.(2020)Aghanim \& Akrami et~al.}]{aghanim2020planck}
Aghanim, N., Akrami, Y., Ashdown, M., et~al., 2020, Astronomy \& Astrophysics,
  641, A6

\bibitem[{{Aihara} et~al.(2018){Aihara} \& {Arimoto} et~al.}]{Aihara2018}
{Aihara}, H., {Arimoto}, N., {Armstrong}, R., et~al., 2018, \pasj, 70, S4

\bibitem[{{Ajani} et~al.(2020){Ajani} \& {Peel} \& {Pettorino} \& {Starck} \&
  {Li} \& {Liu}}]{ajani_peaks}
{Ajani}, V., {Peel}, A., {Pettorino}, V., {Starck}, J.-L., {Li}, Z., {Liu}, J.,
  2020, \prd, 102, 10, 103531

\bibitem[{{Amon} et~al.(2021){Amon} \& {Gruen} et~al.}]{y3-cosmicshear1}
{Amon}, A., {Gruen}, D., {Troxel}, M.~A., et~al., 2021, arXiv e-prints,
  arXiv:2105.13543

\bibitem[{{Asgari} et~al.(2021){Asgari} \& {Lin} et~al.}]{Asgari2021}
{Asgari}, M., {Lin}, C.-A., {Joachimi}, B., et~al., 2021, \aap, 645, A104

\bibitem[{{Bernardeau} et~al.(2002){Bernardeau} \& {Colombi} \& {Gazta{\~n}aga}
  \& {Scoccimarro}}]{Bernardeau2002}
{Bernardeau}, F., {Colombi}, S., {Gazta{\~n}aga}, E., {Scoccimarro}, R., 2002,
  \physrep, 367, 1

\bibitem[{{Blazek} et~al.(2019){Blazek} \& {MacCrann} \& {Troxel} \&
  {Fang}}]{blazek19}
{Blazek}, J.~A., {MacCrann}, N., {Troxel}, M.~A., {Fang}, X., 2019, \prd, 100,
  10, 103506

\bibitem[{{Bridle} \& {King}(2007)}]{Bridle2007}
{Bridle}, S., {King}, L., 2007, New Journal of Physics, 9, 444

\bibitem[{{Brown} et~al.(2005){Brown} \& {Castro} \& {Taylor}}]{Brown2005}
{Brown}, M.~L., {Castro}, P.~G., {Taylor}, A.~N., 2005, \mnras, 360, 1262

\bibitem[{{Cardone} et~al.(2013){Cardone} \& {Camera} et~al.}]{Cardone2013}
{Cardone}, V.~F., {Camera}, S., {Mainini}, R., et~al., 2013, \mnras, 430, 4,
  2896

\bibitem[{{Chang} et~al.(2018){Chang} \& {Pujol} et~al.}]{Chang2018}
{Chang}, C., {Pujol}, A., {Mawdsley}, B., et~al., 2018, \mnras, 475, 3165

\bibitem[{{Chang} et~al.(2015){Chang} \& {Vikram} et~al.}]{Chang2015}
{Chang}, C., {Vikram}, V., {Jain}, B., et~al., 2015, \prl, 115, 5, 051301

\bibitem[{{Chang} et~al.(2019){Chang} \& {Wang} et~al.}]{Chang2019}
{Chang}, C., {Wang}, M., {Dodelson}, S., et~al., 2019, \mnras, 482, 3, 3696

\bibitem[{{Clerkin} et~al.(2017){Clerkin} \& {Kirk} et~al.}]{Clerkin2017}
{Clerkin}, L., {Kirk}, D., {Manera}, M., et~al., 2017, \mnras, 466, 1444

\bibitem[{{Coles} \& {Jones}(1991)}]{Coles1991}
{Coles}, P., {Jones}, B., 1991, \mnras, 248, 1

\bibitem[{{Cordero} et~al.(2021){Cordero} \& {Harrison} et~al.}]{y3-hyperrank}
{Cordero}, J.~P., {Harrison}, I., {Rollins}, R.~P., et~al., 2021, arXiv
  e-prints, arXiv:2109.09636

\bibitem[{{Crocce} et~al.(2006){Crocce} \& {Pueblas} \&
  {Scoccimarro}}]{Crocce2006}
{Crocce}, M., {Pueblas}, S., {Scoccimarro}, R., 2006, \mnras, 373, 369

\bibitem[{{Dark Energy Survey Collaboration}(2016)}]{DES2016}
{Dark Energy Survey Collaboration}, 2016, \mnras, 460, 2, 1270

\bibitem[{{Dark Energy Survey Collaboration}(2018)}]{DES_DR1}
{Dark Energy Survey Collaboration}, 2018, \apjs, 239, 2, 18

\bibitem[{{DES Collaboration}(2021)}]{y3-3x2ptkp}
{DES Collaboration}, 2021, arXiv e-prints, arXiv:2105.13549

\bibitem[{{Dietrich} \& {Hartlap}(2010)}]{Dietrich2010}
{Dietrich}, J.~P., {Hartlap}, J., 2010, \mnras, 402, 2, 1049

\bibitem[{{Dodelson} \& {Schneider}(2013)}]{Dodelson2013}
{Dodelson}, S., {Schneider}, M.~D., 2013, \prd, 88, 6, 063537

\bibitem[{{Doux} et~al.(2021){Doux} \& {Baxter} et~al.}]{Doux2021}
{Doux}, C., {Baxter}, E., {Lemos}, P., et~al., 2021, \mnras, 503, 2, 2688

\bibitem[{{Einstein}(1936)}]{Einstein1936}
{Einstein}, A., 1936, Science, 84, 2188, 506

\bibitem[{{Flaugher} et~al.(2015){Flaugher} \& {Diehl} et~al.}]{Flaugher2015}
{Flaugher}, B., {Diehl}, H.~T., {Honscheid}, K., et~al., 2015, \aj, 150, 150

\bibitem[{{Fluri} et~al.(2019){Fluri} \& {Kacprzak} et~al.}]{Fluri2019}
{Fluri}, J., {Kacprzak}, T., {Lucchi}, A., et~al., 2019, \prd, 100, 6, 063514

\bibitem[{{Fluri} et~al.(2018){Fluri} \& {Kacprzak} \& {Refregier} \& {Amara}
  \& {Lucchi} \& {Hofmann}}]{Fluri2018}
{Fluri}, J., {Kacprzak}, T., {Refregier}, A., {Amara}, A., {Lucchi}, A.,
  {Hofmann}, T., 2018, \prd, 98, 12, 123518

\bibitem[{{Fosalba} et~al.(2008){Fosalba} \& {Gazta{\~n}aga} \& {Castander} \&
  {Manera}}]{Fosalba2008}
{Fosalba}, P., {Gazta{\~n}aga}, E., {Castander}, F.~J., {Manera}, M., 2008,
  \mnras, 391, 435

\bibitem[{{Friedrich} et~al.(2020){Friedrich} \& {Andrade-Oliveira}
  et~al.}]{y3-covariances}
{Friedrich}, O., {Andrade-Oliveira}, F., {Camacho}, H., et~al., 2020, arXiv
  e-prints, arXiv:2012.08568

\bibitem[{{Friedrich} \& {Eifler}(2018)}]{Friedrich2018b}
{Friedrich}, O., {Eifler}, T., 2018, \mnras, 473, 4150

\bibitem[{{Fu} et~al.(2014){Fu} \& {Kilbinger} et~al.}]{Fu2014}
{Fu}, L., {Kilbinger}, M., {Erben}, T., et~al., 2014, \mnras, 441, 3, 2725

\bibitem[{{Gatti} et~al.(2020{\natexlab{a}}){Gatti} \& {Chang} et~al.}]{G20}
{Gatti}, M., {Chang}, C., {Friedrich}, O., et~al., 2020{\natexlab{a}}, \mnras,
  498, 3, 4060

\bibitem[{{Gatti} et~al.(2020{\natexlab{b}}){Gatti} \& {Giannini}
  et~al.}]{y3-sourcewz}
{Gatti}, M., {Giannini}, G., {Bernstein}, G.~M., et~al., 2020{\natexlab{b}},
  arXiv e-prints, arXiv:2012.08569

\bibitem[{{Gatti} et~al.(2021){Gatti} \& {Sheldon} et~al.}]{y3-shapecatalog}
{Gatti}, M., {Sheldon}, E., {Amon}, A., et~al., 2021, \mnras, 504, 3, 4312

\bibitem[{{Gaztanaga} \& {Bernardeau}(1998)}]{Gaztanaga1998}
{Gaztanaga}, E., {Bernardeau}, F., 1998, \aap, 331, 829

\bibitem[{{Giblin} et~al.(2020){Giblin} \& {Heymans} et~al.}]{Giblin2020}
{Giblin}, B., {Heymans}, C., {Asgari}, M., et~al., 2020, arXiv e-prints,
  arXiv:2007.01845

\bibitem[{{G{\'o}rski} et~al.(2005){G{\'o}rski} \& {Hivon} et~al.}]{GORSKI2005}
{G{\'o}rski}, K.~M., {Hivon}, E., {Banday}, A.~J., et~al., 2005, \apj, 622, 759

\bibitem[{{Gualdi} et~al.(2018){Gualdi} \& {Manera} \& {Joachimi} \&
  {Lahav}}]{Gualdi2018}
{Gualdi}, D., {Manera}, M., {Joachimi}, B., {Lahav}, O., 2018, \mnras, 476,
  4045

\bibitem[{{Habib} et~al.(2007){Habib} \& {Heitmann} \& {Higdon} \& {Nakhleh} \&
  {Williams}}]{Habib2007}
{Habib}, S., {Heitmann}, K., {Higdon}, D., {Nakhleh}, C., {Williams}, B., 2007,
  \prd, 76, 8, 083503

\bibitem[{{Hamana} et~al.(2015){Hamana} \& {Sakurai} \& {Koike} \&
  {Miller}}]{Hamana2015}
{Hamana}, T., {Sakurai}, J., {Koike}, M., {Miller}, L., 2015, \pasj, 67, 34

\bibitem[{{Hamana} et~al.(2019){Hamana} \& {Shirasaki} et~al.}]{Hamana2019}
{Hamana}, T., {Shirasaki}, M., {Miyazaki}, S., et~al., 2019, arXiv e-prints

\bibitem[{{Hamana} et~al.(2020){Hamana} \& {Shirasaki} et~al.}]{Hamana2020}
{Hamana}, T., {Shirasaki}, M., {Miyazaki}, S., et~al., 2020, \pasj, 72, 1, 16

\bibitem[{{Hamilton}(2001)}]{Hamilton2001}
{Hamilton}, A.~J.~S., 2001, \mnras, 322, 2, 419

\bibitem[{{Handley} et~al.(2015{\natexlab{a}}){Handley} \& {Hobson} \&
  {Lasenby}}]{poly1}
{Handley}, W.~J., {Hobson}, M.~P., {Lasenby}, A.~N., 2015{\natexlab{a}},
  \mnras, 450, L61

\bibitem[{{Handley} et~al.(2015{\natexlab{b}}){Handley} \& {Hobson} \&
  {Lasenby}}]{poly2}
{Handley}, W.~J., {Hobson}, M.~P., {Lasenby}, A.~N., 2015{\natexlab{b}},
  \mnras, 453, 4, 4384

\bibitem[{{Hartlap} et~al.(2007){Hartlap} \& {Simon} \&
  {Schneider}}]{Hartlap2007}
{Hartlap}, J., {Simon}, P., {Schneider}, P., 2007, \aap, 464, 399

\bibitem[{{Heavens} et~al.(2000){Heavens} \& {Jimenez} \&
  {Lahav}}]{Heavens2000}
{Heavens}, A.~F., {Jimenez}, R., {Lahav}, O., 2000, \mnras, 317, 965

\bibitem[{{Heavens} et~al.(2017){Heavens} \& {Sellentin} \& {de Mijolla} \&
  {Vianello}}]{Heavens2017}
{Heavens}, A.~F., {Sellentin}, E., {de Mijolla}, D., {Vianello}, A., 2017,
  \mnras, 472, 4244

\bibitem[{{Heitmann} et~al.(2006){Heitmann} \& {Higdon} \& {Nakhleh} \&
  {Habib}}]{Heitmann2006}
{Heitmann}, K., {Higdon}, D., {Nakhleh}, C., {Habib}, S., 2006, \apjl, 646, L1

\bibitem[{{Heymans} et~al.(2021){Heymans} \& {Tr{\"o}ster}
  et~al.}]{Heymans2021}
{Heymans}, C., {Tr{\"o}ster}, T., {Asgari}, M., et~al., 2021, \aap, 646, A140

\bibitem[{{Hikage} et~al.(2019){Hikage} \& {Oguri} et~al.}]{Hikage2019}
{Hikage}, C., {Oguri}, M., {Hamana}, T., et~al., 2019, \pasj, 71, 2, 43

\bibitem[{{Hikage} et~al.(2011){Hikage} \& {Takada} \& {Hamana} \&
  {Spergel}}]{Hikage2011}
{Hikage}, C., {Takada}, M., {Hamana}, T., {Spergel}, D., 2011, \mnras, 412, 65

\bibitem[{{Hildebrandt} et~al.(2017){Hildebrandt} \& {Viola}
  et~al.}]{Hildebrandt2017}
{Hildebrandt}, H., {Viola}, M., {Heymans}, C., et~al., 2017, \mnras, 465, 1454

\bibitem[{{Hirata} \& {Seljak}(2004)}]{Hirata2004}
{Hirata}, C.~M., {Seljak}, U., 2004, \prd, 70, 6, 063526

\bibitem[{{Hubble}(1934)}]{Hubble1934}
{Hubble}, E., 1934, \apj, 79, 8

\bibitem[{{Huff} \& {Mandelbaum}(2017)}]{HuffMcal2017}
{Huff}, E., {Mandelbaum}, R., 2017, arXiv e-prints, 1702.02600

\bibitem[{{Jain} \& {Seljak}(1997)}]{Jain1997}
{Jain}, B., {Seljak}, U., 1997, \apj, 484, 2, 560

\bibitem[{{Jarvis} et~al.(2016){Jarvis} \& {Sheldon} et~al.}]{Jarvis2016}
{Jarvis}, M., {Sheldon}, E., {Zuntz}, J., et~al., 2016, \mnras, 460, 2245

\bibitem[{{Jeffrey} et~al.(2021{\natexlab{a}}){Jeffrey} \& {Alsing} \&
  {Lanusse}}]{jeffrey_lfi}
{Jeffrey}, N., {Alsing}, J., {Lanusse}, F., 2021{\natexlab{a}}, \mnras, 501, 1,
  954

\bibitem[{{Jeffrey} et~al.(2021{\natexlab{b}}){Jeffrey} \& {Gatti}
  et~al.}]{y3-massmapping}
{Jeffrey}, N., {Gatti}, M., {Chang}, C., et~al., 2021{\natexlab{b}}, \mnras,
  505, 3, 4626

\bibitem[{{Kacprzak} et~al.(2016){Kacprzak} \& {Kirk} et~al.}]{Kacprzak2016}
{Kacprzak}, T., {Kirk}, D., {Friedrich}, O., et~al., 2016, \mnras, 463, 3653

\bibitem[{{Kaiser} \& {Squires}(1993)}]{KaiserSquires}
{Kaiser}, N., {Squires}, G., 1993, \apj, 404, 441

\bibitem[{{Kratochvil} et~al.(2010){Kratochvil} \& {Haiman} \&
  {May}}]{Kratochvil2010}
{Kratochvil}, J.~M., {Haiman}, Z., {May}, M., 2010, \prd, 81, 4, 043519

\bibitem[{{Kratochvil} et~al.(2012){Kratochvil} \& {Lim} \& {Wang} \& {Haiman}
  \& {May} \& {Huffenberger}}]{Kratochvil2012}
{Kratochvil}, J.~M., {Lim}, E.~A., {Wang}, S., {Haiman}, Z., {May}, M.,
  {Huffenberger}, K., 2012, \prd, 85, 10, 103513

\bibitem[{{Krause} et~al.(2021)}]{y3-generalmethods}
{Krause}, E., et~al., 2021, To be submitted to MNRAS

\bibitem[{{Kuijken} et~al.(2015){Kuijken} \& {Heymans} et~al.}]{Kuijken2015}
{Kuijken}, K., {Heymans}, C., {Hildebrandt}, H., et~al., 2015, \mnras, 454, 4,
  3500

\bibitem[{{Laureijs} et~al.(2011){Laureijs} \& {Amiaux} et~al.}]{Laureijs2011}
{Laureijs}, R., {Amiaux}, J., {Arduini}, S., et~al., 2011, arXiv e-prints,
  arXiv:1110.3193

\bibitem[{{Liu} et~al.(2015){Liu} \& {Petri} \& {Haiman} \& {Hui} \&
  {Kratochvil} \& {May}}]{Liu2015}
{Liu}, J., {Petri}, A., {Haiman}, Z., {Hui}, L., {Kratochvil}, J.~M., {May},
  M., 2015, \prd, 91, 6, 063507

\bibitem[{{LSST Science Collaboration} et~al.(2009){LSST Science Collaboration}
  \& {Abell} et~al.}]{Abell2009}
{LSST Science Collaboration}, {Abell}, P.~A., {Allison}, J., et~al., 2009,
  arXiv e-prints, arXiv:0912.0201

\bibitem[{{MacCrann} et~al.(2020){MacCrann} \& {Becker} et~al.}]{y3-imagesims}
{MacCrann}, N., {Becker}, M.~R., {McCullough}, J., et~al., 2020, arXiv
  e-prints, arXiv:2012.08567

\bibitem[{{Martinet} et~al.(2018){Martinet} \& {Schneider}
  et~al.}]{Martinet2018}
{Martinet}, N., {Schneider}, P., {Hildebrandt}, H., et~al., 2018, \mnras, 474,
  1, 712

\bibitem[{{Morganson} et~al.(2018){Morganson} \& {Gruendl}
  et~al.}]{Morganson2018}
{Morganson}, E., {Gruendl}, R.~A., {Menanteau}, F., et~al., 2018, \pasp, 130,
  989, 074501

\bibitem[{{Muir} et~al.(2020){Muir} \& {Bernstein} et~al.}]{y3-blinding}
{Muir}, J., {Bernstein}, G.~M., {Huterer}, D., et~al., 2020, \mnras, 494, 3,
  4454

\bibitem[{{Myles} et~al.(2021){Myles} \& {Alarcon} et~al.}]{y3-sompz}
{Myles}, J., {Alarcon}, A., {Amon}, A., et~al., 2021, \mnras, 505, 3, 4249

\bibitem[{{Oguri} et~al.(2018){Oguri} \& {Miyazaki} et~al.}]{Oguri2018}
{Oguri}, M., {Miyazaki}, S., {Hikage}, C., et~al., 2018, \pasj, 70, S26

\bibitem[{{Parroni} et~al.(2020){Parroni} \& {Cardone} \& {Maoli} \&
  {Scaramella}}]{Parroni2020}
{Parroni}, C., {Cardone}, V.~F., {Maoli}, R., {Scaramella}, R., 2020, \aap,
  633, A71

\bibitem[{{Peel} et~al.(2018){Peel} \& {Pettorino} \& {Giocoli} \& {Starck} \&
  {Baldi}}]{Peel2018}
{Peel}, A., {Pettorino}, V., {Giocoli}, C., {Starck}, J.-L., {Baldi}, M., 2018,
  \aap, 619, A38

\bibitem[{{Petri} et~al.(2015){Petri} \& {Liu} \& {Haiman} \& {May} \& {Hui} \&
  {Kratochvil}}]{Petri2015}
{Petri}, A., {Liu}, J., {Haiman}, Z., {May}, M., {Hui}, L., {Kratochvil},
  J.~M., 2015, \prd, 91, 10, 103511

\bibitem[{{Porredon} et~al.(2021){Porredon} \& {Crocce}
  et~al.}]{y3-2x2maglimforecast}
{Porredon}, A., {Crocce}, M., {Fosalba}, P., et~al., 2021, \prd, 103, 4, 043503

\bibitem[{{Porredon} et~al.(in prep.)}]{y3-2x2ptaltlensresults}
{Porredon}, A., et~al., in prep., To be submitted to PRD

\bibitem[{Potter et~al.(2017)Potter \& Stadel \& Teyssier}]{potter2017pkdgrav3}
Potter, D., Stadel, J., Teyssier, R., 2017, Computational Astrophysics and
  Cosmology, 4, 1, 2

\bibitem[{{Pujol} et~al.(2016){Pujol} \& {Chang} et~al.}]{Pujol2016}
{Pujol}, A., {Chang}, C., {Gazta{\~n}aga}, E., et~al., 2016, \mnras, 462, 35

\bibitem[{{Pyne} et~al.(2022){Pyne} \& {Tenneti} \& {Joachimi}}]{Pyne2022}
{Pyne}, S., {Tenneti}, A., {Joachimi}, B., 2022, arXiv e-prints,
  arXiv:2204.10342, \eprint arXiv:{2204.10342}

\bibitem[{{Raveri} \& {Hu}(2019)}]{Raveri2019}
{Raveri}, M., {Hu}, W., 2019, \prd, 99, 4, 043506

\bibitem[{Raveri et~al.(2020)Raveri \& Zacharegkas \& Hu}]{Raveri:2019gdp}
Raveri, M., Zacharegkas, G., Hu, W., 2020, Phys. Rev. D, 101, 10, 103527

\bibitem[{{Ribli} et~al.(2019){Ribli} \& {Pataki} \& {Csabai}}]{Ribli2018}
{Ribli}, D., {Pataki}, B.~{\'A}., {Csabai}, I., 2019, Nature Astronomy, 3, 93

\bibitem[{{S{\'a}nchez} et~al.(2021){S{\'a}nchez} \& {Prat}
  et~al.}]{y3-shearratio}
{S{\'a}nchez}, C., {Prat}, J., {Zacharegkas}, G., et~al., 2021, arXiv e-prints,
  arXiv:2105.13542

\bibitem[{{Scoccimarro} \& {Couchman}(2001)}]{Scoccimarro2001}
{Scoccimarro}, R., {Couchman}, H.~M.~P., 2001, \mnras, 325, 1312

\bibitem[{{Secco} et~al.(2021){Secco} \& {Samuroff} et~al.}]{y3-cosmicshear2}
{Secco}, L.~F., {Samuroff}, S., {Krause}, E., et~al., 2021, arXiv e-prints,
  arXiv:2105.13544

\bibitem[{{Semboloni} et~al.(2011){Semboloni} \& {Schrabback} \& {van Waerbeke}
  \& {Vafaei} \& {Hartlap} \& {Hilbert}}]{Semboloni2011}
{Semboloni}, E., {Schrabback}, T., {van Waerbeke}, L., {Vafaei}, S., {Hartlap},
  J., {Hilbert}, S., 2011, \mnras, 410, 1, 143

\bibitem[{Sevilla et~al.(2011)}]{Sevilla2011}
Sevilla, I., et~al., 2011, in {Meeting of the APS Division of Particles and
  Fields}

\bibitem[{{Sevilla-Noarbe} et~al.(2021){Sevilla-Noarbe} \& {Bechtol}
  et~al.}]{y3-gold}
{Sevilla-Noarbe}, I., {Bechtol}, K., {Carrasco Kind}, M., et~al., 2021, \apjs,
  254, 2, 24

\bibitem[{{Shan} et~al.(2018){Shan} \& {Liu} et~al.}]{Shan2018}
{Shan}, H., {Liu}, X., {Hildebrandt}, H., et~al., 2018, \mnras, 474, 1, 1116

\bibitem[{{Sheldon} et~al.(2020){Sheldon} \& {Becker} \& {MacCrann} \&
  {Jarvis}}]{SheldonMetadetect2019}
{Sheldon}, E.~S., {Becker}, M.~R., {MacCrann}, N., {Jarvis}, M., 2020, \apj,
  902, 2, 138

\bibitem[{{Sheldon} \& {Huff}(2017)}]{SheldonMcal2017}
{Sheldon}, E.~S., {Huff}, E.~M., 2017, \apj, 841, 24

\bibitem[{{Springel}(2005)}]{springel2005}
{Springel}, V., 2005, \mnras, 364, 1105

\bibitem[{{Takada} \& {Jain}(2003)}]{Takada2003}
{Takada}, M., {Jain}, B., 2003, \mnras, 344, 3, 857

\bibitem[{{Takada} \& {Jain}(2004)}]{Takada2004}
{Takada}, M., {Jain}, B., 2004, \mnras, 348, 3, 897

\bibitem[{{Takahashi} et~al.(2017){Takahashi} \& {Hamana}
  et~al.}]{Takahashi2017}
{Takahashi}, R., {Hamana}, T., {Shirasaki}, M., et~al., 2017, \apj, 850, 24

\bibitem[{{Takahashi} et~al.(2012){Takahashi} \& {Sato} \& {Nishimichi} \&
  {Taruya} \& {Oguri}}]{Takahashi2012}
{Takahashi}, R., {Sato}, M., {Nishimichi}, T., {Taruya}, A., {Oguri}, M., 2012,
  \apj, 761, 152

\bibitem[{{Takahashi} et~al.(2014){Takahashi} \& {Soma} \& {Takada} \&
  {Kayo}}]{Takahashi2014}
{Takahashi}, R., {Soma}, S., {Takada}, M., {Kayo}, I., 2014, \mnras, 444, 3473

\bibitem[{{Tegmark} et~al.(1997){Tegmark} \& {Taylor} \&
  {Heavens}}]{Tegmark1997}
{Tegmark}, M., {Taylor}, A.~N., {Heavens}, A.~F., 1997, \apj, 480, 22

\bibitem[{{Troxel} et~al.(2018){Troxel} \& {MacCrann} et~al.}]{Troxel2017}
{Troxel}, M.~A., {MacCrann}, N., {Zuntz}, J., et~al., 2018, \prd, 98, 4, 043528

\bibitem[{{Vafaei} et~al.(2010){Vafaei} \& {Lu} \& {van Waerbeke} \&
  {Semboloni} \& {Heymans} \& {Pen}}]{Vafaei2010}
{Vafaei}, S., {Lu}, T., {van Waerbeke}, L., {Semboloni}, E., {Heymans}, C.,
  {Pen}, U.-L., 2010, Astroparticle Physics, 32, 6, 340

\bibitem[{{Van Waerbeke} et~al.(2013){Van Waerbeke} \& {Benjamin}
  et~al.}]{VanWaerbeke2013}
{Van Waerbeke}, L., {Benjamin}, J., {Erben}, T., et~al., 2013, \mnras, 433,
  3373

\bibitem[{{Van Waerbeke} et~al.(2001){Van Waerbeke} \& {Hamana} \&
  {Scoccimarro} \& {Colombi} \& {Bernardeau}}]{VanWaerbeke2001}
{Van Waerbeke}, L., {Hamana}, T., {Scoccimarro}, R., {Colombi}, S.,
  {Bernardeau}, F., 2001, \mnras, 322, 918

\bibitem[{{Vicinanza} et~al.(2016){Vicinanza} \& {Cardone} \& {Maoli} \&
  {Scaramella} \& {Er}}]{Vicinanza2016}
{Vicinanza}, M., {Cardone}, V.~F., {Maoli}, R., {Scaramella}, R., {Er}, X.,
  2016, arXiv e-prints, arXiv:1606.03892

\bibitem[{{Vicinanza} et~al.(2018){Vicinanza} \& {Cardone} \& {Maoli} \&
  {Scaramella} \& {Er}}]{Vicinanza2018}
{Vicinanza}, M., {Cardone}, V.~F., {Maoli}, R., {Scaramella}, R., {Er}, X.,
  2018, \prd, 97, 2, 023519

\bibitem[{{Vicinanza} et~al.(2019){Vicinanza} \& {Cardone} \& {Maoli} \&
  {Scaramella} \& {Er} \& {Tereno}}]{Vicinanza2019}
{Vicinanza}, M., {Cardone}, V.~F., {Maoli}, R., {Scaramella}, R., {Er}, X.,
  {Tereno}, I., 2019, \prd, 99, 4, 043534

\bibitem[{{Vikram} et~al.(2015){Vikram} \& {Chang} et~al.}]{Vikram2015}
{Vikram}, V., {Chang}, C., {Jain}, B., et~al., 2015, \prd, 92, 2, 022006

\bibitem[{{Wallis} et~al.(2017){Wallis} \& {McEwen} \& {Kitching} \& {Leistedt}
  \& {Plouviez}}]{Wallis2017}
{Wallis}, C.~G.~R., {McEwen}, J.~D., {Kitching}, T.~D., {Leistedt}, B.,
  {Plouviez}, A., 2017

\bibitem[{{Wild} et~al.(2005){Wild} \& {Peacock} et~al.}]{Wild2005}
{Wild}, V., {Peacock}, J.~A., {Lahav}, O., et~al., 2005, \mnras, 356, 247

\bibitem[{{Xavier} et~al.(2016){Xavier} \& {Abdalla} \&
  {Joachimi}}]{Xavier2016}
{Xavier}, H.~S., {Abdalla}, F.~B., {Joachimi}, B., 2016, \mnras, 459, 3693

\bibitem[{{Zuntz} et~al.(2018){Zuntz} \& {Sheldon} et~al.}]{ZuntzY1}
{Zuntz}, J., {Sheldon}, E., et~al., 2018, \mnras, 481, 1149

\bibitem[{{Z{\"u}rcher} \& et~al.(2021)}]{Zuercher2021b}
{Z{\"u}rcher}, D., et~al., 2021, in prep.

\bibitem[{{Z{\"u}rcher} et~al.(2021){Z{\"u}rcher} \& {Fluri} \& {Sgier} \&
  {Kacprzak} \& {Refregier}}]{Zuercher2021}
{Z{\"u}rcher}, D., {Fluri}, J., {Sgier}, R., {Kacprzak}, T., {Refregier}, A.,
  2021, \jcap, 2021, 1, 028

\end{thebibliography}
\bibliographystyle{mn2e_2author_arxiv_amp.bst}

\end{document}